\documentclass[twocolumn]{aastex631}
\usepackage[T1]{fontenc}
\usepackage[utf8]{inputenc}
\usepackage{lmodern}
\usepackage{amsmath,amsfonts,amssymb}
\usepackage{graphicx,xcolor}
\usepackage{esint}
\usepackage{nicefrac}
\usepackage{mathtools}
\usepackage{marginnote}
\DeclareMathAlphabet{\mathbfsf}{\encodingdefault}{\sfdefault}{bx}{sl} 

\newcommand{\vek}[1]{\boldsymbol{#1}}
\newcommand{\tens}[1]{\mathbfsf{#1}}
\newcommand{\tdq}[2]{\frac{\mathrm{d}#1}{\mathrm{d}#2}}
\newcommand{\pdq}[2]{\frac{\partial #1}{\partial #2}}
\newcommand{\I}{\mathrm{i}}
\newcommand{\e}[1]{\mathrm{e}^{#1}}

\newcommand{\kB}{k_\mathrm{B}}

\shorttitle{COCONUT, a novel MHD coronal model}
\shortauthors{Perri, Leitner et al.}

\begin{document}

\title{COCONUT, a novel fast-converging MHD model for solar corona simulations:\\
I.~Benchmarking and optimization of polytropic solutions}

\author[0000-0002-2137-2896]{Barbara Perri*}
\affiliation{Centre for mathematical Plasma Astrophysics, \\
KU Leuven, 3001 Leuven, Belgium 
}

\author[0000-0003-3792-0452]{Peter Leitner*}
\affiliation{Centre for mathematical Plasma Astrophysics, \\
KU Leuven, 3001 Leuven, Belgium}
\affiliation{Institute of Physics, University of Graz \\
Universit\"atsplatz~5, 8010 Graz, Austria 
}

\author[0000-0003-0874-2669]{Michaela Brchnelova}
\affiliation{Centre for mathematical Plasma Astrophysics, \\
KU Leuven, 3001 Leuven, Belgium}

\author[0000-0002-1986-4496]{Tinatin Baratashvili}
\affiliation{Centre for mathematical Plasma Astrophysics, \\
KU Leuven, 3001 Leuven, Belgium}

\author[0000-0001-9438-9333]{B{\l}a{\.{z}}ej Ku{\'{z}}ma}
\affiliation{Centre for mathematical Plasma Astrophysics, \\
KU Leuven, 3001 Leuven, Belgium}

\author[0000-0002-9425-994X]{Fan Zhang}
\affiliation{Centre for mathematical Plasma Astrophysics, \\
KU Leuven, 3001 Leuven, Belgium}

\author[0000-0003-4017-215X]{Andrea Lani}
\affiliation{Centre for mathematical Plasma Astrophysics, \\
KU Leuven, 3001 Leuven, Belgium}

\author[0000-0002-1743-0651]{Stefaan Poedts}
\affiliation{Centre for mathematical Plasma Astrophysics, \\
KU Leuven, 3001 Leuven, Belgium}
\affiliation{Institute of Physics, University of Maria Curie-Sk{\l}odowska,\\
Pl.\ M.\ Curie-Sk{\l}odowskiej 5, 20-031 Lublin, Poland \\
*These authors contributed equally to this work.}

\correspondingauthor{Barbara Perri}
\email{barbara.perri@kuleuven.be}

\begin{abstract}
We present a novel global 3-D coronal MHD model called COCONUT, polytropic in its first stage and based on a time-implicit backward Euler scheme. Our model boosts run-time performance in comparison with contemporary MHD-solvers based on explicit schemes, which is particularly important when later employed in an operational setting for space weather forecasting. It is data-driven in the sense that we use synoptic maps as inner boundary input for our potential field initialization as well as an inner boundary condition in the further MHD time evolution. The coronal model is developed as part of the EUropean Heliospheric FORecasting Information Asset (EUHFORIA) and will replace the currently employed, more simplistic, empirical Wang-Sheeley-Arge (WSA) model. At $21.5\ R_\odot$ where the solar wind is already supersonic, it is coupled to EUHFORIA's heliospheric model. We validate and benchmark our coronal simulation results with the explicit-scheme Wind-Predict model and find good agreement for idealized limit cases as well as real magnetograms, while obtaining a computational time reduction of up to a factor 3 for simple idealized cases, and up to 35 for realistic configurations, and we demonstrate that the time gained increases with the spatial resolution of the input synoptic map. We also use observations to constrain the model and show that it recovers relevant features such as the position and shape of the streamers (by comparison with eclipse white-light images), the coronal holes (by comparison with EUV images) and the current sheet (by comparison with WSA model at 0.1 AU).
\end{abstract}


\section{Introduction} \label{sec:introduction}
With our increasing exposure to the Sun through our expanding spaceborne technological infrastructure such as communication systems as well as space flight but also earthbound through the vulnerability of our power grids, the need for a prediction of solar storms has gained significant importance within the last decades. The source of these storms has been realized to originate from eruptive events in the solar atmosphere and to be associated with free magnetic energy released during the reconfiguration of its field lines. A deeper understanding of the solar corona, while of fundamental scientific interest in its own regard, is also of vital importance for our civilization's technological progress and expansion into space.

With the increasing availability of accurate measurements of the photospheric magnetic field from instruments like the Helioseismic and Magnetic Imager (HMI) on SDO and the Spectropolarimeter (SP) on Hinode, it has become possible to investigate the coronal magnetic field, which has been accomplished to varying degree of sophistication: from potential field extrapolations and force-free approximations to sophisticated full magnetohydrodynamic (MHD) and multifluid numerical models. The first two are based on the realization that in the low plasma-beta regime non-magnetic forces are negligible and the assumption of a vanishing Lorentz force $\vek j \times \vek B = \vek 0$. This so-called force-free field approximation immediately implies the magnetostatic field equations $(\nabla \times \vek B) \times \vek B = \vek 0$ together with $\nabla \cdot \vek B = 0$. The former relation is either true for a vanishing curl (and thus current) allowing the derivation of the $\vek B$-field from a scalar potential $\phi$, that, together with the zero-convergence condition fulfills the Laplace equation $\Delta \phi = 0$. It is equally realized when the curl of $\vek B$ is aligned to the field itself, $\nabla \times \vek B \parallel \vek B$, i.e. $\nabla \times \vek B = \alpha \vek B$ with a so-called force-free parameter $\alpha$ satisfying $\vek B \cdot \nabla \alpha = 0$. A detailed review of linear (LFFF) and non-linear Force-Free fields (NLFFF) is given by \cite{Wiegelmann2012}. While these two models give  better results than the simpler potential field approximation that lacks accuracy in reproducing the global field structure when compared to TRACE images \citep{Schrijver2005}, we will nonetheless focus on the potential field approximations to compute the magnetic field to first order. This is justified for our purposes in so far as the obtained field solution is only used as an initialization for our MHD coronal model, which anyways converges to the same solution no matter the initial condition, however the better the choice, the faster the convergence should be obtained. Potential-field source surface models have been studied for decades and are routinely computed for observed photospheric field distributions, see e.g.\ \cite{Wiegelmann2004}.

Besides potential- and force-free field models, full 3-D MHD models of the solar corona are crucial for improving our understanding of the interaction between the coronal plasma and its field and the complex dynamics in form of plasma wave propagation, energy transfer and magnetic field reconfiguration. Numerical MHD models might be key in solving the question of the detailed processes giving rise to the coronal heating. From an economic perspective even more pressing than pure scientific inquiry is the need for reliable space weather forecasts to be able to protect our exposed navigation and communication systems from severe damage (see e.g.\ \cite{Hapgood2011,Green2015,Schrijver2015}).

The first analytical solutions of the solar wind were derived by \cite{Parker1958} for the hydrodynamic case and \cite{Weber1967} for the magnetized case; \cite{Sakurai1985} generalized this magnetic solution to 2-D. These solutions are however usually derived for simple configurations, hence calling for models to take over for more realistic solar-like configurations. The first coronal MHD models (e.g.\ \cite{Endler1971,Pneuman1971}) did not incorporate observational data into their boundary conditions but were based on idealized plasma conditions for reproducing helmet streamer configurations. One of the first validations of a polytropic wind based on a dipolar magnetic field is given by \cite{Keppens1999}. Data-driven models have been devised by \cite{Mikic1996b,Usmanov1996}, that for the first time could be validated through eclipse and interplanetary observations \citep{Mikic1996b}. During the Whole Sun Month campaign from August 10 to September 8, 1996 an extensive set of coronal data, both ground- and space based has been collected during solar minimum and used for testing and tuning the data-driven 3-D MHD model of \cite{Linker1999}. The Magnetohydrodynamics Around a Sphere (MAS) code, developed at Predictive Science, Inc.\ to date represents one of the most matured and advanced coronal models based on time-dependent resistive MHD and including comprehensive energy transport mechanisms from radiation and thermal conduction to Alfv\'en wave heating (e.g.\ \cite{Mikic1999,Mikic2018}). Although this model is one of the most advanced in terms of physics and yields very good results compared to observations \citep{Lionello2009}, it is not the only approach available to model the solar corona. 1-D wind models have also proved to be interesting for their capacity of including realistic physics at reasonable computational costs \citep{Lionello2001, Suzuki2005, Grappin2010}, and can be used to map the entire sphere to simulate the lower corona as in the MULTI-VP model \citep{Pinto2017}. Other 3-D MHD models have been developed and improved over the years with systematic comparisons to both in-situ and remote-sensing data for validation, such as AWSoM \citep{vanderHolst2010, vanderHolst2014, Sachdeva2019}, the model from \cite{Usmanov2014, Usmanov2016, Usmanov2018, Chhiber2021} or Wind-Predict \citep{Reville2015, Perri2018, Reville2020, Parenti2022}. Yet other models have focused on boundary conditions to provide the most physical conditions for time-dependent models \citep{Wu2006}, such as in \cite{Yalim2017, Singh2018}. Some of them have even already been included in frameworks dedicated to space weather \citep{Toth2012}.

Global 3-D space weather forecasts on typically large-scale meshes pose a further problem in that they are computationally extraordinarily expensive while converged solutions need to be obtained fast and reliably. Upon the obvious speed-up gain through partitioning heliospheric forecasting into coupled models for the coronal domain of the low corona\footnote{This is possible because for $r \gtrsim 0.1~\mathrm{AU}$ where this coupling boundary is usually set, no information is travelling radially inward towards the Sun because the solar wind plasma is supersonic and super-Alfv\'enic beyond $0.1\ \mathrm{AU}$. As a consequence, the heliospheric model is agnostic of the coronal model and allows for a free choice of the inner coronal model.} and using an approximate empirical model for the first, one is left with the numerical constraints on the simulation run-time. This is where the COOLFluiD platform with its massively parallel architecture, implicit time scheme algorithms for MHD on unstructured meshes, also including more efficient Adaptive Mesh Refinement algorithms currently under development \citep{BENAMEUR2021107700}, prove particularly useful in cutting down significantly on computation time and thus occupying an important niche besides the multitude of existing models of varying degree of model sophistication.

The COOLFluiD data-driven coronal model COolfluid COroNa UnsTructured (COCONUT) we have been developing over the past years, while providing fast stand-alone MHD relaxations based on magnetogram maps in its own right, has been intended from the very beginning to serve as a support model for the EUropean Heliospheric FORecasting Information Asset (EUHFORIA) 2.0 heliospheric space weather forecasting tool \citep{Pomoell2018,Poedts2020}. EUHFORIA combines an MHD solar wind and a coronal mass ejection (CME) evolution model with solar energetic particle (SEP) transport and acceleration models. To date, EUHFORIA employs the empirical Wang-Sheeley-Arge (WSA) model for the corona, which is based on an observed correlation between the solar wind speed and the coronal magnetic field line expansion \citep{Wang1990,Sheeley2017}. This is also the case for the other active forecasting models such as ENLIL \citep{Odstrcil2003} or SUSANOO \citep{Shiota2014}. It has however been shown that coupling to an MHD model instead of the WSA provides better forecasts, especially for specific structures such as the high-speed streams \citep{Samara2021}. The empirical model provides plasma parameters to serve as boundary conditions for the MHD variables of the heliospheric model. In the case of propagating CMEs, the associated transient disturbances of these parameters are passed to the heliospheric model provided by the cone CME model of \cite{Odstrcil1999}. Currently, to compute the fields, a PFSS model is used for the lower corona and a Schatten current sheet (SCS) model in the adjacent upper domain \citep{Schatten1969}, introducing an additional Laplace equation with a vanishing field boundary condition for $r \to \infty$. The SCS extends the magnetic field close to radially outwards while retaining a thin structure for the heliospheric current sheet \citep{Pomoell2018}. Based on the computed 3-D field distribution, a field line tracing for each pixel on the inlet map is performed to determine open and closed field line regions and the associated field in the upper corona, passing through a point $P$, so that the flux tube area expansion factor (e.g.\ \cite{Riley2015}) can be computed as:
\begin{equation}
    f = \left( \frac{R_\odot}{R_P} \right)^2 \left| \frac{B_r(R_\odot,\theta,\varphi)}{B_r(R_P,\theta_P,\varphi_P)} \right|.
\end{equation}
In the empirical model, by assumption the radial solar wind speed at Earth is approximated by the specific functional relation:
\begin{equation}
    V_r(f,d) = V_{r,0} + \frac{V_{r,1}}{(1+f)^\alpha} \left\{1 - 0.8\,\exp [-(d/w)^\beta]\right\}^3,
\end{equation}
of which several are discussed in the literature (e.g.\ \cite{Arge2003,Detman2006,Owens2008,McGregor2011,Wiengarten2014}), with $d$ denoting the great-circle angular distance between an open field footpoint and the closest coronal hole boundary and parameter values as given in \cite{vanderHolst2010,McGregor2011} with the exception of $w=0.02\ \mathrm{rad}$. Based on this solar wind speed estimate, the radial magnetic field component is determined as $B_r = \mathrm{sgn}(B_\mathrm{cor})B_\mathrm{fsw}(V_r/V_\mathrm{fsw})$ with a fast solar wind speed of $V_\mathrm{fsw} = 675\ \mathrm{km/s}$ carrying a magnetic field of $B_\mathrm{fsw} = 3 \times 10^{-3}\ \mathrm{G}$. With this prescription, rather than using $B_r$ from the coronal magnetic field, the so-called open flux problem denoting an underestimation of the field strength in the interplanetary space when deduced from coronal models can be avoided \citep{Linker2017}. The azimuth component $B_\varphi = -(B_r/V_r) R_P \varOmega \sin \theta$ is modeled such that the electric field in a frame co-rotating with angular velocity $\varOmega$ is zero as is required from a steady-state solution in that frame. Setting the number density as $n = n_\mathrm{fsw}(V_\mathrm{fsw}/V_r)^2$, the kinetic energy density on the spherical surface $r=R_P$ is held constant with parameter value $n_\mathrm{fsw} = 300\ \mathrm{cm}^{-3}$ for the fast solar wind. The pressure is set to a constant value of $P=3.3 \times 10^{-8}\ \mathrm{dyn/cm^2}$ on the boundary. Different empirical prescriptions exist for the density and pressure such as empirical fits to in-situ data \citep{Hayashi2003}. The COOLFluiD coronal MHD model presented here is aimed at putting the input parameters that are passed from the empirical to the heliospheric model on a more solid physical basis. In particular, no assumptions have to be imposed on the wind speed profiles or currents but solution and convergence depend solely on observed photospheric fields prescribed as boundary conditions and reasonably good choices for the initialization of the MHD variables. The goal of this paper is to show a benchmarking procedure for COCONUT, both numerical and physical, before demonstrating its run-time advantages for space weather forecasting. For the numerical benchmark, we compare it to a modified version of Wind-Predict which is as close as possible to the COCONUT set-up to check our solutions and quantify how close the two codes can be compared. For the physical benchmark, we have selected several observations that are relevant to compare to our code, and we also compare it to the optimized version of Wind-Predict to see what accuracy can be expected from a polytropic version of the code.

The outline of this paper is as follows: In Sect.~\ref{sec:code} we present COCONUT and the MHD equations in conservation form as used by the solver. We continue to discuss the numerical scheme and discretization in Sect.~\ref{sec:FVM}, and say a few words about the Hyperbolic Divergence Cleaning (HDC) method used to numerically ensure the zero-divergence constraint for the $\vek B$-field. Sect.~\ref{sec:PFSS} gives an outline of the potential field source-surface model and shows a validation of the COCONUT Poisson-solver with a direct implementation of the spherical harmonics solution of the Laplace equation in spherical coordinates, based on a particular photospheric field map as boundary condition. The model equations for the spherical harmonics solution and a rough sketch of their derivation are presented in Appendix~\ref{sec:potential-field-extrapolation}. In Sect.~\ref{sec:BCs} the boundary conditions (BCs) for all the MHD variables are stated and motivated on both ends of the computational domain. We conclude our discussion about the used methods by briefly discussing the mesh generation for the computational spherical shell domain (Sect. \ref{subsec:mesh}) and the numerical details of the Wind-Predict code (Sect. \ref{subsec:wp-description}). The Results section is organized into two paragraphs on idealized limit cases, namely I) the dipole and II) the quadrupole approximation for the magnetic field, which we studied in detail for testing our model. In Sect.~\ref{sec:data-driven-model} we present a case study based on an actual photospheric magnetogram. We then validate the model through a direct comparison with the polytropic Wind-Predict MHD model that we discuss in Sect. \ref{sec:comparison-with-wind-predict}. We also validate the model by comparison with various observations, as discussed in Sect. \ref{sec:validation_obs}. Finally, we demonstrate the timing efficiency of our new code by doing a run-time benchmark for all these different cases in Sect. {\ref{subsec:runtime_bench}}.

\section{Methods} \label{sec:methods}
\subsection{Code \& MHD Model}\label{sec:code}
While coronal models based on the full set of MHD equations constitute a self-consistent description of the magnetic field and the plasma, providing an accurate description of the plasma dynamics, the large CPU requirements for global 3-D models, especially when employing explicit schemes typically demands simplifications to obtain solutions within a reasonable simulation time. Especially for space weather applications, crude approximations such as neglecting the plasma pressure in linear or non-linear force-free models or potential fields are not sufficient while computation time poses a key constraint. To address this problem, we developed a coronal MHD model based on a fully implicit solver for Finite Volume Methods (FVM) on unstructured grids. The solver is part of COOLFluiD (Computational Object-Oriented Libraries for Fluid Dynamics) \citep{Lani2005, Lani2006, Kimpe2005, Lani2013}, a world-class framework for scientific heterogeneous high-performance computing of multi-physics applications, including compressible flows \citep{Vandenhoeck2019}, space re-entry \citep{Panesi2007, Lani2011, Zhang2016}, radiation \citep{Santos2016} and astrophysical plasmas \citep{LaniGPU, Laguna2016, Maneva2017, Laguna2018, Asensio2019}. COOLFluiD has been developed within a collaboration between the Von Karman Institute for Fluid Dynamics and the KU Leuven Centre for mathematical Plasma Astrophysics since 2003. Due to the implicit-in-time nature of the solver, steady-state solutions can be obtained considerably faster than with explicit solvers, allowing Courant-Friedrichs-Lewy (CFL) numbers up to 10,000 in some applications.

We solve the ideal MHD equations in conservation form in Cartesian coordinates:
\begin{multline}
\frac{\partial}{\partial t}\left(\begin{array}{c}
\rho \\
\rho \vek{V} \\
\vek{B} \\
E
\end{array}\right)+\vek{\nabla} \cdot \left(\begin{array}{c}
\rho \vek{V} \\
\rho \vek{V} \vek{V}+\tens I\left(p+\frac{1}{2}|\vek{B}|^{2}\right)-\vek{B} \vek{B} \\
\vek{V} \vek{B}-\vek{B} \vek{V}+\underline{\tens I \phi} \\
\left(E+p+\frac{1}{2}|\vek{B}|^{2}\right) \vek{V}-\vek{B}(\vek{V} \cdot \vek{B})
\end{array}\right) \\ =\left(\begin{array}{c}
0 \\
\rho \vek{g}\\
0 \\
\rho \vek{g} \cdot \vek{V}
\end{array}\right),
\end{multline}
in which ${E}$ is the total energy $\rho \frac{\vek V^2}{2} + \rho \mathcal E + \frac{\vek B^2}{8\pi}$, $\vek{B}$ is the magnetic field, $\vek{V}$ the velocity, $\vek{g}$ the gravitational acceleration, $\rho$ the density, and $p$ is the thermal gas pressure. The gravitational acceleration is given by $\vek{g}(r) = -(G M_\odot/r^2)\, \hat{\vek{e}}_r$ and the identity dyadic $ \tens I = \hat{\vek{e}}_x \otimes \hat{\vek{e}}_x + \hat{\vek{e}}_y \otimes \hat{\vek{e}}_y + \hat{\vek{e}}_z \otimes \hat{\vek{e}}_z$.
The closure is given by the ideal equation of state, thus giving for the internal energy density $\rho \mathcal E = P/(\gamma - 1)$ with a reduced adiabatic index of 1.05 rather than $\gamma = \nicefrac {f+2} f = \nicefrac 5 3$ with number of degrees of freedom $f=3$. While our used value is clearly not physical for a fully ionized gas\footnote{The effect of partial ionization however also reduces the adiabatic index to $\gamma \gtrapprox 1$, see e.g.\ \cite{Aschwanden2005}.}, the model corona effectively becomes approximately isothermal, which is why this procedure can be considered a temporary fix in the model development before further empirical heating terms and Alfv\'en wave pressure are added to the energy and momentum equation, respectively \citep{Mikic1999,Riley2001}. The Hall term has been neglected in the induction equation
We do not include the Coriolis or centrifugal forces as we do not the equations in a co-rotating frame. This choice of not including rotation is motivated by the scope of this study. Indeed, we want for this first study to include as little physics as necessary to better narrow down the sources of difference between the codes during the benchmark. We also only focus on the solar corona where the effect of rotation is limited, both by the size of the computational domain (the Parker spiral angle is only 5.5 degrees at 0.1 AU), and the slow solar rotation (magneto-centrifugal forces are negligible, as shown in \cite{Reville2015}). It is hence a possible approximation for this benchmark that we consider rotation to be negligible.

\subsection{Numerical Finite Volume Scheme}\label{sec:FVM}
The conservation form equation
\begin{equation}\label{FVM1}
    \pdq{\underline{U}(\underline{P})}{t} + \nabla \cdot \underline{F} = \underline{S}
\end{equation}
considers state vectors of conservative and primitive variables $\underline{U}$ and $\underline{P} \in \mathbb{R}^9$, respectively in Cartesian coordinates,
\begin{equation}
    \underline{U} = \begin{pmatrix}
\rho\\
\rho \vek V\\
\boldsymbol B\\
E\\
\psi
\end{pmatrix},
\qquad
\underline{P} = \begin{pmatrix}
\rho\\
\vek V\\
\boldsymbol B\\
P\\
\psi
\end{pmatrix},
\end{equation}
an inviscid flux divergence $\nabla \cdot \underline{F}$ and a source term on the right hand side. While the ideal, hyperbolic conservation equations 
are solved in the conservative variables $\underline{U}$, the solution is mapped $\underline{U} \mapsto \underline{P}(\underline{U})$, stored and updated in the primitive state vector $\underline{P}$.

$\psi$ denotes a Lagrange multiplier that is introduced to numerically ensure the divergence constraint $\nabla \cdot \vek B = 0$. While there exists a number of different methods for that purpose, we use the Artificial Compressibility Analogy \citep{chorin1997}, which is very similar to the Hyperbolic Divergence Cleaning (HDC) method originally developed by \cite{Dedner2002} (but better suited to unstructured meshes and implicit solvers):
\begin{equation}\label{eqn:hdc}
    \begin{rcases}
        \begin{aligned}
            \pdq{\vek B}{t} - \nabla \times \Big(\vek V \times \vek B\Big) + \nabla \psi &= \vek 0\\
            \pdq \psi t + c_h^2 \nabla \cdot \vek B &= - \frac{\tau}{\psi}
        \end{aligned}
    \end{rcases}
\end{equation}
which couples the zero-divergence constraint to the evolution equations
, ensuring that the whole system remains purely hyperbolic. $c_h$ denotes the propagation speed of the numerical divergence error. The term on the right hand side of Eq.~(\ref{eqn:hdc}b) is  a source of parabolic diffusion allowing further damping and control of the divergence error on a timescale $\tau = h/\sigma c_h$ with $h$, the smoothing length and a dimensionless constant $\sigma$ whose values are to be determined empirically and is $\approx 1.0$ in 3-D \citep{Tricco2012}. In the MHD version of COOLFluiD, the parameters $c_h$ and $\tau$ are set to 1 and 0, respectively. For a more detailed discussion about the influence of these parameters for COCONUT, we refer the reader to appendix \ref{app:num_cf_ch}.

For the discretization the computational domain is divided into non-overlapping cells of volume $\varOmega_i$. Then, upon flux splitting into convective and diffusive parts for instance, the application of the Gauss divergence theorem on a cell $\varOmega_i$ gives:
\begin{equation}
    \tdq{}{t} \int_{\varOmega_i}\!\!\! \mathrm{d}V \, \underline{U}(\underline{P}) + \oint_{\partial \varOmega_i}\!\!\! \mathrm{d}\underline {\mathcal S} \cdot \underline F_\mathrm{c} = \oint_{\partial \varOmega_i}\!\!\! \mathrm{d}\underline {\mathcal S} \cdot \underline F_\mathrm{d} + \int_{\varOmega_i} \!\!\! \mathrm{d}V \, \underline{S},
\end{equation}
which in discrete form is:
\begin{equation}
    \tdq {\underline U(\underline P)} t \varOmega_i + \sum_{j \in D_i} \underline H_{ij} \partial \varOmega_{ij} = \sum_{j \in D_i} \underline G_{ij} \partial \varOmega_{ij} + \underline S_i \varOmega_i,
\end{equation}
with inviscid and diffusive fluxes $\underline H_{ij}$ and $\underline G_{ij}$ at the interface $\partial \varOmega_{ij}$ bordering cells $i$ and $j$ and $D_i$ representing the set of neighboring cells to $\varOmega_i$. The state variables are evolved in time using a one-point and three-point implicit Backward Euler scheme for steady and unsteady cases \citep{Yalim}, respectively, solving the resultant linear system with the Generalized Minimal RESidual (GMRES) method  \citep{Saad1986} implemented within the PETSc library \citep{petsc-web-page, petsc-user-ref, petsc-efficient}.

Since the ideal MHD equations are scale independent, i.e. do not depend on length-, magnetic field amplitude- or mass density scale, they are implemented in COCONUT in dimensionless form, see e.g.\ \cite{Goedbloed2019}, by virtue of which they also become numerically more stable by avoiding over- and underflow issues. The following basis set $\{\ell_0,\rho_0,B_0\}$ of code units $Q_0$ is used to adimensionalize any physical quantity $Q$ as $\tilde Q = Q/Q_0$: the unit length $\ell_0 = R_\odot =6.95\times10^{10}\,\rm{cm}$, unit mass density $\rho_0 = \rho_\odot=1.67\times10^{-16}\,\rm{g\,cm^{-3}}$, and $B_0 = 2.2\ \mathrm{G}$, a typical value for the background solar dipole field all represent solar surface values. The triplet of code units also implies independence of time scales $t_0 = \ell_0/V_0$ as these are linked via the Alfv\'en speed $V_{\mathrm{A},0} = B_0/\sqrt{4\pi \rho_0} = V_0$ to $B_0$ and $\rho_0$. All other code units are composed of combinations of the three base units, such as unit pressure $P_0 = \rho_0 V_0^2$ and gravitational acceleration $g_0 = V_0^2/\ell_0$. To arrive at the scale-independent form also the differential operators are adimensionalized as $\nabla = \tilde{\nabla}/\ell_0$ and $\partial_t = \partial_{\tilde{t}}/t_0$. For convenience however, all equations and numerical values presented here are given in dimensional form in cgs units.

\subsection{Poisson solver for the magnetic field initialization}\label{sec:PFSS}
In order to be able to pass an initial condition for the magnetic field distribution to the MHD solver, we compute a potential field approximation based on a particular magnetic map as inner boundary condition.

The origins of the potential field model date back to the 1960s, when it was observed that the interplanetary magnetic field shows a 27-day periodicity and is well aligned to the average photospheric magnetic field and it was thus realized that it has to be of solar origin \citep{Ness1964}. A model, consistent with these observational findings was developed by \cite{Schatten1969} who accounted for three regions with distinctive magnetic field properties: close to the Sun the field dynamics follows the plasma motion in the photosphere. Above the photosphere in region~2, the plasma becomes rapidly diluted with distance from the Sun and the magnetic energy density starts to dominate over the plasma energy density and to define the configuration. To first order the plasma may then be considered force-free $\vek j \times \vek B = 0$ and, by imposing an even stronger restriction on the plasma, even current free, $\nabla \times \vek B = 0$. The force- but not current-free approximation would still allow for field line twisting without affecting the large-scale magnetic structure \citep{Schatten1969}. Considerably higher above, the magnetic energy density falls off below the plasma density so that the field no longer dominates the plasma structure but gets dragged along with the solar wind flow. As was pointed out by \cite{Davis1965}, even before that limit is reached, the field has aligned itself close to the radial direction, such that the transverse magnetic energy density is overtaken by the plasma energy density even sooner. In this region currents eliminate the transverse magnetic field component that is effectively interacting with the plasma. This is where a surface boundary is defined, acting as a source for the interplanetary magnetic field that is carried outwards with the solar wind. Closed field line loops only exist below that surface. It is usually thought of being located somewhere between $0.7$ and $2.5\ R_\odot$ although some studies show that it can be higher up, depending on the open magnetic flux \citep{Reville2015b}. It should be noted however that also its spherical shape is an approximation considerably simplifying the mathematical treatment. Deviations from sphericity and the overall reproducibility of the global magnetic field structure obtained from MHD have been studied by \cite{Riley2006b}. From $r>20\ R_\odot$ the flow is entirely dominated by the solar wind.

Through the force- and current-free approximation in region~2 on which we focus our attention in the coronal modeling, the field can be derived from a scalar potential $\vek B = - \nabla \phi$, which, together with the constraint $\nabla \cdot \vek B = 0$ results in a Laplace equation for the scalar potential, $\nabla^2 \phi = 0$. The well-known general solution in spherical coordinates, together with a Neumann condition at the inner boundary provided by the magnetogram, i.e. a radial magnetic field map at the solar surface, $\left.\nabla \phi(\vek r)\right|_{r=R_\odot} = - B_r(R_\odot, \theta,\varphi) \, \hat{\vek e}_r$ and a source-surface condition, forcing the field to be purely radial at the outer boundary, $\vek B(R_\mathrm{S},\theta,\varphi) = B_r(R_\mathrm{S}, \theta,\varphi)\,\hat{\vek e}_r$ can be expanded in spherical harmonics, giving analytical expressions for the spherical magnetic field components, see App.~\ref{sec:potential-field-extrapolation}.

\begin{figure*}
   \centering
   \gridline{\fig{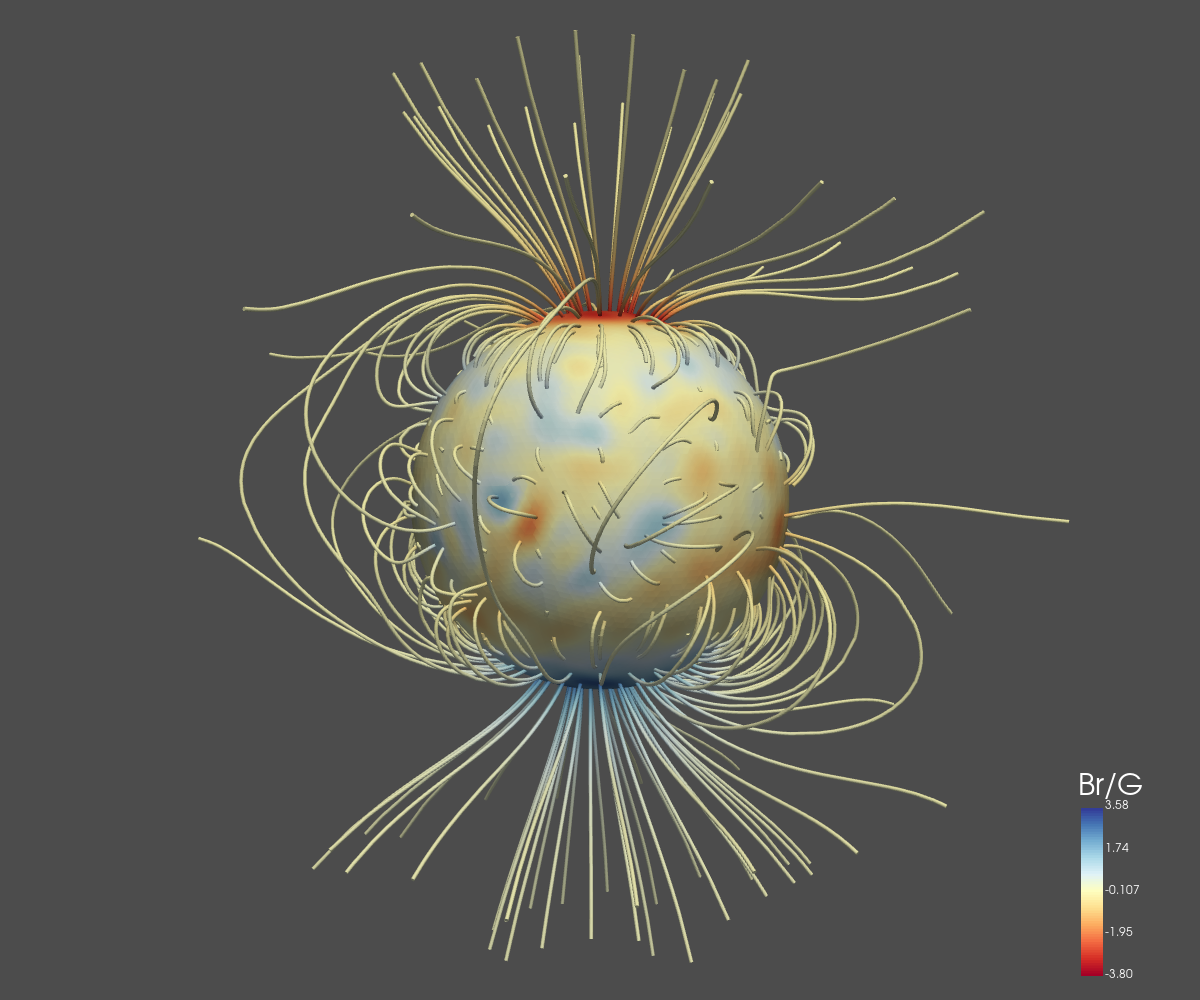}{0.25\textwidth}{(a)}
             \fig{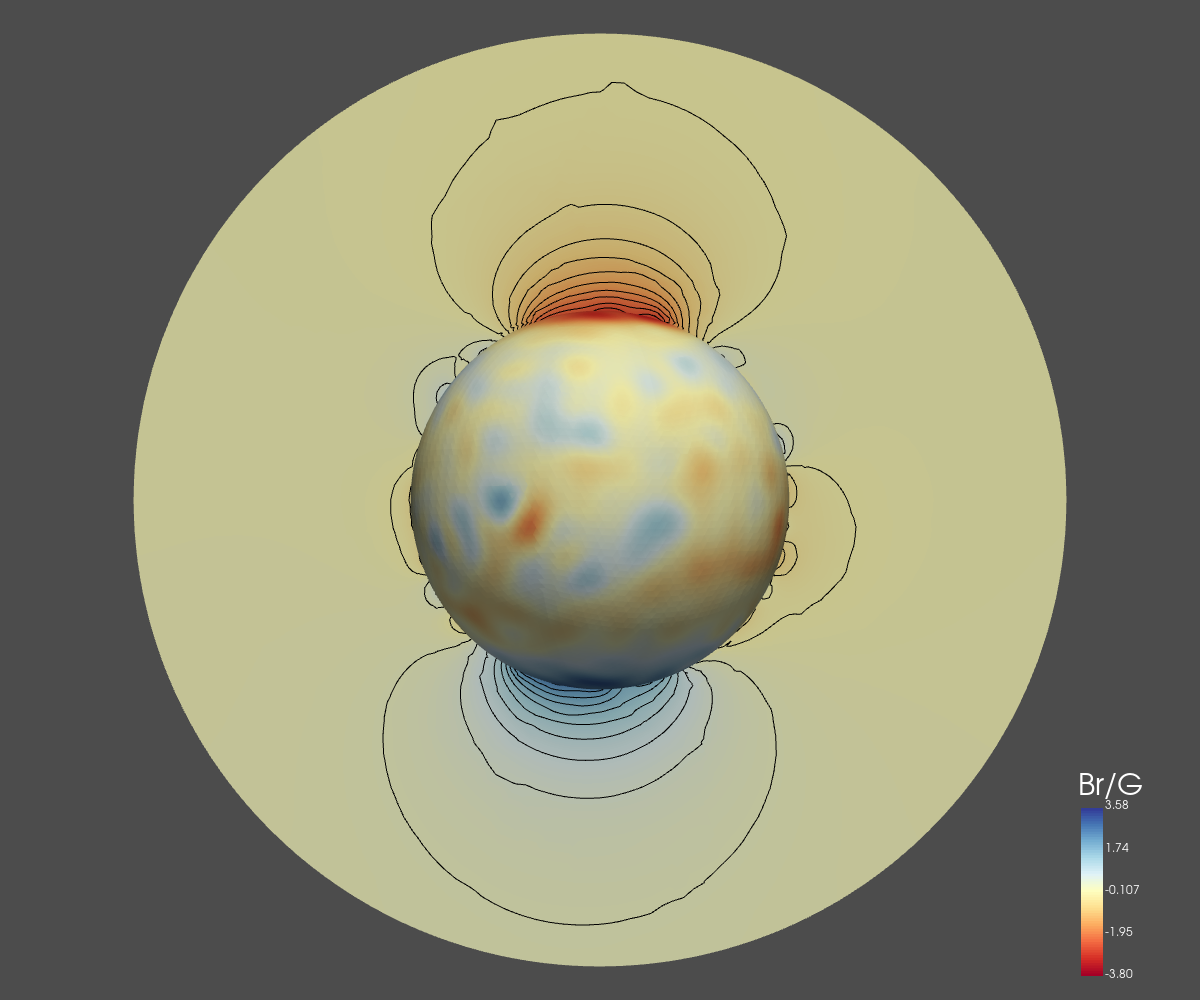}{0.25\textwidth}{(b)}
             \fig{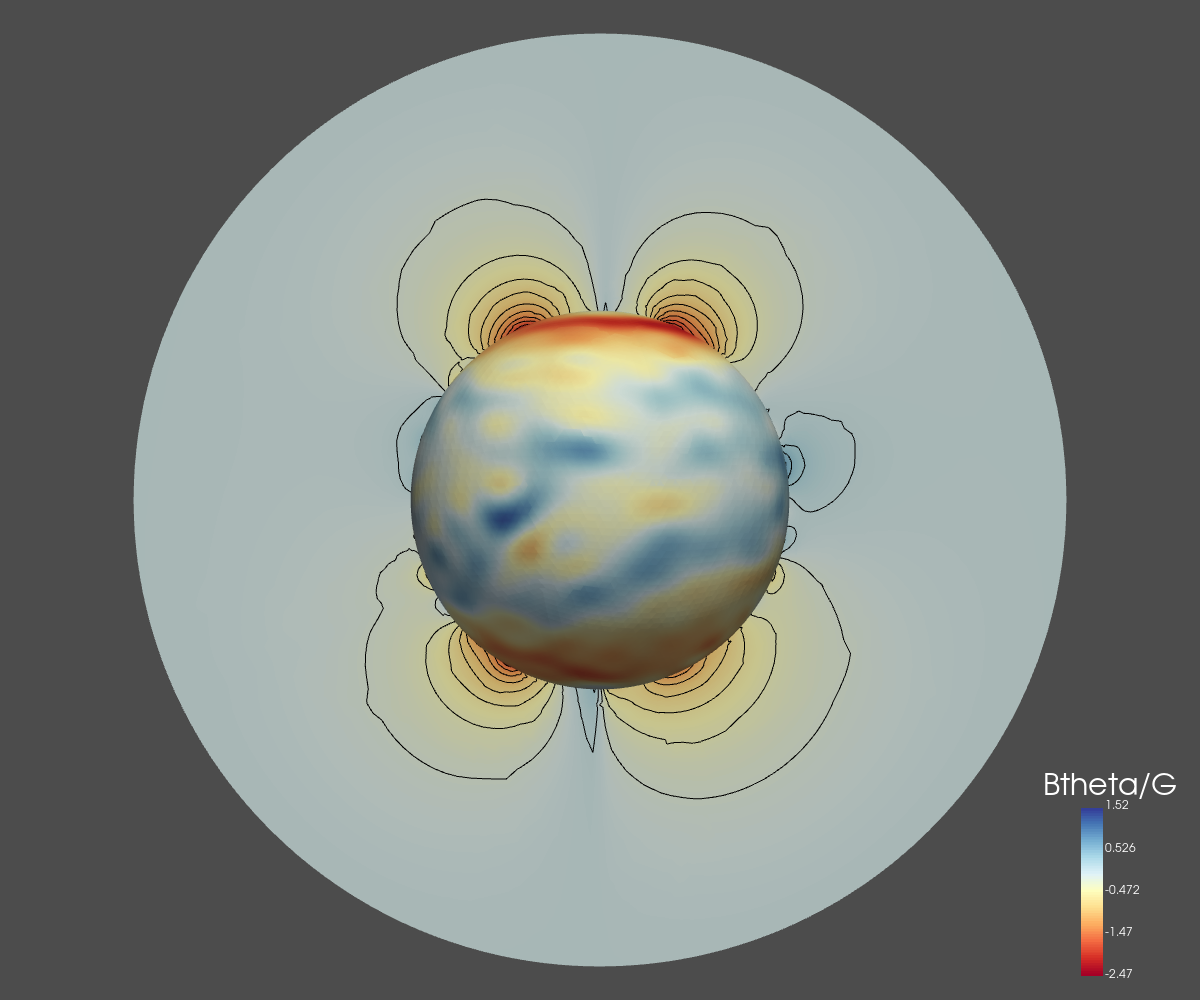}{0.25\textwidth}{(c)}
             \fig{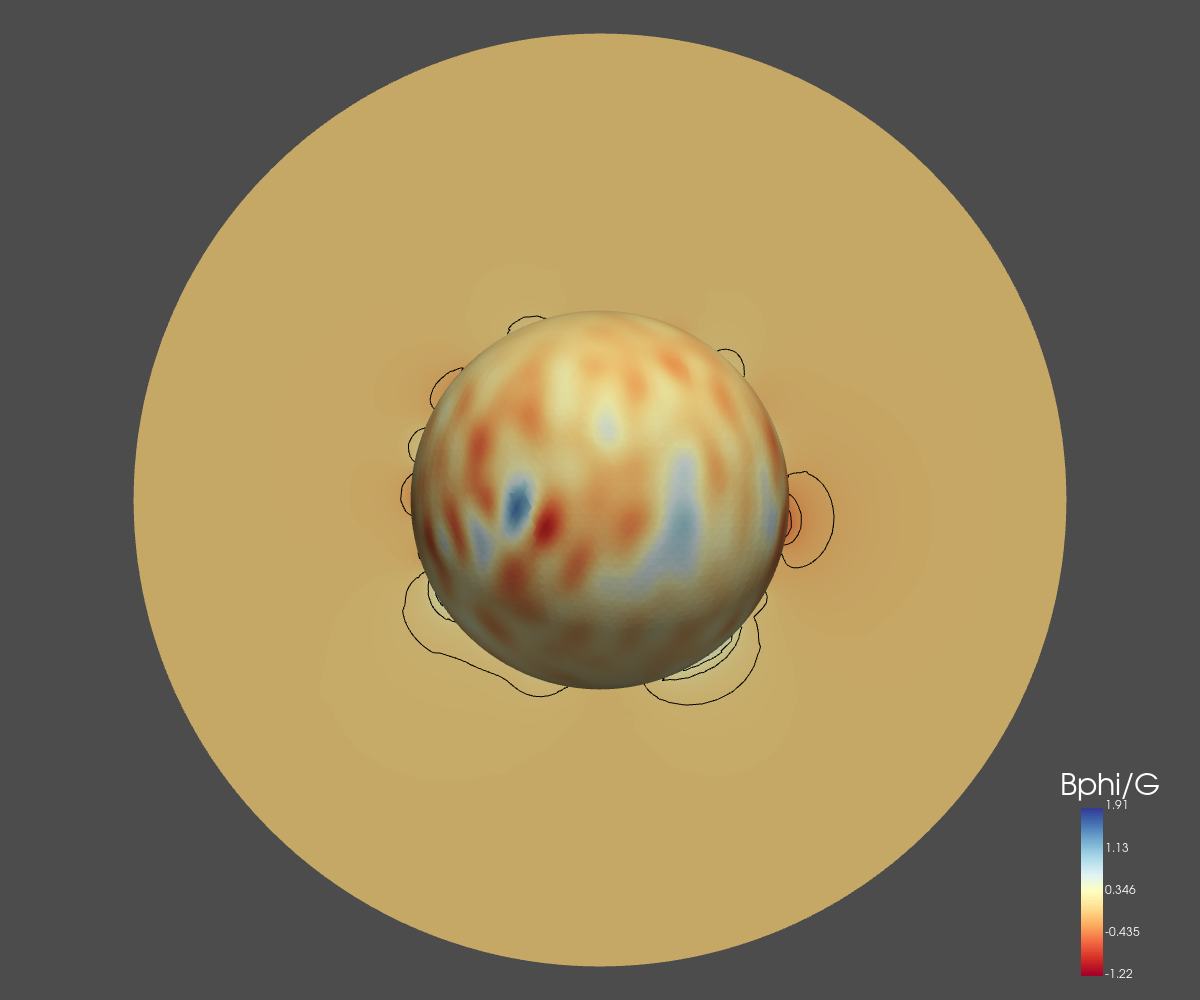}{0.25\textwidth}{(d)}}
   \gridline{\fig{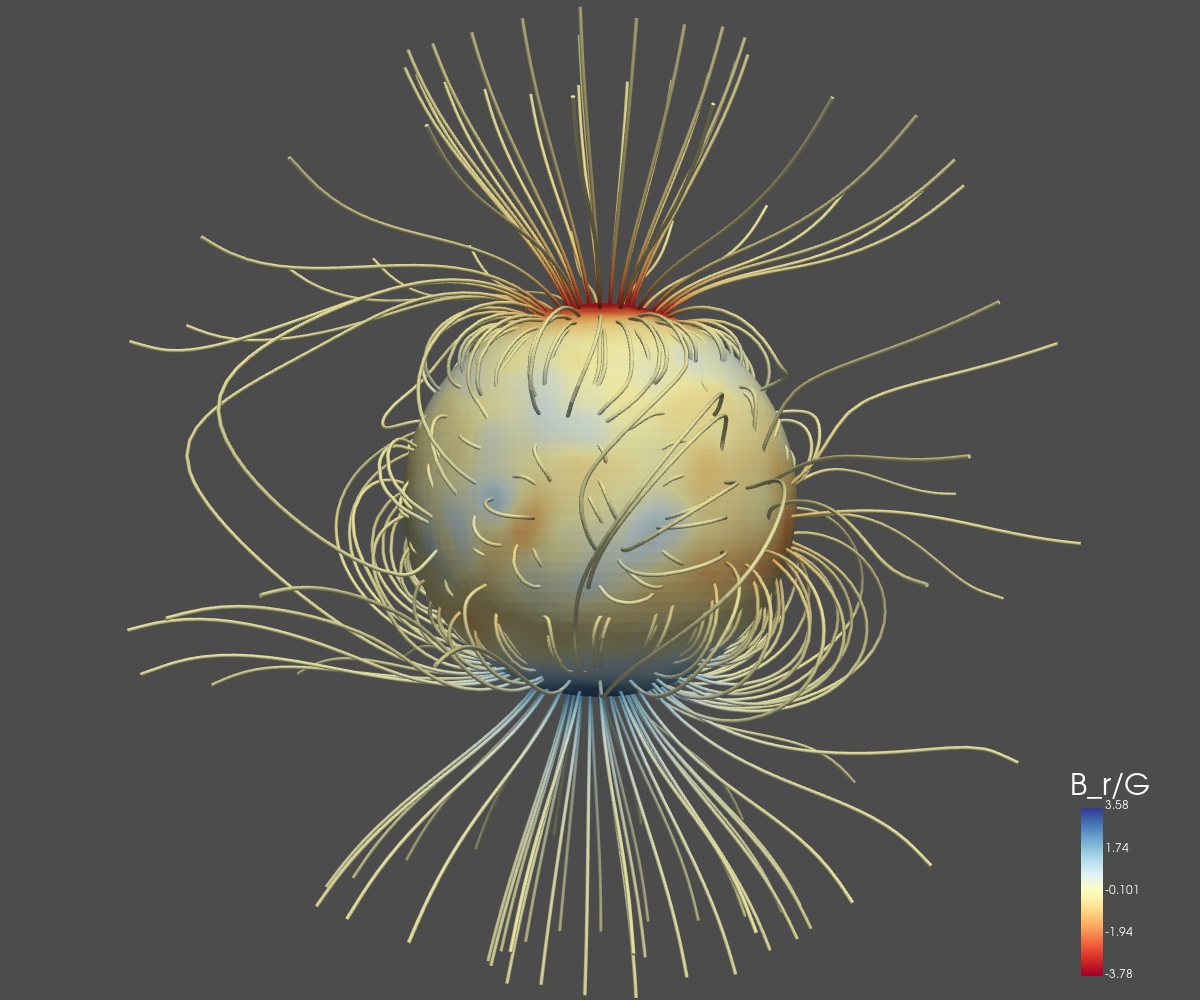}{0.25\textwidth}{(e)}
             \fig{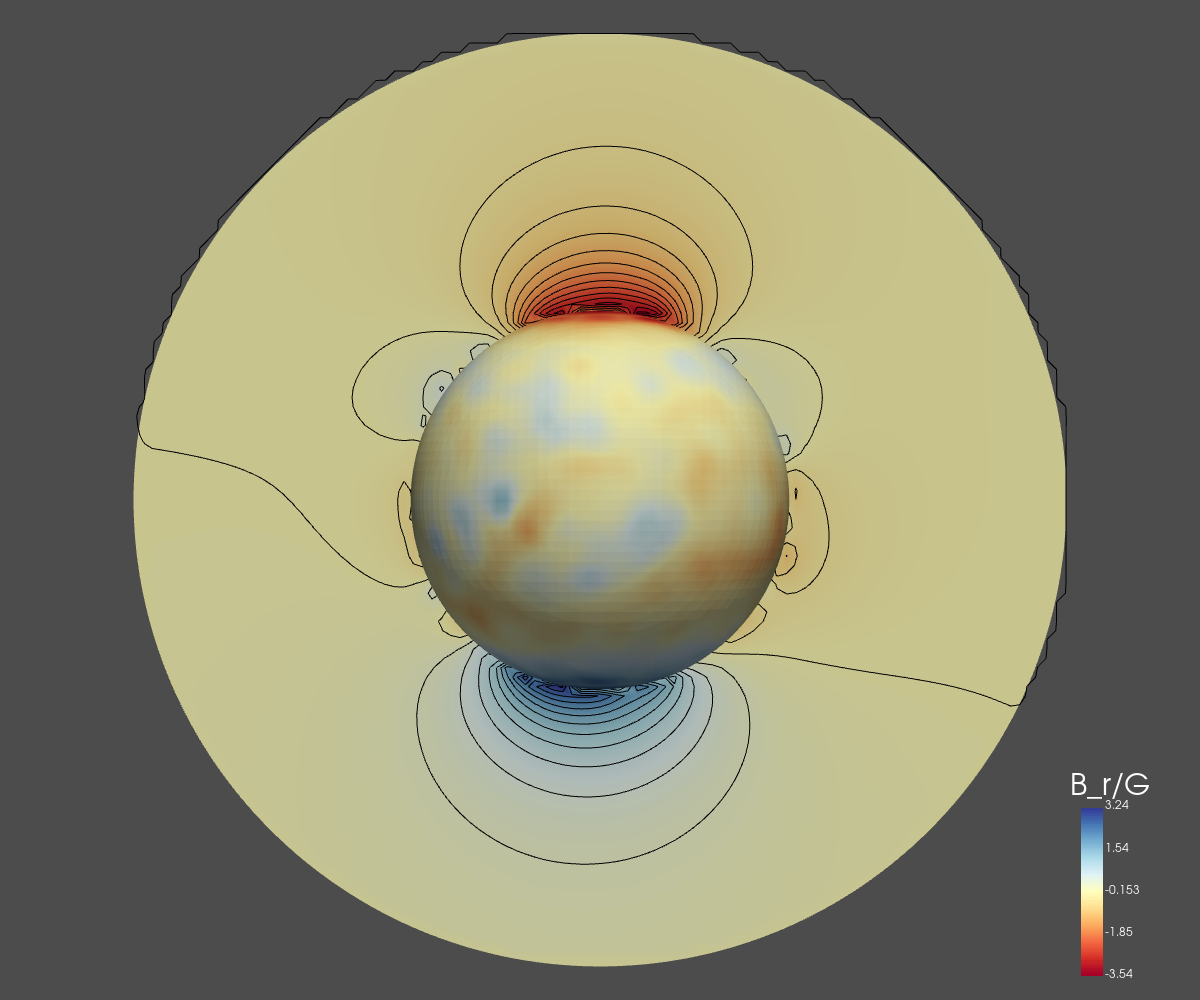}{0.25\textwidth}{(f)}
             \fig{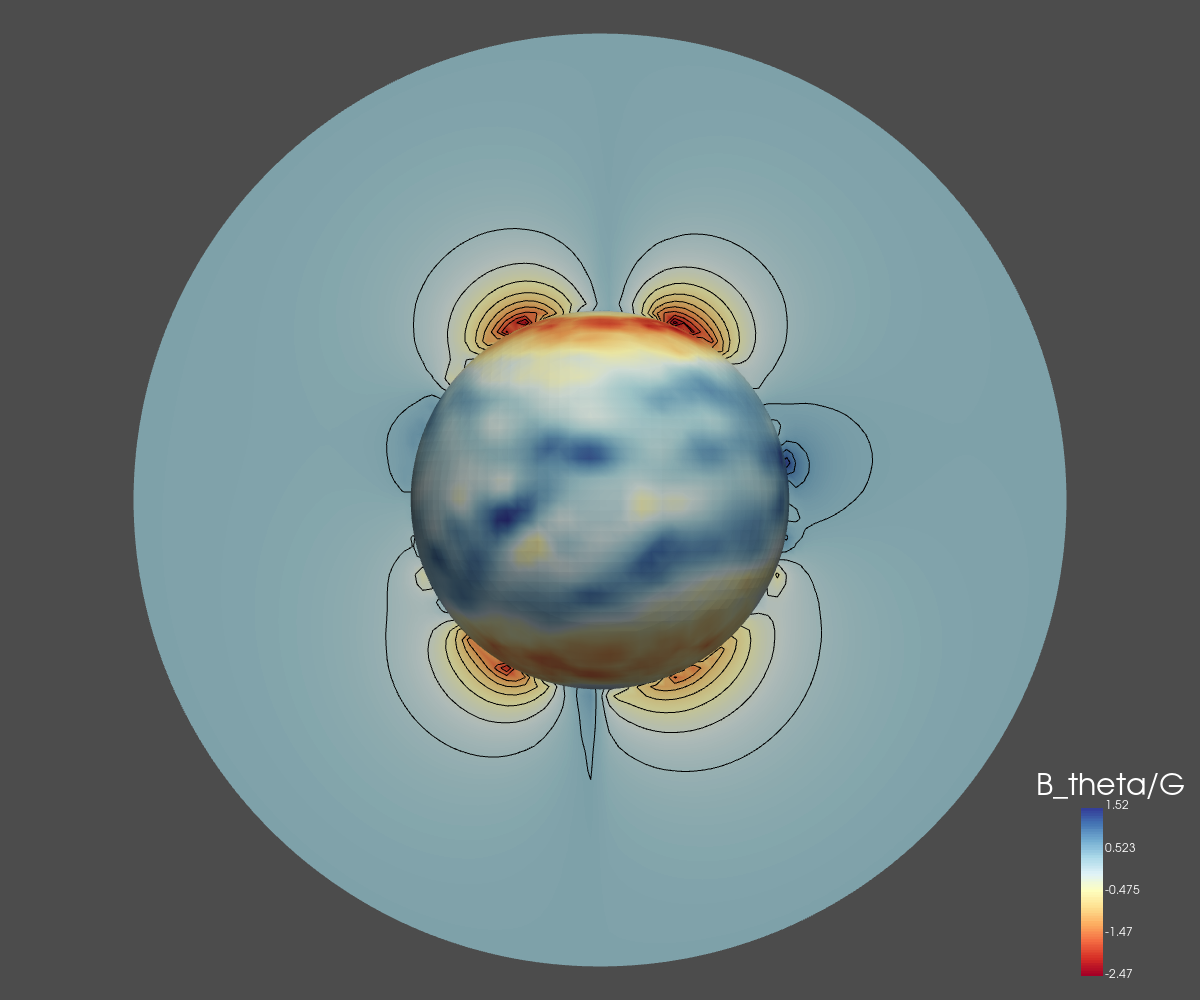}{0.25\textwidth}{(g)}
             \fig{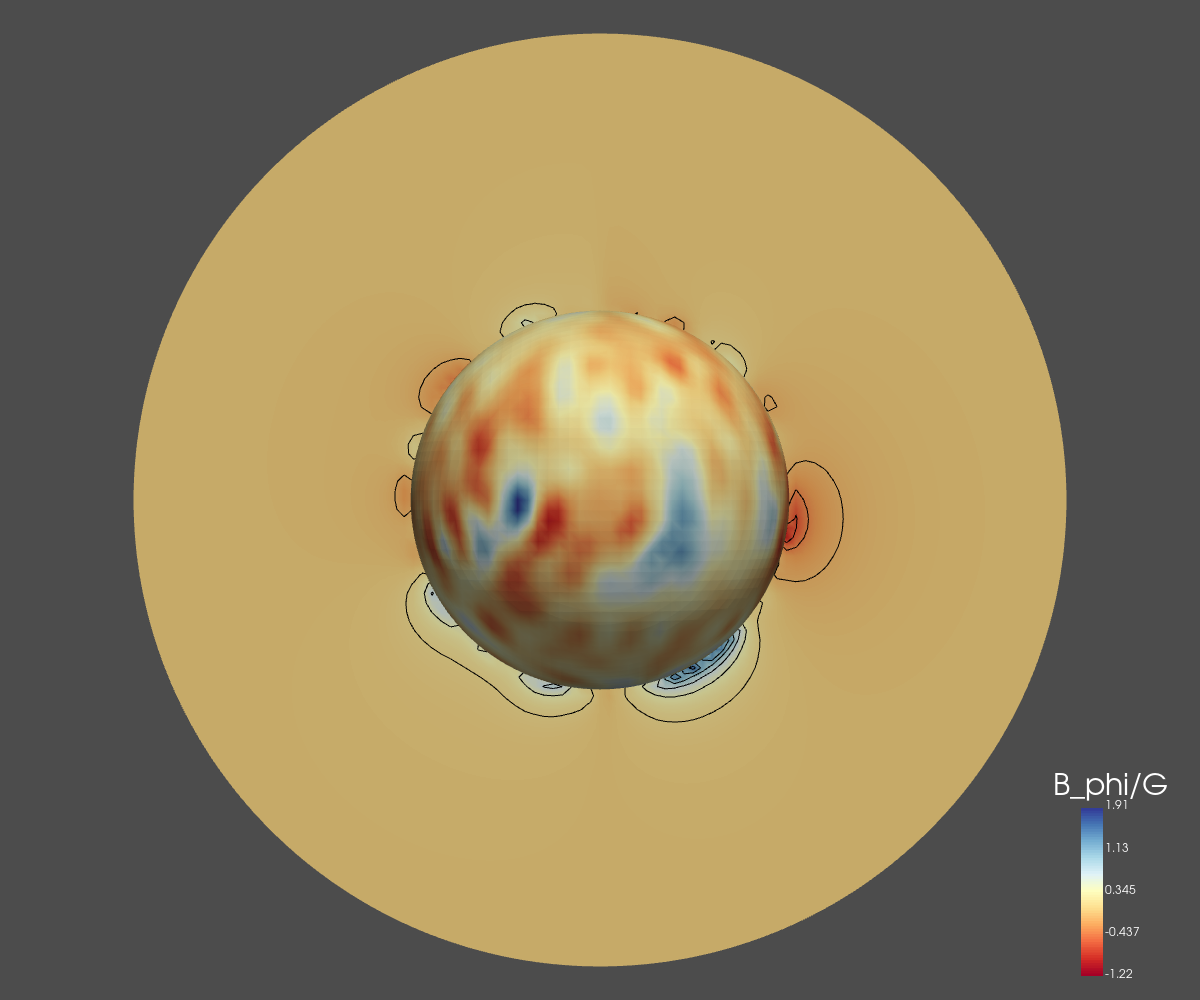}{0.25\textwidth}{(h)}}
   \caption{COCONUT finite volume solution of the Laplace equation (row 1) compared to a straightforward implementation of the spherical harmonics solution presented in App.~A using $\ell_\mathrm{max}=30$ (row 2). In Column 1 we show a field-line tracing based on the field solution. Columns 2, 3 and 4 show the respective radial, polar and azimuthal magnetic field components.}
   \label{fig:spherical-harmonics}
\end{figure*}

Our numerical model is based on a fast Finite Volume (FV) solver for the Laplace equation. For this data-driven part of our model we employ either Global Oscillation Network Group (GONG) maps from the National Solar Observatory or ADAPT\footnote{based on the Air Force Data Assimilative Photospheric Flux Transport model by \cite{Worden2000}} magnetograms. The synoptic GONG magnetograms for instance represent full-surface maps of the radial magnetic flux density with an angular resolution of $1^\circ$ in azimuthal and polar direction.

For validation of the field result, a parallelized PYTHON code, directly evaluating the spherical harmonics expansions Eqns.~(\ref{eqn:spherical-harmonics-Br})-(\ref{eqn:spherical-harmonics-Bphi}) was written, giving good agreement with the considerably faster numerical solutions of the FV Poisson equation solver implemented within COOLFluiD, see Fig.~\ref{fig:spherical-harmonics}. We have selected the most complicated case of this paper (which will be detailed in future sections) with a GONG zero-corrected magnetogram of the 1st of August 2008, including a lot of small structures as the cut-off harmonics frequency has been set to $\ell_{max}=30$.  The two initial solutions show very good agreement, for both the vertical slices and contours, but also the 3D magnetic field lines. The few differences in the contour placements may come from the fact that we must use different grids to do these computations, as COCONUT only works with unstructured meshes while the PFSS code works with regular grids.

\subsection{Boundary conditions}\label{sec:BCs}

Solar wind models use a wide variety of boundary conditions depending on the physical properties that they want to conserve the most \citep{Zanni2013, Yeates2018}. In \cite{Brchnelova2022b}, we have experimented with many of them for the COCONUT model, and have come to the conclusion that the most important aspect was to minimize the currents at the surface of the star, which is achieved through the boundary conditions described below.

From the input synoptic map, we derive a Dirichlet condition based on the radial magnetic field:
\begin{equation}\label{eqn:B-BC-inlet}
    B_{r\mathrm{G}} = 2 B_{r\mathrm{PF}}\Big|_{\partial \varOmega_\mathrm{i}} - B_{r\mathrm{I}}
    .
\end{equation}
Here and in the following index~``G'' is supposed to indicate a value evaluated at a particular ghost cell center, while index ``I'' refers to the corresponding inner cell, adjacent to the ghost cell. The field value at the ghost cell center is assigned such that the exact boundary value at the cell face bordering ghost- and inner state symmetrically, e.g.\ $B_{r\mathrm{PF}}|_{\varOmega_\mathrm{i}}$ is the arithmetic mean of the quantity in question as evaluated on the ghost- and inner state cell centers. $\partial \varOmega_\mathrm{i} = \{(r,\theta,\varphi)|r=R_\odot\}$ denotes the solar surface boundary and $\partial \varOmega_\mathrm{o}$ the outer spherical shell at $r=21.5\ R_\odot$. 
Because the other components of the magnetic field are not derived from data, we use simple zero gradient conditions across the inner boundary ($\partial B_\theta/\partial r = \partial B_\varphi/\partial r = 0$), which is proven effective to reduce the generation of currents at the surface of the star (see \cite{Matt2008, Zanni2009, Reville2015} for more discussions about the $B_\varphi$ conditions and their implications for the model).
The most physical boundary condition would however consist in using all three components of the magnetic field as provided by vector magnetograms (SDO/HMI for example), as well as transverse velocity components $V_\theta$ and $V_\phi$. The number of Dirichlet conditions to use is then determined by the directions of the characteristic waves going in and out of the photosphere \citep{Wu2006, Yalim2017, Singh2018}. The applicability of such methods to unstructured meshes and implicit solvers is however questionable and out of the scope of this paper.

We use typical solar surface values for the mass density $\rho_\odot = 1.67 \times 10^{-16}\ \mathrm{g/cm^3}$ and $T_\odot = 1.9 \times 10^6\ \mathrm{K}$ for fixed-value Dirichlet conditions of density and pressure,
\begin{equation}
    \rho_\mathrm{G} = 2 \rho_\odot - \rho_\mathrm{I}, \quad P_\mathrm{G} = 2 P_\odot - P_\mathrm{I}.
\end{equation}
The pressure at the inner boundary follows from the solar surface temperature by application of the ideal gas law,
\begin{equation}
    P_\odot = \frac{2 \rho_\odot \kB T_\odot}{\mu m_\mathrm{H}} = 4.15 \times 10^{-2} \, \mathrm{dyn/cm^2}
\end{equation}
with $\mu \approx 1.27$ signifying the mean molecular mass in the corona \citep{Aschwanden2005} and factor 2 in the numerator originating from the electron pressure. 
The inner velocity is set to 0 at the inner boundary by following the prescription:
\begin{equation}
    V_{{x,y,z}G} = - V_{{x,y,z}I}.
\end{equation}
This means that we do not include rotation yet in our code (this will be left for development in a following paper). This may seem at first contradictory in order to produce a wind solution and not a breeze solution \citep{Parker1958, Velli1994}. We must however point out that due to numerical precision, the boundary outflow is never exactly 0, which means that our boundary condition is actually equivalent to setting a minimal outflow at the surface of the star. We have also tested aligning the velocity with the magnetic field to limit the generation of currents close to the star, but in the end we selected this boundary condition due to the fact that it helps streamers be better collimated \citep{Brchnelova2022b}. We will show more about that in section \ref{sec:results}.

Due to the solar wind being supersonic at $r = 21.5\ R_\mathrm{S}$, we can extrapolate the spherical field components $r^2 B_r$, $B_\theta$, $B_\varphi$, as well as $\rho$, $V_r$, $V_\theta$, $V_\varphi$ and $P$ from the outermost cell centers to the ghost cells with a zero gradient. We extrapolate $r^2 B_r$ instead of $B_r$ to comply with the divergence-free constraint for the magnetic field (see \cite{Perri2018} for more details).

\subsection{Meshing of the spherical shell domain}
\label{subsec:mesh}

The mesh used for the simulation is a spherical shell domain defined by:

\begin{equation*}
 \varOmega = \{(r,\theta,\varphi)|R_\odot < r < 21.5 R_\odot\}    
\end{equation*}
 
\noindent where the inner BC is applied on $r = R_\odot$ and the outer BC on $r = 21.5R_\odot$. 

Since we operate close to the hydrostatic equilibrium and the currently applied scheme is not well-balanced, spurious fluxes due to numerical errors might appear in the solution if the mesh is not sufficiently uniform (see e.g.\ \cite{Fuchs}, \cite{Popov} and \cite{Krause}). Creating a perfectly uniform spherical mesh is not possible with the conventional meshing technique in which the spherical surface is discretised according to a given number of lines of latitude and longitude (the so-called UV mapping). This is due to the fact that in that case, degenerate regions near the poles arise, which, on the spherical surface, form triangles instead of quadrilaterals. Several different mesh topologies were tested to evaluate which one of them has the optimum performance in terms of accuracy and ease of convergence. In the end, the mesh used was derived from a geodesic polyhedron. The surface mesh of a level-6 subdivided geodesic polyhedron (consisting of triangular elements) was generated to represent the inner boundary and then extended radially outwards in layers until the outer boundary was reached, resulting in a 3-D domain consisting of prism elements. The default mesh used a 6th-level subdivision of the geodesic polyhedron with 20,480 surface elements, resulting in a grid with 3.9M elements which matched the radial discretization of the Wind-Predict simulations. One advantage of this mesh is that it does not produce any polar singularity contrary to most spherical structured meshes, as will be discussed in further details in Sect.~\ref{sec:comparison-with-wind-predict}. The mesh topology more similar to the Wind-Predict simulation, defined according to lines of latitude and longitude, was also tested, in which the degenerate polar prism elements were turned into hexahedrons of very high aspect ratio. While these simulations generally had worse convergence, they proved that the final results were mostly mesh-topology-independent and thus that the differences observed between the CF and Wind-Predict solutions were due to other numerical aspects. For more details about the impact of the mesh on the solution, see \cite{Brchnelova2022JPlPh}.

\begin{figure*}[!t]
    \centering
    \gridline{\fig{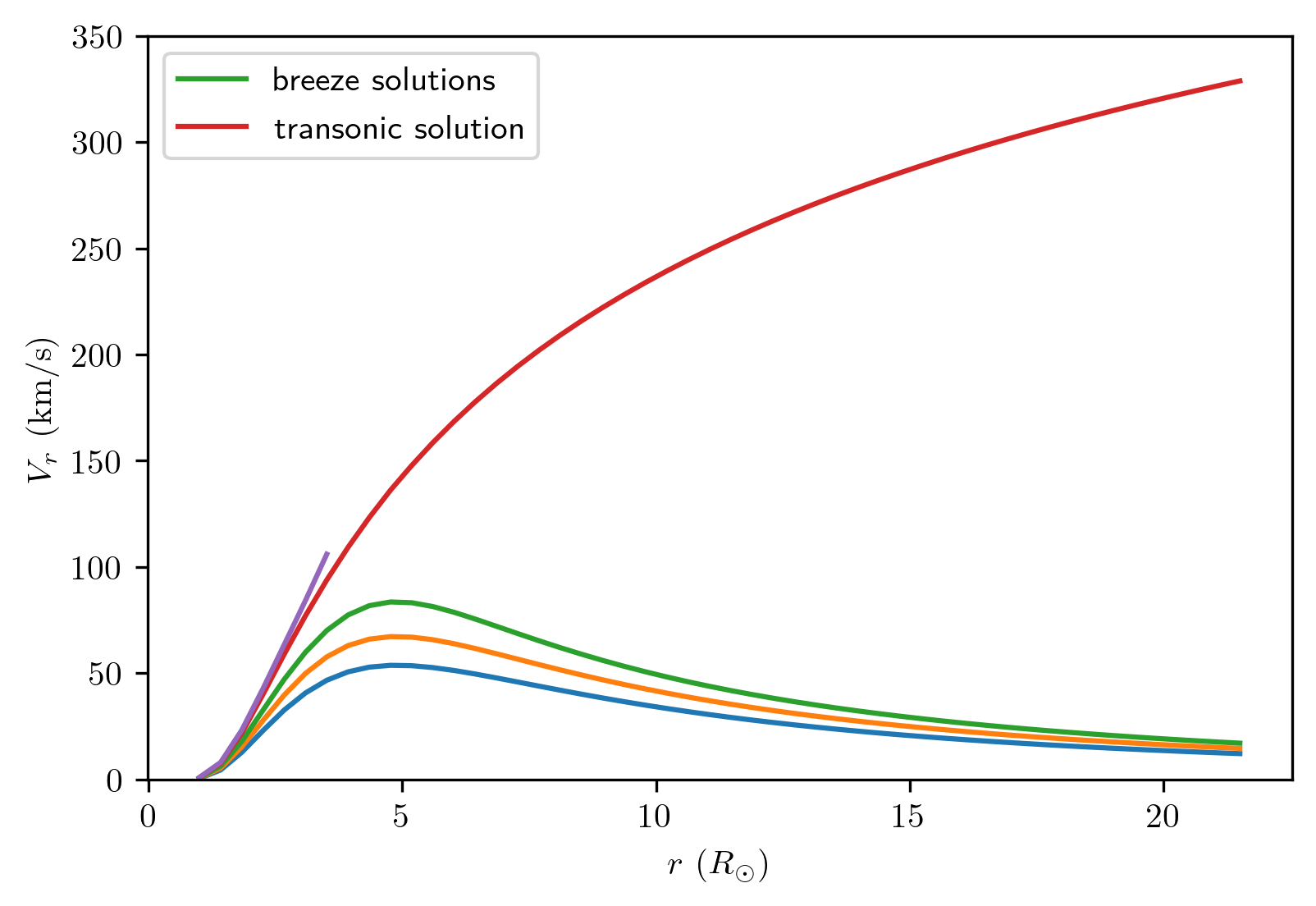}{0.49\textwidth}{(a)}
        \fig{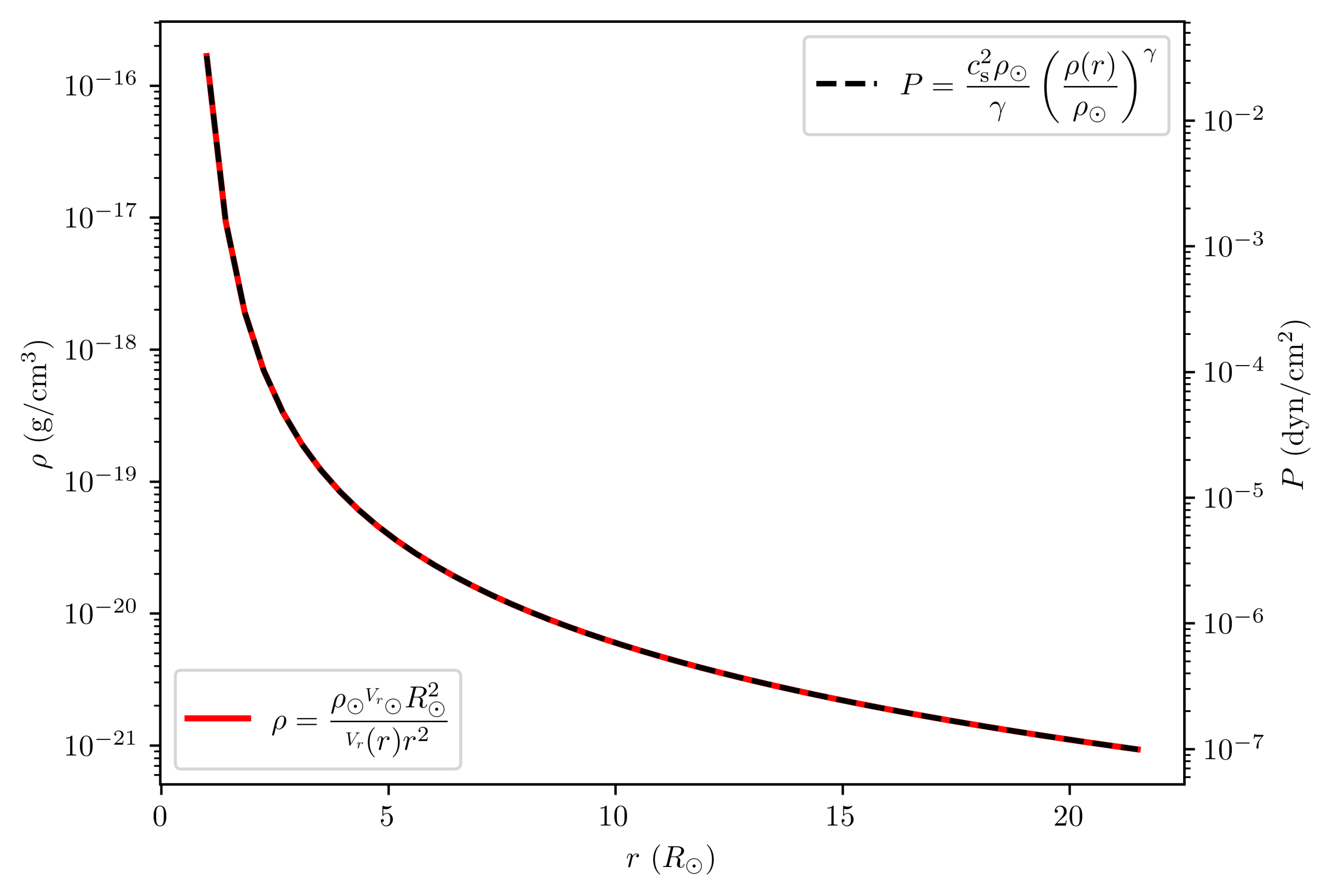}{0.49\textwidth}{(b)}}
    \caption{Solar wind initial conditions: (a) Numerical solution of the Parker solar wind differential equation. The transonic solution at the critical flow speed is used to initialize the initially isotropic velocity field $\vek V(r) = V_\mathrm{Parker}(r)\ \hat{\vek e}_r$. (b) Based on the Parker solution $V(r)$, the initial density profile is evaluated satisfying the continuity equation. Finally, a polytropic initial profile was chosen for the pressure.}
    \label{fig:Parker}
\end{figure*}

\subsection{Wind-Predict polytropic model}
\label{subsec:wp-description}

We will explain briefly here the main characteristics of the Wind-Predict model which is used extensively in this paper to provide validation for the solutions found by COCONUT. For more details, readers can refer themselves to the following papers \citep{Reville2015, Perri2018}. The Wind-Predict model is based on the PLUTO code which is a multi-physics and multi-solver code \citep{Mignone2007}. The model solves the set of ideal MHD equations in conservative form. It uses its own normalization to adimensionalize the equations by setting $R_0=R_\odot=6.96\times10^{10}\,\rm{cm}$, $\rho_0=\rho_\odot=1.67\times10^{-16}\,\rm{g\,cm^{-3}}$ and $V_0=V_{\mathrm{kep},\odot}=4.37\times 10^7\,\rm{cm\,s^{-1}}$ (Keplerian speed at the solar surface). To match as closely as possible the results of COCONUT, we have removed some options that we use normally, such as the inclusion of rotation or a background field decomposition. The simulation is then controlled by the temperature of the corona, set by the adiabatic index $\gamma=1.05$ and the sound speed normalized by the escape velocity $c_\mathrm{s}/V_\mathrm{esc}=0.261$.
We use the HLL solver \citep{Einfeldt1988} in addition to a finite-volume method in space, and a Runge-Kutta scheme in time. Also here, a generalized Lagrange multiplier to enforce the divergence-free property of the magnetic field \citep{Dedner2002} is applied. We use a limiter for the piecewise TVD linear reconstruction of the primitive variables, more specifically the monotonized central difference limiter (which is the least diffusive available in PLUTO).
The polytropic version is based on spherical coordinates, the numerical domain being a spherical shell with radii within $r\in[1.0,20.0]\,R_\odot$, polar angle $\theta\in[0,\pi]$ and azimuth $\varphi\in[0,2\pi]$. The grid is stretched in radial direction with an initial grid spacing of $\delta r/R_\odot=0.001$ at the solar surface. 
For the outer boundary conditions, they are set to outflow except for the magnetic field where $r^2B_r$ is continuous. In the optimized version of Wind-Predict, the inner boundary conditions are different from the ones for COCONUT: the density and pressure are fixed, the rotation is set to the solar mean value, the poloidal speed is aligned on the poloidal magnetic field and $\partial_r^2 B_{\varphi}=0$. In addition, the two first layers of the computational domain have also the condition $V_\mathrm{pol}\parallel B_\mathrm{pol}$ enforced for stability.

\section{Results} \label{sec:results}

During the early stage of the model development, we tested our model against simple limit cases, more specifically the dipole and quadrupole approximation of the solar magnetic field, for which solutions are well known. A first preliminary test that our coronal model had to pass successfully before even considering full MHD scenarios, is the hydrodynamic limit with zero magnetic field prescribed at the inner boundary. We shortly discuss the initial conditions applied there, as they are also employed in the more general scenarios discussed in detail in the following sections, for which also benchmarks and model comparisons are provided.

The initial radial flow velocity is based on a numerical solution of Parker's isothermal solar wind model,
revealing three solution branches depending on the numerical value $V_r(R_\odot)$ of the initial value problem. 
In the left panel of Fig.~\ref{fig:Parker}, a family of curves representing the radial steady flow solution is shown. The breeze solutions at the bottom are decelerated by a shock wave. A supersonic solution at the critical speed separates the subsonic and nonphysical flows \citep{Cargill1980}. With the Parker velocity profile $V_r(r)$ at hand, one can evaluate the initial density profile in such a manner that the continuity equation is fulfilled \citep{Perri2018}:
\begin{equation}
    \rho(r) = \frac{\rho_\odot V_r(R_\odot) R_\odot^2}{V_r(r) r^2}.
\end{equation}
We further assume an isotropic polytropic pressure profile:
\begin{equation}
    P(r) = \frac{c_\mathrm{s}^2 \rho_\odot}{\gamma} \left( \frac{\rho(r)}{\rho_\odot} \right)^\gamma = \frac{2 \rho_\odot \kB T_\odot}{\mu m_\mathrm{H}} \left( \frac{\rho(r)}{\rho_\odot} \right)^{2\gamma-1},
\end{equation}
with speed of sound $c_\mathrm{s} = \sqrt{2 \gamma \kB T/\mu m_\mathrm{H}}$. With this preparatory check undertaken, we focus in the following on important idealized limit test cases for coronal models, namely the magnetic dipole and quadrupole, for which we also draw a detailed comparison with the Wind-Predict model. For all the results presented in this section, we have tried to make the two codes as similar as possible (same physical parameters, same boundary conditions, same scheme) to check how well they can compare.

\begin{figure*}[!t]
    \centering
    \gridline{\fig{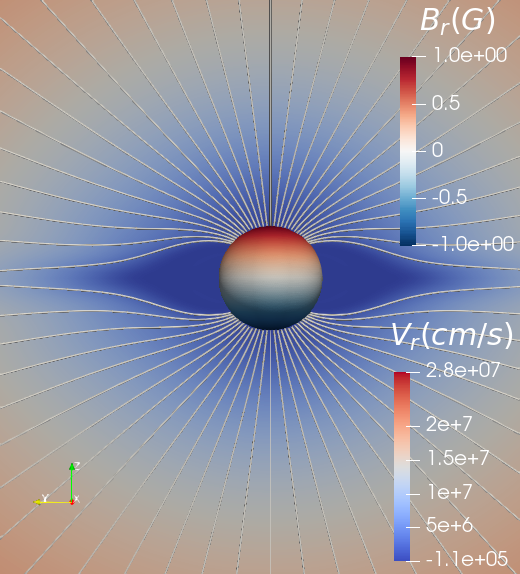}{0.49\textwidth}{(a)}
              \fig{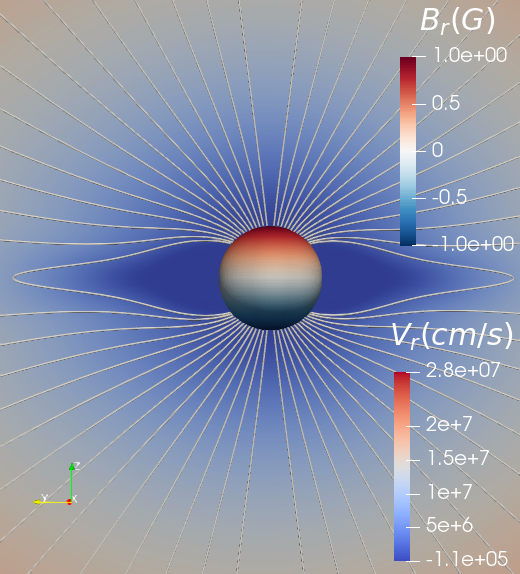}{0.49\textwidth}{(b)}}
    \caption{Comparison for the dipole between the COCONUT solution (a) and the Wind-Predict solution (b). The colorscale shows the radial velocity $V_r$ displayed in the background in $\rm{cm.s^{-1}}$ (colorscale on the bottom-right). The surface of the star is shown via a spherical slice showing the radial magnetic field $B_r$ in G (colorscale on the top-right). Magnetic field lines are traced as white tubes.}
    \label{fig:dip}
\end{figure*}

\subsection{Physical limit cases I: Dipole approximation of the solar magnetic field} \label{sec:dipole-approximation}

We approximate the solar magnetic field by a simple dipole expression
$\vek B = \nicefrac{1}{r^5} [ 3 (\vek m \cdot \vek r) \vek r - r^2 \vek m]$ with $\vek m = \nicefrac 1 {2c} \int \mathrm{d}V'\, \vek r' \times \vek j(\vek r')$, the magnetic dipole moment. An observational upper limit for the solar magnetic dipole moment is e.g.\ given by $m < 5 \times 10^{32}\ \mathrm{G\,cm^3}$ \citep{Treimann1954}.

The results for the dipole are shown in Fig.~\ref{fig:dip}. The comparison with Wind-Predict begins to become essential at this point because there is no general analytical solution to an MHD wind relaxing with a dipolar magnetic field (though some comparisons can be made with the 1-D Weber \& Davies and 2-D Sakurai models for open field regions, see \cite{Perri2018} for examples). 
The lower color map shows the radial velocity in cgs units, corresponding to the meridional slice displayed in the picture. The upper color map shows the radial magnetic field in units of G, corresponding to the spherical slice displayed on the picture. This allows us to show the solar surface, characterized here by the input magnetic field derived from the input synoptic map. The white tubes correspond to the magnetic field lines, selected by a prescribed number of source seeds implemented following a sphere with a prescribed radius. Due to the fact that the two codes use different geometries on different meshes, it is difficult to provide exactly the same location for the source seeds. We have however selected the most relevant seedpoints in the meridional plane to get similar magnetic field lines. We recover the expected structure for a dipolar solution, described for example in \cite{Keppens1999} and being close to a solar minimum of activity \citep{McComas2008}. In this configuration, we have two helmet streamers at the equator where the magnetic field lines are closed and causes the wind to slow down. On the contrary, at the poles we can find open field lines\footnote{As split magnetic monopoles do not exist, this means that they reconnect outside the computational domain.} associated with a faster wind. The amplitude reached (around 300 km/s at 20 $R_\odot$) would correspond to a 500 km/s wind at Earth distance, thus corresponding more to the slow component of the wind, as is expected from a polytropic model at this temperature. Although we have the same initial and boundary conditions for the two codes, we can notice a few differences in the final results. For example, the maximum value of the velocity reached is different, as can be seen in the meridional slice with the velocity reaching higher values sooner in the case of COCONUT. We can also see that the streamers have a slightly different shape, appearing more sharpened and contracted in COCONUT. We will discuss the differences between the two models and explain their causes in more details in Sect.~\ref{sec:comparison-with-wind-predict}. We can also notice some negative values for $V_r$, which is expected in a coronal model due to the interaction between the Riemann solver and the boundary conditions used (such negative values occur only very close to the star surface while the radial velocity is positive for the most part of the domain as physically sensible and to be expected).

\subsection{Physical limit cases II: A synoptic quadrupole magnetic field}
\label{sec:quadrupole}

The quadrupolar magnetic field can be derived from the second spherical harmonic mode of the magnetic field (like in \cite{Reville2015} for instance) and thus is expressed by the following analytical formula:
\begin{equation}
B_r(r,\theta,\varphi) = \frac{B_m}{2}\frac{\left(3\rm{cos}^2 \theta - 1 \right)}{r^4}.
\end{equation}
We set the amplitude $B_m$ so that the maximum amplitude reached by the quadrupole (which is going to be at mid-latitudes in the two hemispheres, i.e. at $45^\circ$ and $135^\circ$) is 1 G. 

\begin{figure*}
    \centering
    \gridline{\fig{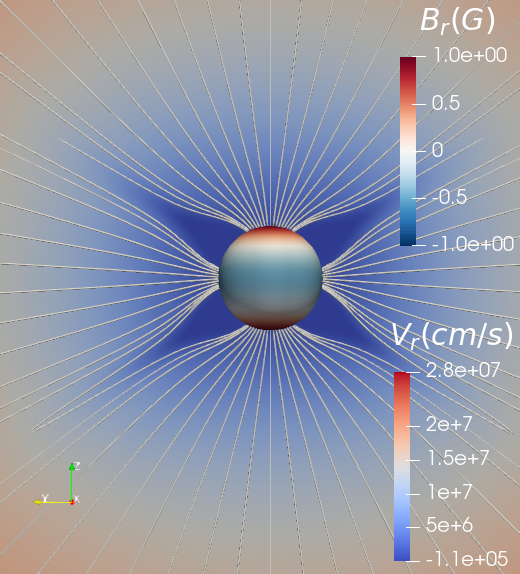}{0.45\textwidth}{(a)}
              \fig{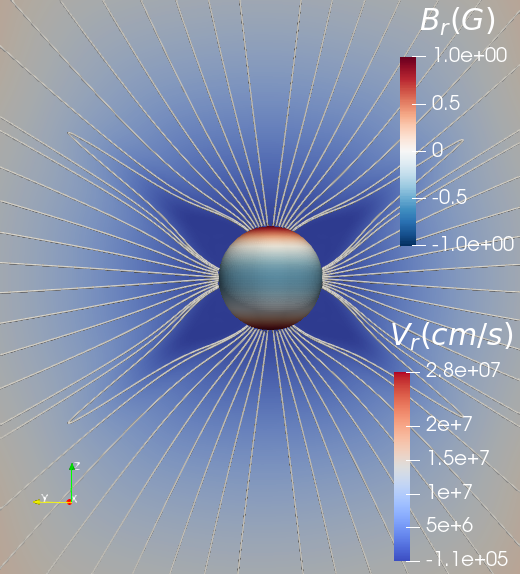}{0.45\textwidth}{(b)}}
    \caption{Comparison for the quadrupole between the COCONUT solution (a) and the Wind-Predict solution (b). The color scale shows the radial velocity $V_r$ in units of $\rm{cm\ s^{-1}}$. The surface of the star is shown via a spherical slice showing the radial magnetic field $B_r$ in units of G. Magnetic field lines are traced in white.}
    \label{fig:quad}
\end{figure*}

The results are shown in Fig.~\ref{fig:quad}. This case illustrates the ability of the code to adapt to more complex topologies, which is going to be essential for realistic solar configurations. Because of the new geometry, the positions of the streamers are now relocated at mid-latitudes, and their extension is smaller (up to 3 solar radii for the dipole, here rather 1.5). The same differences appear again between COCONUT and Wind-Predict, with the outflow velocity being slightly different and the shape of the streamers being slightly altered. This proves that this was not an effect due only to the dipolar configuration, but something more generalized that has to do with the numerics of the codes. Once again, this is going to be discussed in more details in Sect.~\ref{sec:comparison-with-wind-predict} once we have presented all the reference cases. 

\subsection{The general case: Data-driven coronal models}\label{sec:data-driven-model}

To be able to do space-weather forecasts, our new coronal model has to be data-driven to be adjusted to the latest data received by the satellites monitoring the Sun. As explained before, we have added this feature to the model and present here an example of the relaxed state reached for such a synoptic map test case. We have selected the zero-corrected GONG synoptic map corresponding to the 1st of August 2008. This specific map was selected for a certain number of reasons. First, because it is a minimum of activity which is known for being quiet, thus guaranteeing good convergence without active regions, whose modeling and consequence on the convergence will be dealt in a following paper. But it is also known for not being completely dipolar, 
thus making it an interesting test case for our model. 
It has also been extensively studied for comparison of coronal models in other studies \citep{rusin2010, wiegelmann2017}.
It is finally close to the time intervals selected by the International Space Weather Action Team (ISWAT) COSPAR initiative to validate ambient wind models. The input magnetogram is pre-processed by a projection on spherical harmonics and a selection of a maximum frequency for the reconstruction (noted as $\ell_{\rm{max}}$ and fixed to 15 in this case), which is equivalent to a smoothing of the map to remove the small intense structures.
These structures are indeed numerically more challenging, while their contribution to the overall structure of the wind at 0.1 AU is not clear: for the velocities, they can have a strong impact on the distribution between slow and fast wind (but this is not very relevant for a polytropic wind); for the magnetic field, the dipolar mode is going to become more and more dominant further away from the star, thus reducing the impact of small scale magnetic structures \citep{Samara2021, Stansby2021}.

\begin{figure*}
    \centering
    \gridline{\fig{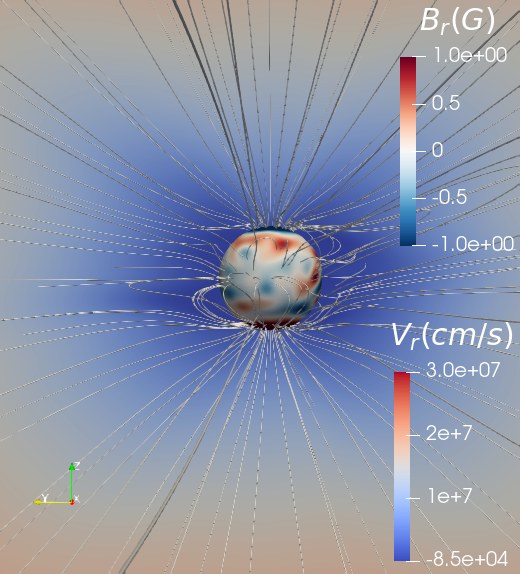}{0.45\textwidth}{(a)}
              \fig{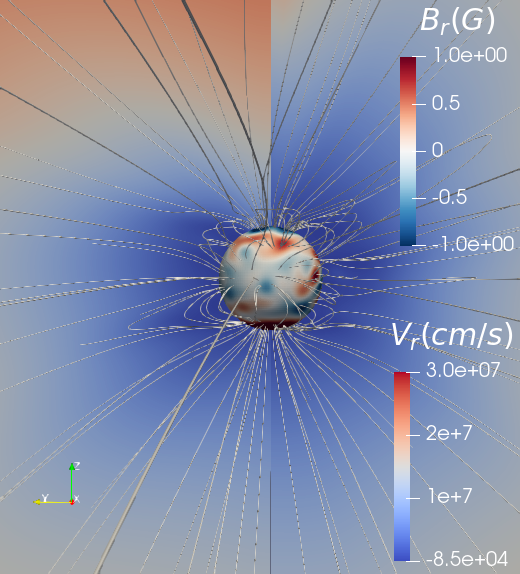}{0.45\textwidth}{(b)}}
    \caption{Comparison for the 1st of August 2008 daily zero-corrected GONG synoptic magnetogram case between the COCONUT solution (a) and the Wind-Predict solution (b). The color scale shows the radial velocity $V_r$ in units of $\rm{cm\ s^{-1}}$. The surface of the star is shown via a spherical slice showing the radial magnetic field $B_r$ in G. Magnetic field lines are traced in white.} 
    \label{fig:map08}
\end{figure*}

The results are displayed in figure \ref{fig:map08}. The magnetic field used as inner boundary condition is displayed for the two cases as a spherical slice (the minimum and maximum values have been manually adjusted to -1 and 1~G for a better visualization of the structures, although in reality they fall between -3.3 and 3.8 G). 
We can see that the inner boundary condition is mostly the same for the two models, using a similar pre-processing with $\ell_{max}=15$. A small supplementary adjustment had to be made for Wind-Predict to remove noise at the poles, as in spherical coordinates the code is more sensitive to the polar boundary condition. This is also due to the fact that Wind-Predict was not intended to run with the boundary conditions from COCONUT, as it has its own set of optimized boundary conditions and numerical parameters described in section \ref{sec:BCs}. For the sake of the numerical benchmark, we however made the necessary adjustments to run it in conditions as close as possible to COCONUT, which can result in more numerical instabilities. This will be discussed in more details in sections \ref{sec:comparison-with-wind-predict} and \ref{sec:validation_obs}.
Contrary to the previous results, these are full 3-D simulations without an axisymmetric approximation, which means that the slice we are displaying is not representative of the entire volume. However, for the purpose of the benchmark and the comparison with the previous cases, we have kept the same layout.
We can see that we recover four helmet streamers distributed over the sphere: a first one on the top left corner, second one on the bottom left, third one on the bottom right and fourth one on the top right. These positions are in good agreement with the reference case given by Wind-Predict. We can see however that there is a better agreement close to the star for the magnetic field lines, as the streamers' termination can show deviations at around 2 solar radii. This can be a result of the modification of the poles needed by Wind-Predict, as well as numerical effects. We can also notice a patch of higher velocities at the north pole for Wind-Predict, which can be a consequence of the polar boundary condition. As expected, the velocities and exact edges of the streamers are slightly different, as foreseen in the dipolar and quadrupolar cases. A difference which is more prevalent here is the influence of the polar boundary condition: because the Wind-Predict model is in spherical coordinates, it has naturally a singularity at $\theta=0$ and $\theta=\pi$, thus creating the slight mismatch observable in the figure. COCONUT on the other hand, thanks to its unstructured mesh and Cartesian geometry, does not exhibit any singularity in its computational domain. 
This will be discussed in more details in the next section.

\section{Benchmarking procedure}
\label{sec:benchmark}

Now that we have a qualitative agreement between COCONUT and Wind-Predict wind solutions, we want to quantify better these differences and try to understand where they come from (Sect. \ref{sec:comparison-with-wind-predict}). We also want to use a more physical validation with direct comparison with observations (Sect. \ref{sec:validation_obs}). Finally, we compare the computation times to show the improvement brought by COCONUT in terms of speed (Sect. \ref{subsec:runtime_bench}).

\subsection{Model validation with Wind Predict}\label{sec:comparison-with-wind-predict}

We have demonstrated in the previous sections the ability of our new coronal code to adapt to both simple and complex magnetic topologies to produce a realistic wind solution. In the benchmarking procedure that we have described, we have adapted the two coronal models to have the same initial conditions, boundary conditions and physical parameters. We have also tried to make the numerical options as close as possible (by disabling the rotation in Wind-Predict, for example). That being said, the models are undeniably different, and since no official benchmark for polytropic coronal models has ever been used, we wanted to assess more quantitatively how close the models can be expected to be. We can note however that some previous attempts to collect polytropic wind models results (with indicators such as the Alfvén radius or the mass loss) have shown a great sensibility to physics and numerics, and in result a great variety of results for similar cases \citep{Pneuman1971, Durney1975}. 

We do such a comparison with a point by point comparison between the two models for the dipolar and quadrupolar cases, which already exhibit differences that are going to repeat themselves for more complex topologies as seen for the map case. This already requires a specific pipeline of post-processing, because COCONUT uses an unstructured Cartesian mesh while Wind-Predict uses a structured spherical grid. To be able to do this comparison, we use options from the ParaView visualization software that allows us to resample one dataset using another one; thus we resample COCONUT using the Wind-Predict grid. Because of this procedure, the reader has to bear in mind that some differences may come from the interpolating process, as we will explain in the next paragraph.

\begin{figure*}
    \centering
    \gridline{\fig{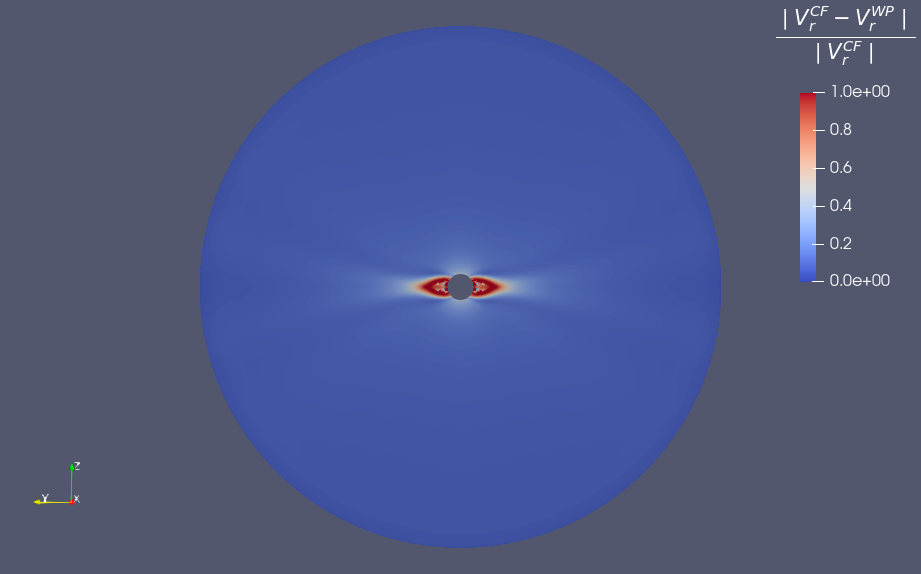}{0.45\textwidth}{(a)}
              \fig{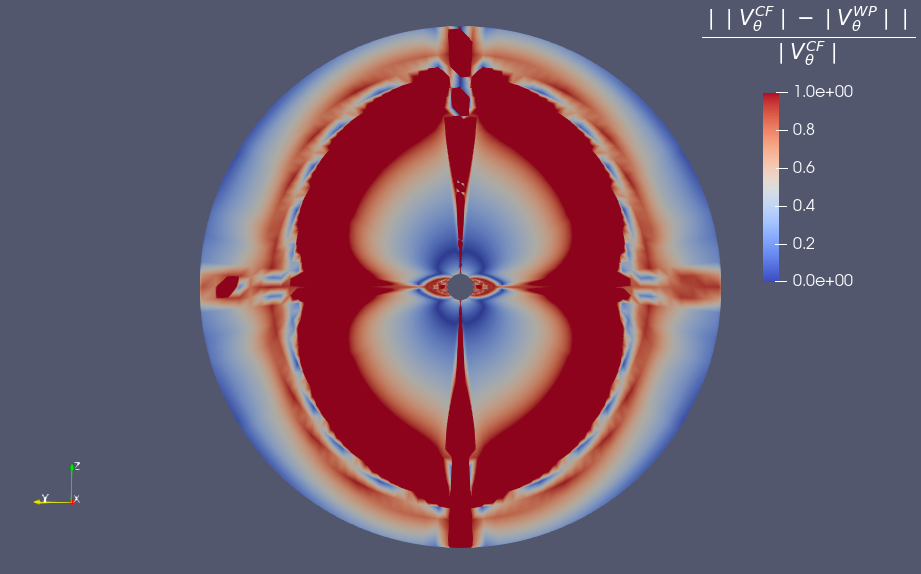}{0.45\textwidth}{(b)}}
    \gridline{\fig{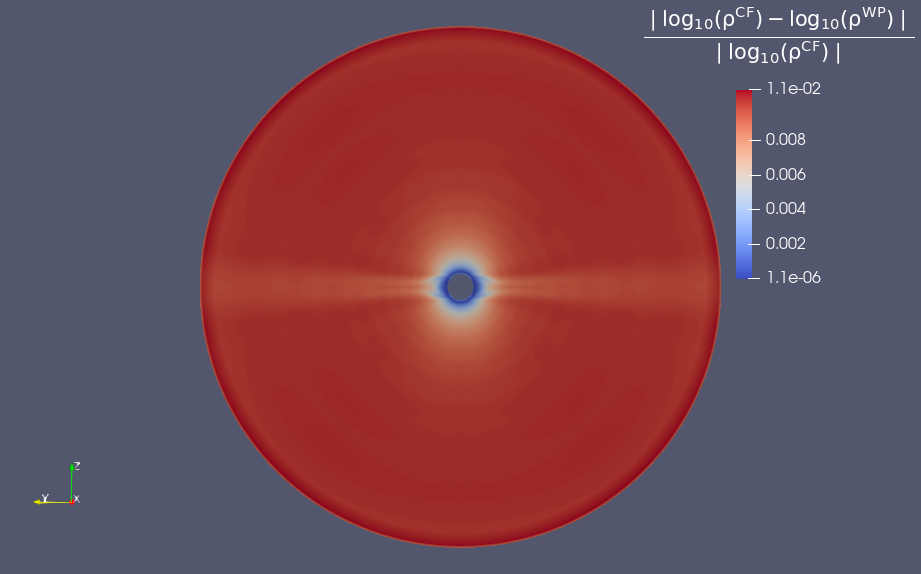}{0.45\textwidth}{(c)}
              \fig{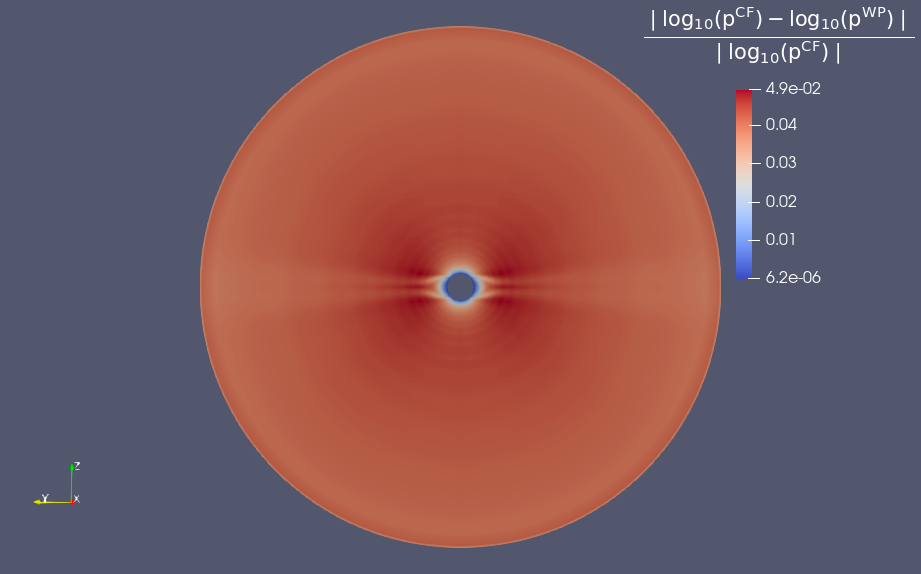}{0.45\textwidth}{(d)}}
    \caption{Comparison for the dipole between the COCONUT solution and the Wind-Predict solution. The color scale shows the relative difference for (a) $V_r$, (b) $V_\theta$, (c) $\rm{log}_{10}(\rho)$, and (d) $\rm{log}_{10}(p)$}.
    \label{fig:diff_dip}
\end{figure*}

 We have selected the quantities that seem the most relevant to compare for us, which are the radial velocity $V_r$, the latitudinal velocity $V_\theta$, the density $\rho$ and the pressure $p$. Indeed, there is little interest in comparing the magnetic field because we use exactly the same field as a boundary condition, and the PFSS reconstruction is only used at the initialization before the MHD relaxation and thus has little to no impact over the final state. The density and pressure have logarithmic profiles that decrease very fast from the star, and are thus difficult to compare precisely (because their values reach the numerical precision of the model). Because of this, we compare them not directly, but using their common logarithm. Because we do not include rotation yet in this model, this leaves the two first components of the velocity as the best indicators to show the differences between the models. To be able to do this comparison, we also have to convert the Cartesian vectors from COCONUT to spherical. 

We have used two different kind of point to point differences to display various sources of differences between the two codes. The first one is the traditional relative difference, defined as:
\begin{align}
    ||V_r|| & = \frac{|V_r^\mathrm{CF} - V_r^\mathrm{WP}|}{|V_r^\mathrm{CF}|}, \\
    \nonumber ||\rho|| & = \frac{|\mathrm{log}_{10}(\rho)^\mathrm{CF} - \mathrm{log}_{10}(\rho)^\mathrm{WP}|}{|\mathrm{log}_{10}(\rho)^\mathrm{CF}|},\\ 
    \nonumber ||p|| & = \frac{|\mathrm{log}_{10}(p)^\mathrm{CF} - \mathrm{log}_{10}(p)^\mathrm{WP}|}{|\mathrm{log}_{10}(p)^\mathrm{CF}|}. 
\end{align}
However, because in a coronal model we have significantly smaller velocities closer to the solar surface, we are aware this may enhance the differences in this region.

We also use a modified relative difference:
\begin{equation}
    ||V_\theta|| = \frac{||V_\theta^{CF}| - |V_\theta^\mathrm{WP}||}{|V_\theta^\mathrm{CF}|}.
\end{equation}
The definition for $V_\theta$ is slightly different because it is a signed quantity (can be positive or negative depending on position). Although there are some local negative values for $V_r$, as said before they are very localized and limited to the star surface.

The results for the dipole case are summarized in Fig.~\ref{fig:diff_dip}. Fig.~a) shows the relative difference for $V_r$, Fig.~b) the modified relative difference for $V_\theta$, Fig.~c) the relative difference for $\mathrm{log}_{10}(\rho)$ and d) the relative difference for $\mathrm{log}_{10}(p)$. 
It is clear that there are minor differences between the two codes, and we list here all the factors that can explain this. We made sure to use the same inner and outer boundary conditions, the same scheme (HLL), but we still have some differences in numerics. An important factor is that we need to use a limiter in Wind-Predict but not in COCONUT, which is going to smoothen the zones of strong gradients, so especially the edges of the streamer. Unfortunately we cannot turn the limiter off in Wind-Predict without crashing the simulation and it is very difficult to use the same limiter in both cases because limiter algorithms for unstructured and structured grids, respectively are significantly different. We also have a different number of points used for the reconstruction, which affects the accuracy of the solution. Finally, the level of convergence has been taken to be as close as possible based on the evolution of the residuals of the pressure, while it is not possible to ensure that all quantities are converged to the same state when comparing runs of the two codes. 
We would like also to clarify two points which may confuse the reader: we can see an enhancement of the difference at the poles in Fig.~b), which is due to the fact that we have a polar boundary condition in Wind-Predict but not in COCONUT. We also see some deformation at the outer edge of the domain: this is due to the interpolation used to compare the outputs from the two codes, since Wind-Predict uses a stretched grid in radius which leads to the final point of the grid not being placed exactly at the outer boundary. The differences seen here are thus the effects of the numerics, not the physics. 

Fig.~a) shows that for $V_r$, the differences between the two codes are concentrated in the equatorial streamers. This is certainly due to the limiter, which is used to attenuate numerical shock instabilities and thus acts where there are sharp gradients. For more discussion on the Wind-Predict code sensibility to limiters, the reader can refer to appendix \ref{app:num_wp_lim}. The other significant zone of difference is the current sheet, whose size and shape is very sensitive to numerical diffusion. Everywhere else, the relative difference is less than 1\%. Fig.~b) shows that for $V_\theta$, the most significant difference is set at around 10 $R_\odot$. This is because $V_\theta$ is a signed quantity, hence the relative difference is enhanced when crossing from positive to negative values. The other significant zones are the current sheet and the polar regions, due to the difference in boundary conditions. Everywhere else, the relative difference is less than 10\%. Fig.~c) and d) show a remarkable agreement between the density and the pressure for both codes, with the maximum difference in common logarithm reaching 1.1\% for density and 4.9\% in pressure. We have an almost exact agreement at the surface of the star, then the most significant differences come from the edges of the current sheet and the outer boundary condition (due to the stretching of the grid, as explained before).

\begin{figure*}
    \centering
    \gridline{\fig{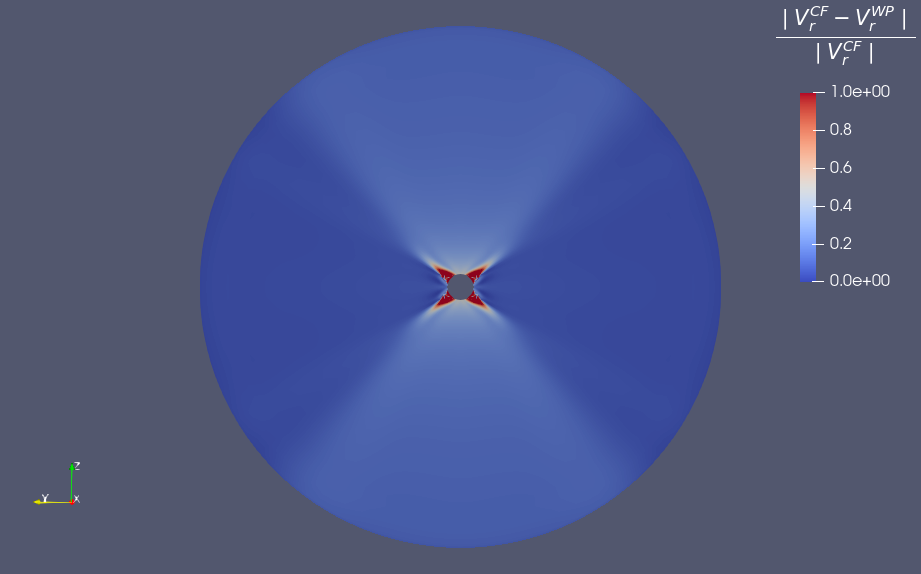}{0.49\textwidth}{(a)}
              \fig{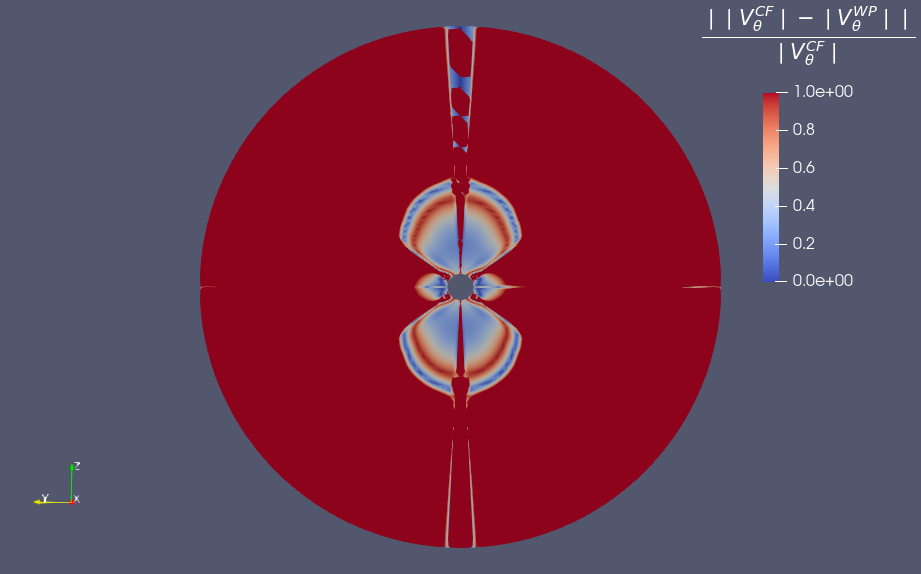}{0.49\textwidth}{(b)}}
    \gridline{\fig{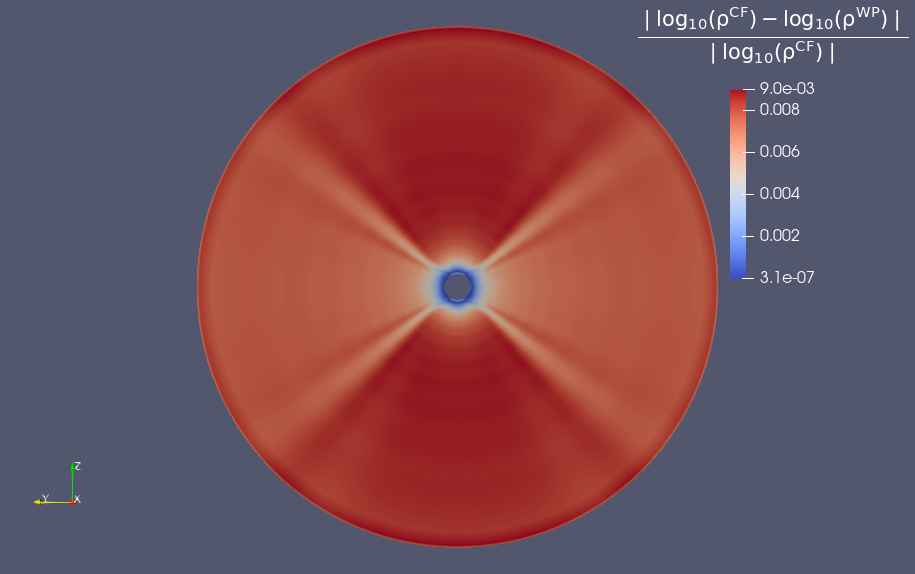}{0.49\textwidth}{(c)}
              \fig{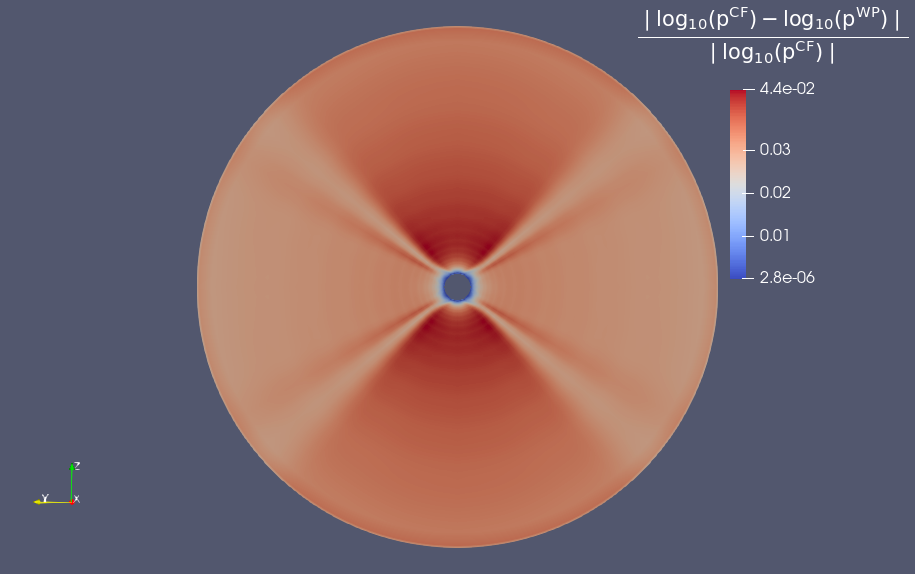}{0.49\textwidth}{(d)}}
    \caption{Comparison for the quadrupole between the COCONUT solution and the Wind-Predict solution. The colorscale shows the relative difference for (a) $V_r$, (b) $V_\theta$, (c) $\rm{log}_{10}(\rho)$, and (d) $\rm{log}_{10}(p)$.}
    \label{fig:diff_quad}
\end{figure*}

We did the same comparison for the quadrupolar case in Fig.~\ref{fig:diff_quad}, where we recover the same trends, thus confirming that these differences are going to repeat themselves for the different input boundary conditions. The change in topology helps us however to better identify the causes of the differences, as the current sheets have now moved to mid-latitudes.
In Fig.~a), we see once again that the difference is maximum inside the streamers and along the current sheets, thus confirming the previous description. However in Fig.~b), we see a strong difference in $V_\theta$ in most of the domain, except for the poles and the streamers. Figs.~c) and d) recover the same trend, with a maximum difference in logarithm of 0.9\% in density and 4.4\% in pressure, again following the edges of the current sheet and the outer boundary.

\begin{figure*}
    \centering
    \gridline{\fig{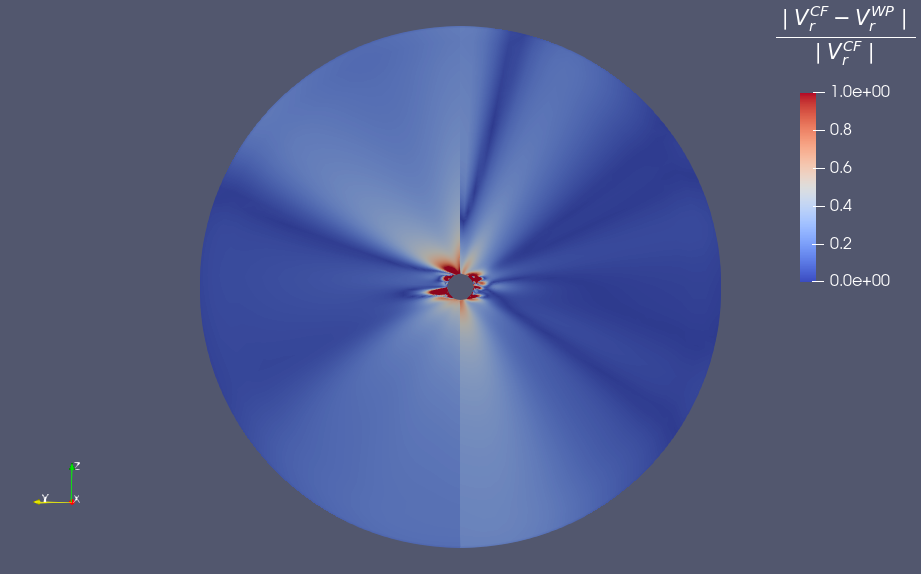}{0.49\textwidth}{(a)}
              \fig{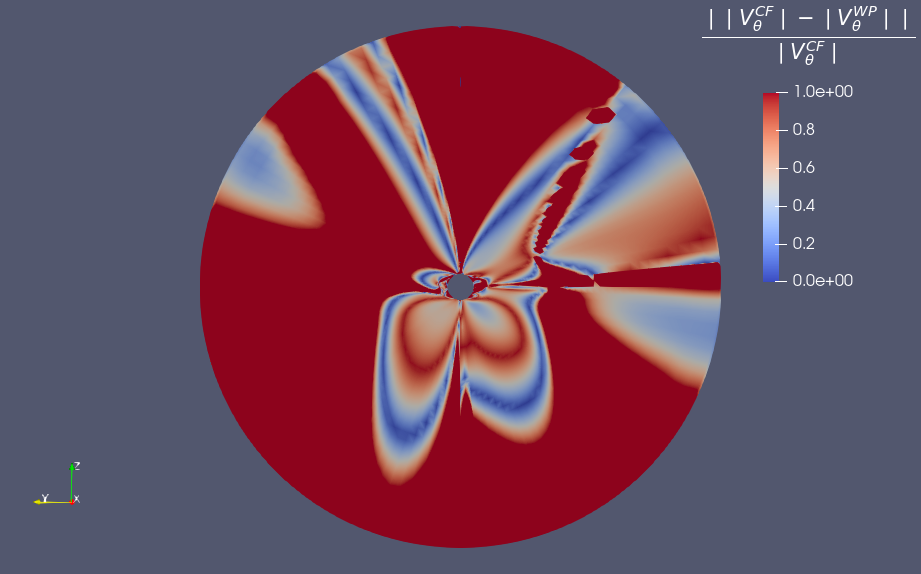}{0.49\textwidth}{(b)}}
    \gridline{\fig{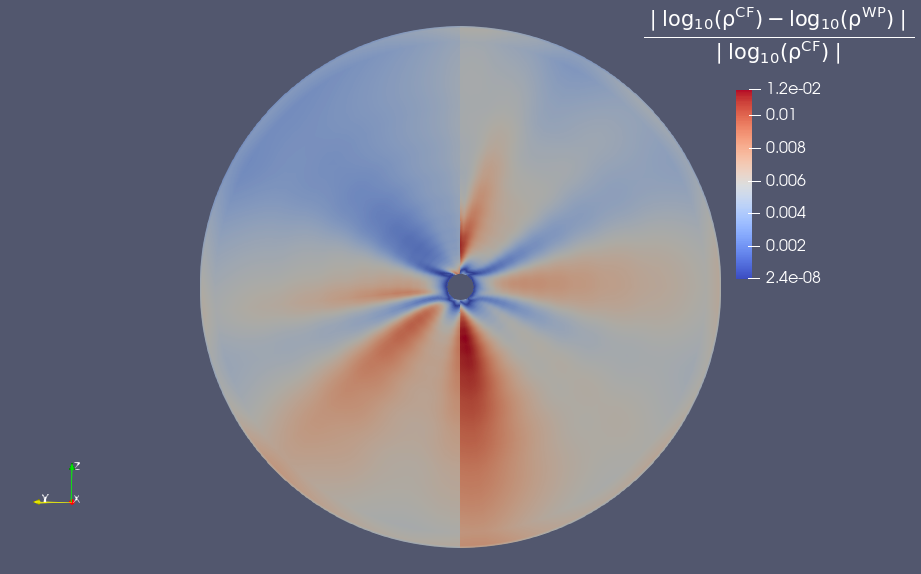}{0.49\textwidth}{(c)}
              \fig{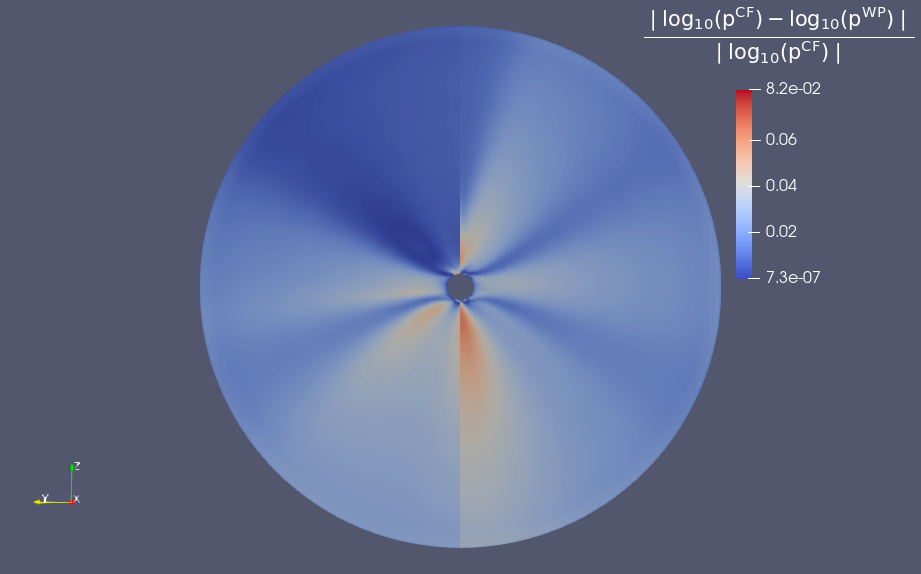}{0.49\textwidth}{(d)}}
    \caption{Comparison for the CR2072 GONG synoptic map case between the COCONUT solution and the Wind-Predict solution. The color scale shows the relative difference for (a) $V_r$, (b) $V_\theta$, (c) $\rm{log}_{10}(\rho)$, and (d) $\rm{log}_{10}(p)$.}
    \label{fig:diff_map}
\end{figure*}

Finally, we did the same comparison for the realistic data-driven case, as can be seen in Fig.~\ref{fig:diff_map}. For this GONG synoptic map, the trends are a bit more complex as the topology is more complicated. In Fig.~a), the difference is still following the streamers, although the polar boundary condition now has some more pronounced influence. In Fig.~b), we do not see an effect in the equatorial plane and the transition between positive and negative values of $V_\theta$ is not so regular anymore. The density and pressure maximum difference is a bit higher, reaching respectively 1.2\% and 8.2\%, but again mostly following the edges of the streamers and the southern polar boundary condition. 

This shows how strong an impact the numerics can have on even a simple polytropic model, and how important it is to proceed to benchmark procedures to quantify these effects as we are heading towards multimodel forecasting for space weather predictions. This numerical benchmark also shows that a set of boundary conditions that are optimal for one model may not for another one. The $\vek{V}=\mathbf{0}$ inner boundary condition seems to interfere with the axisymmetric polar boundary condition of Wind-Predict and generate artifacts for the data-driven case. This exercise ultimately shows how difficult it is to compare numerical codes with the same physics and different numerics, and that ultimately it may be more relevant to compare the results of optimized codes rather than try to use the same computation methods.

\subsection{Model validation with observations}
\label{sec:validation_obs}

On top of the numerical benchmark performed, we have also performed a more physical benchmark with observations. Indeed, the confirmation that we can reproduce results from the Wind-Predict code is not enough to make sure the model is suitable for space weather applications, hence we also need the confrontation with solar data. This also stresses the current need for broad benchmark procedures for coronal models, which is under construction \citep{Wagner2021, Badman2022}. However, we recall that we use here the polytropic approximation for this first version of the model, which means that we do not reproduce accurately the heating of the corona, and hence a number of structures that are typically used for validation such as bimodal distribution of solar wind velocity or EUV coronal hole dimming. We however would like to suggest broad validation metrics that can be used for polytropic wind models, which are usually the first step for coronal models, and are still largely used for solar wind comparisons \citep{Karageorgopoulos2021} and by the stellar community \citep{Ireland2022ApJ}. For this reason, we will focus mostly on magnetic field quantities, as they are the ones best described by our MHD model for the moment. We will also show the results with the Wind-Predict code to estimate what is expected from another polytropic coronal model in these cases. 
Since the numerical benchmark has shown that using the same boundary condition for both codes leads to artifacts at the poles, we will use the optimized version of Wind-Predict with its usual inner boundary condition (described in Sect. \ref{subsec:wp-description}), in order to provide the best comparison to observations.

\begin{figure*}
    \centering
    \gridline{
        \fig{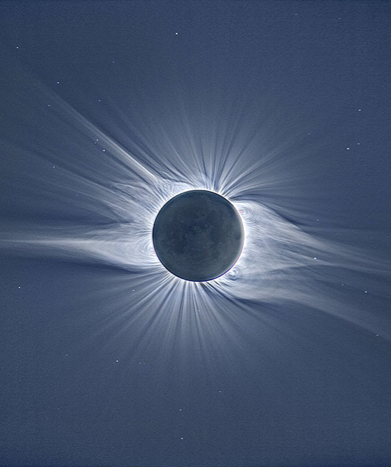}{0.305\textwidth}{(a)}
        \fig{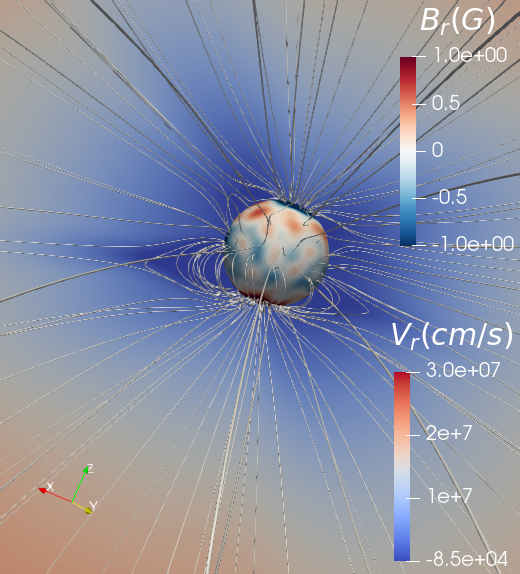}{0.33\textwidth}{(b)}
        \fig{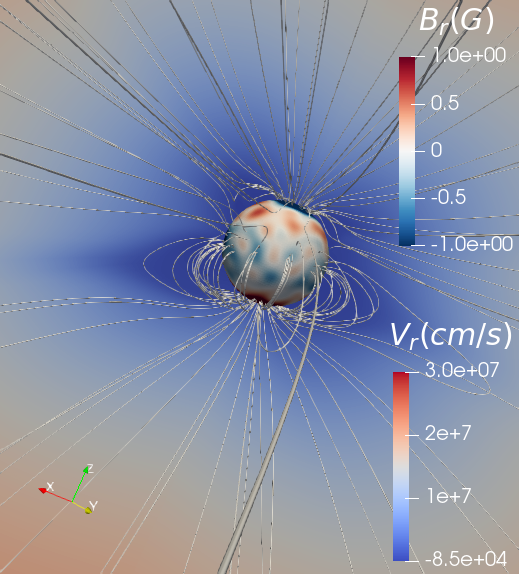}{0.33\textwidth}{(c)}
    }
    \caption{Comparison for the 1st of August 2008 GONG corrected synoptic map case between a white-light composite eclipse picture (a), the COCONUT solution (b) and the Wind-Predict solution with its usual boundary conditions (c). The eclipse picture was taken at Novosibirsk in Russia (credits: J. C. Casado and D. López). For the simulations, the color scale shows the radial velocity $V_r$ in units of $\rm{cm\ s^{-1}}$. The surface of the star is shown via a spherical slice showing the radial magnetic field $B_r$ in G. Magnetic field lines are traced in white.}
    \label{fig:wl}
\end{figure*}

The first type of observations we use is white-light images. They are usually records of polarization brightness (pB) formed by Thomson scattering of photospheric light by coronal free electrons in the K corona \citep{Aschwanden2005}. They are extremely useful to determine the shape of the streamers in the corona, as they reveal the underlying magnetic field structure. Most white-light images are generated using a coronograph from a spacecraft (like SOHO/LASCO) or from ground-based observatories (like COSMO/KCOR). The problem with these technics is that the coronograph extends above 1 solar radius, thus dimming some of the structures. It is actually during the solar eclipses on Earth that the solar disk is perfectly covered by the Moon, and that we can see the most precisely the shape of the streamers. For this reason, white-light pictures of eclipses have been traditionally used to constrain coronal models \citep{Mikic1999}. This is another reason why we have selected the date of the 1st of August 2008, as there was a solar eclipse on Earth at that date. We can see the processed white-light composite image in panel a) of \ref{fig:wl}. This picture has been made by assembling 67 different pictures of the eclipse, in order to recover the finest structures of the corona. We can clearly a minimum of activity structure, close to a dipole, with mostly equatorial streamers. On the left side of the picture, we have two streamers, one in the northern hemisphere and another bigger one in the southern hemisphere. On the right side of the picture, we have a main streamers in the northern hemisphere, and a smaller one in the southern hemisphere, probably located behind the main one. The poles exhibit clear patterns of open magnetic field lines. In panels b) and c), we have rotated the simulations shown in Fig. \ref{fig:map08} to match the position of the Sun as seen by Earth at the date of the eclipse, and show the corresponding streamers directly with the plotting of the magnetic field lines. We can see that we recover the right positions for the streamers for COCONUT, with the only exception being the streamer from the upper right which is a bit narrower. For Wind-Predict, figure (d) is similar to COCONUT, except for the streamer on the bottom-right which is clearly enhanced. Due to the difference for the inner boundary condition, the streamers are more rounded and less collimated, which is actually less close to the white-light picture than the COCONUT model. This proves that our coronal model has an accurate description of the magnetic field closed field-lines configuration for the first solar radii. An even more accurate comparison would be to generate synthetic white-light images from the simulation, but we leave this for future papers. 

\begin{figure*}
    \centering
    \gridline{
        \fig{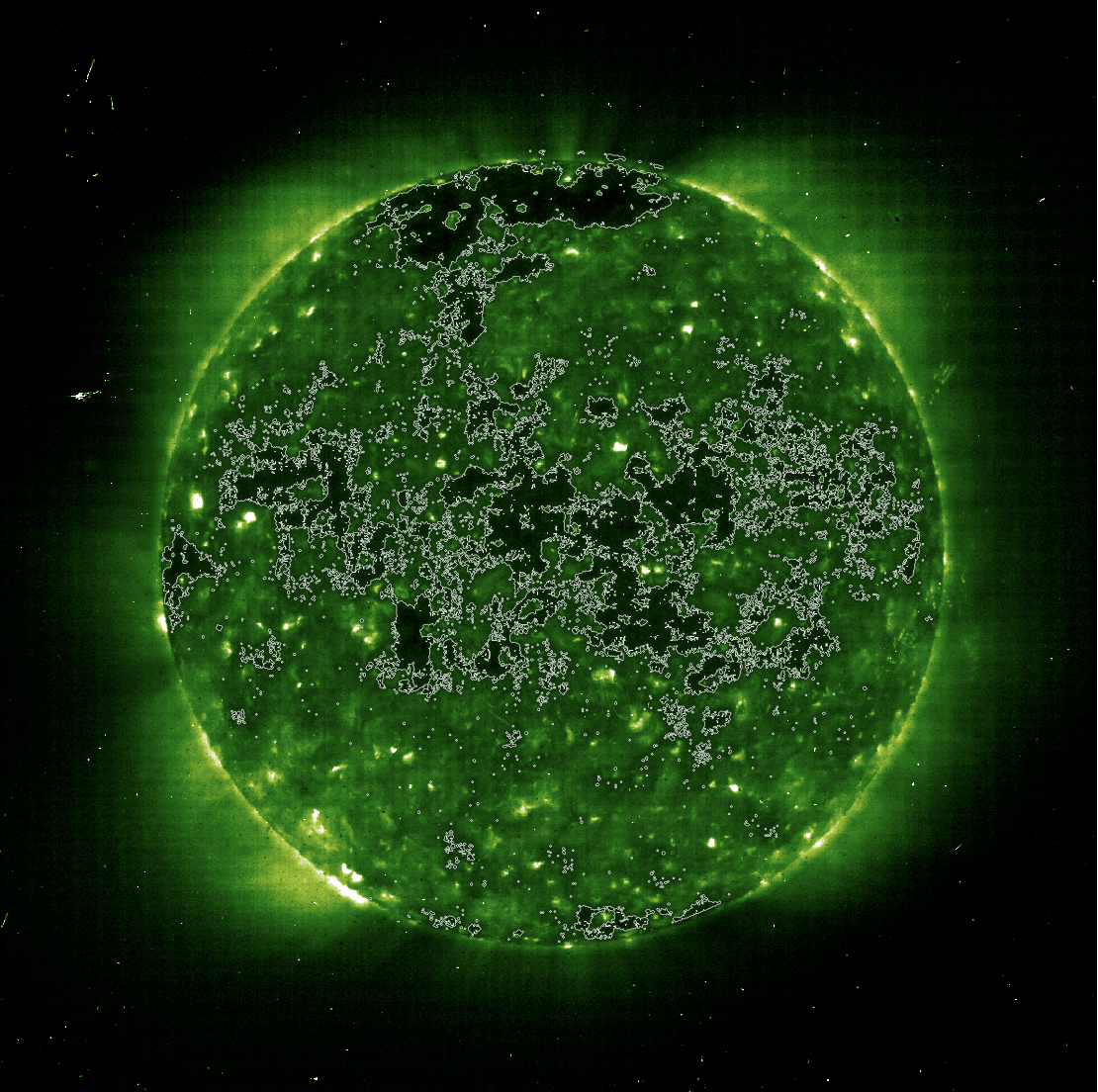}{0.275\textwidth}{(a)}
        \fig{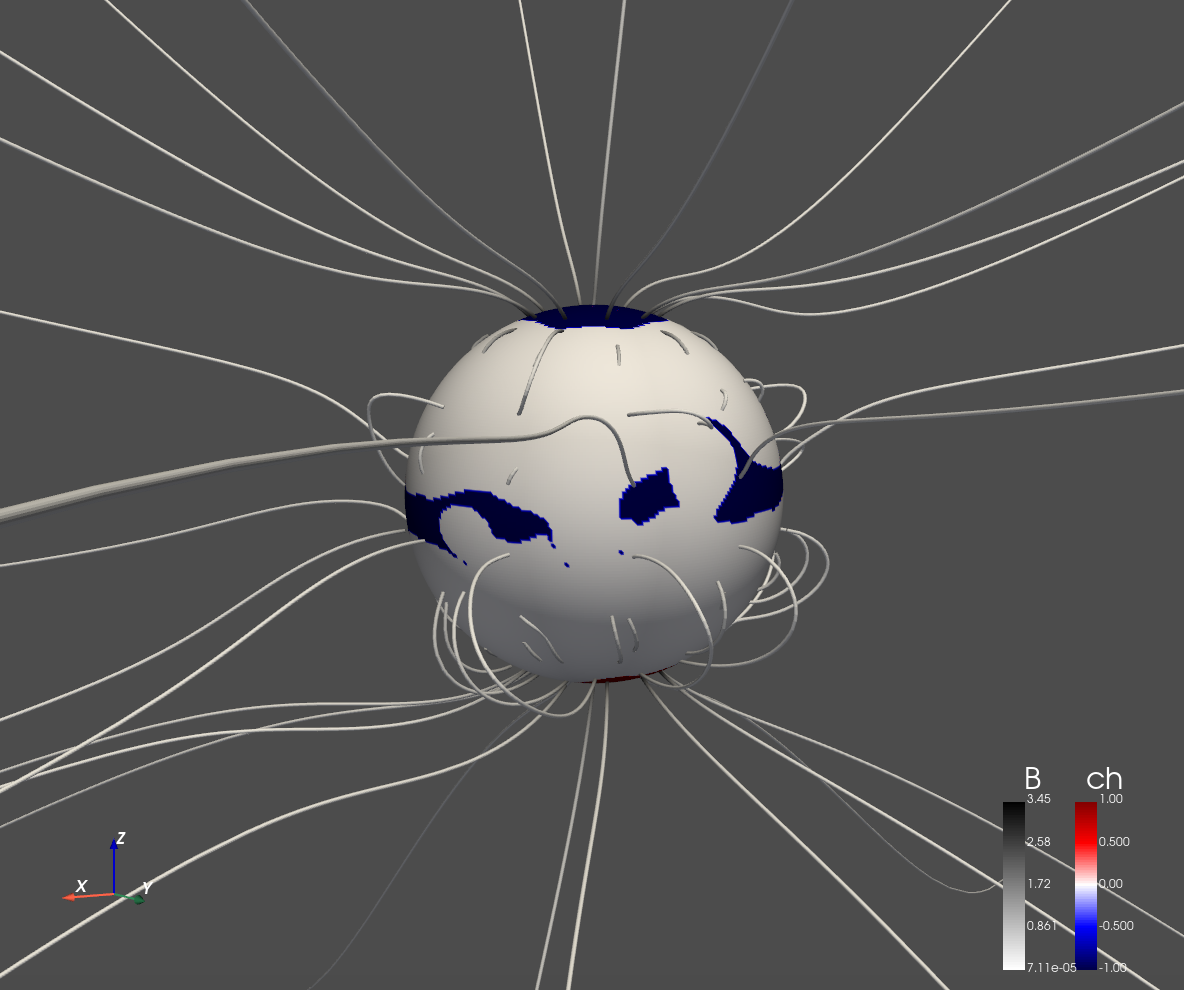}{0.33\textwidth}{(b)}
        \fig{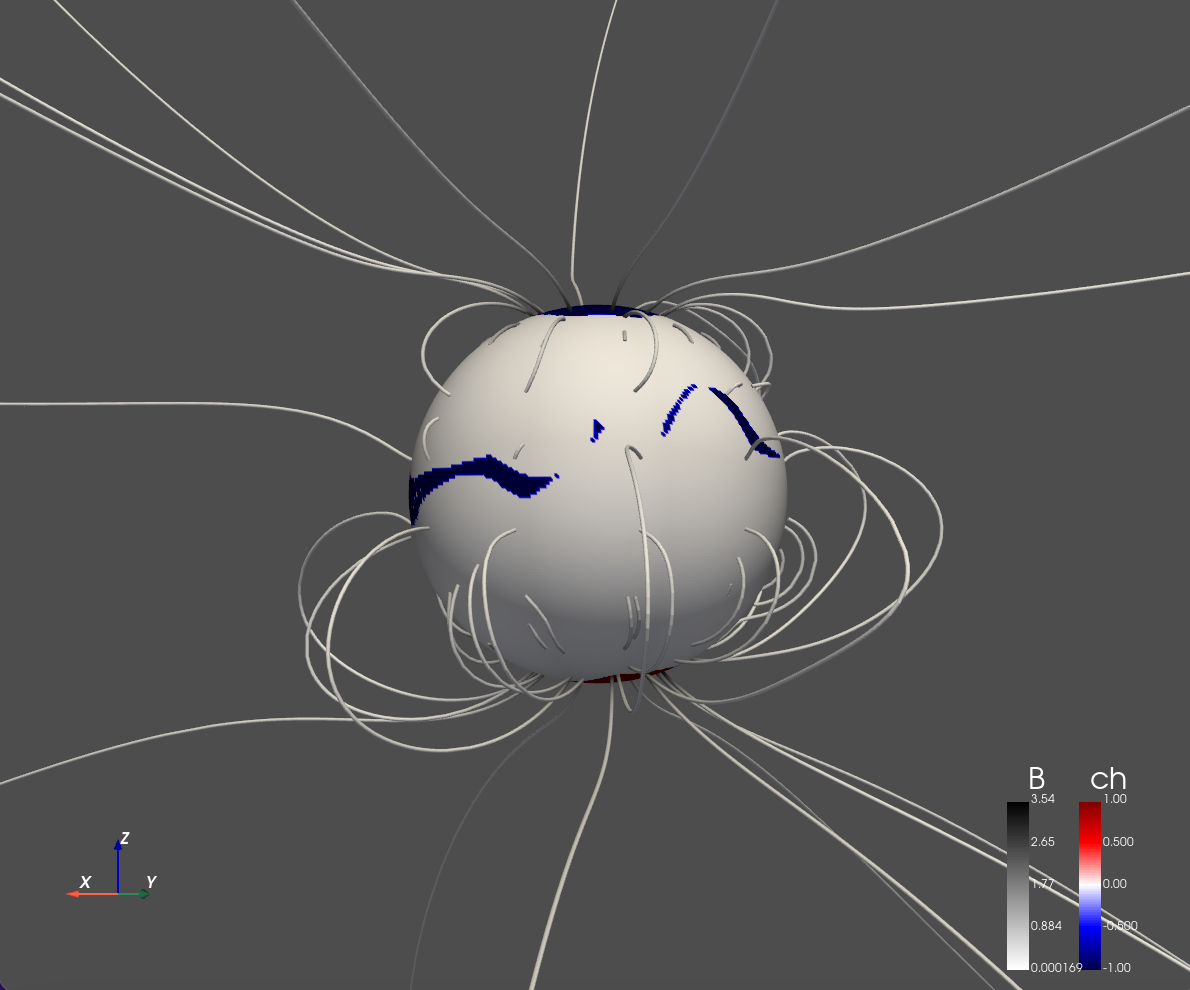}{0.33\textwidth}{(c)} 
    }
    \caption{Comparison for the 1st of August 2008 GONG synoptic map case between the SOHO/EIT EUV observations (a), the COCONUT solution (b) and the Wind-Predict solution with its usual boundary conditions (c). The EUV picture is from the 195 channel and was retrieved using the JSOC database. It has been post-processed to show clearly coronal holes in dark (highlighted by a white contour). For the simulations, we show the solar surface at 1.01 $R_\odot$ with colors indicating coronal holes (detected by open magnetic field lines) with the associated magnetic field polarity (red for positive, blue for negative). Magnetic field lines are traced in white. These results were obtained using the package PyVista \citep{sullivan2019pyvista}.}
    \label{fig:ch}
\end{figure*}

The second observation we want to use is EUV images. In 2008, the only source available is SOHO/EIT, as seen in the panel a) of Fig. \ref{fig:ch}. We chose the channel 195 as it is the best one to see the coronal holes, indicated by a darkening of the plasma. We have applied a simple post-processing using a threshold detection to highlight the coronal holes by contouring them in white. We can clearly see two polar coronal holes, as well as minor equatorial coronal holes (the most visible being on the eastern limb on the left side of the picture and in the middle of the solar disk). Here our goal is to validate the distribution of open magnetic field lines at the surface of the star, rather than the temperature distribution (which we know is roughly approximated because of the polytropic assumption). We show the comparison with the COCONUT and Wind-Predict solutions in respectively panels b) and c) of Fig. \ref{fig:ch}. We can see that we recover in both codes the polar coronal holes. We however show more equatorial coronal holes than detected by SOHO/EIT, enhancing open field lines on the western limb for example. Since this feature is present in both cases (b) and (c) at the same locations, it is likely that it is dependent on the initial synoptic map that we use (whose influence can affect greatly the coronal holes detection, see for example \cite{Li2021} with PFSS extrapolations) and its pre-processing, but also the boundary conditions used. Indeed, case (c) has a reduced enhancement, but fails to reproduce the central coronal holes. The exact impact of the choice of $\ell_{max}$ and the magnetogram source will be discussed in follow-up papers.

\begin{figure*}
    \centering
    \gridline{\fig{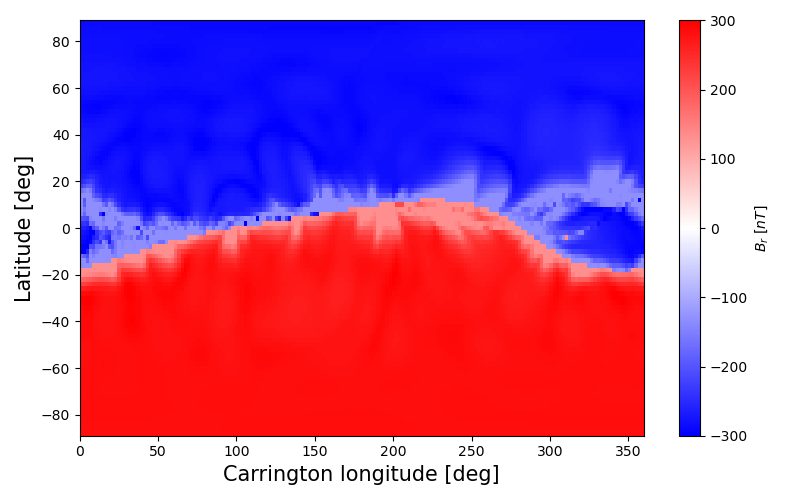}{0.33\textwidth}{(a)}
              \fig{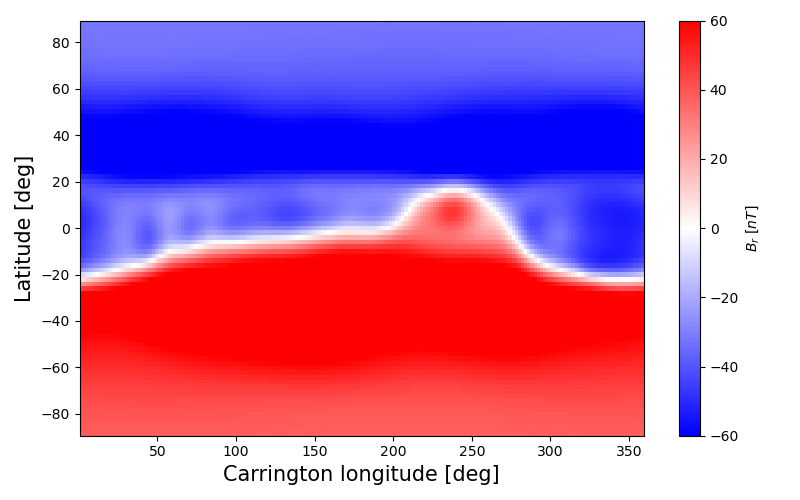}{0.33\textwidth}{(b)}
              \fig{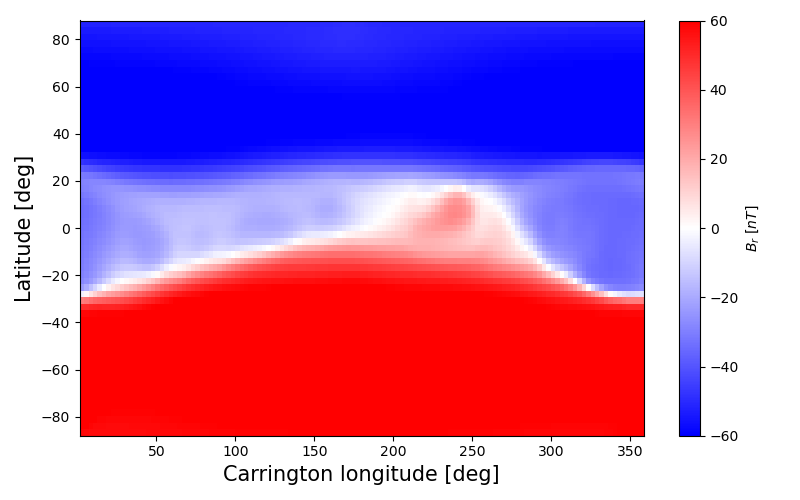}{0.33\textwidth}{(c)}}
    \caption{Comparison for the 1st of August 2008 GONG synoptic map case for the boundary conditions at 0.1 AU between the WSA solution (a), the COCONUT solution (b) and the Wind-Predict solution with its usual boundary conditions (c). 
    We show the radial magnetic field $B_r$ in nT.
    }
    \label{fig:euhforia_bc}
\end{figure*}

Finally, it would also be interesting to compare our model with in situ data. The difficulty here is that there are extremely few measurements of the solar wind in the solar corona to do so. The recent Parker Solar Probe (PSP) is starting to provide such data that are relevant for numerical simulations, but only during its perihelia \citep{Riley2021}. An interesting possibility lies with Interplanetary Scintillations (IPS), but this is still under development \citep{Shoda2022}. As none of these data are available at the date we chose to study, another possibility is to compare to data typically used for boundary conditions for EUHFORIA, that we know yield good results at Earth \citep{Pomoell2018}. EUHFORIA normally uses the WSA+PFSS+SCS approximation to get boundary conditions at 0.1 AU, but it can also be replaced directly by a coronal model \citep{Samara2021}. Because our model for the moment uses the polytropic assumption, the most relevant physical quantity to compare is the magnetic field that we obtain at 0.1 AU. We show it in Fig. \ref{fig:euhforia_bc} for all three models: WSA in panel a), COCONUT in panel b) and Wind-Predict in panel c). We can see that indeed the position of the current sheet is in good agreement with positive polarity in the northern hemisphere (in blue), negative polarity in the southern hemisphere (in red) and a quasi-horizontal current sheet except for a slight bump around Carrington longitude 250. We notice however that it has more structures around the equator in COCONUT and even more in Wind-Predict. This is expected because they are MHD codes. The difference between the two codes comes from the limiter and the resolution, as was previously explained. We can also see the influence of the polar boundary condition as the northern pole as COCONUT is able to show more subtle effects with a light dimming of the magnetic field at the northern pole. The major difference between the empirical and MHD magnetic field is the amplitude: panel (a) has a magnetic field between -300 and 300 G, while panels (b) and (c) show a magnetic field between -60 and 60 G. The comparison with Wind-Predict shows that this is expected from a MHD code, and is more physical since the final amplitude of the magnetic field is an input parameter for empirical models. In fact, both other codes and data suggests that the PFSS can either overestimate or underestimate the magnetic field at 0.1 AU depending on the parameters chosen, which shows that COCONUT is yielding expected results \citep{vanderHolst2010, McGregor2011, Reville2022}. Unfortunately, we don't have observations data for this date to discuss which code is better, but in future studies we can select other events to use PSP \citep{Badman2022} or SoHO/LASCO white-light images \citep{Poirier2021}.

More detailed quantification of the differences with the observations (following procedures like described in \cite{Wagner2021} for example) will be provided in a follow-up paper, also focusing on the impact of the solar activity on these comparisons.

\subsection{Run-time Benchmarks}
\label{subsec:runtime_bench}
The timing of the simulations plays a crucial role for the operational purposes. In the forecasting procedure, the coronal model has to provide reliable plasma conditions at 0.1 AU, from where the inner heliosphere starts. Therefore, in order to model the space weather conditions at Earth, first the plasma values need to be provided at the inner heliospheric boundary, from which the heliosphere is modeled. Each phase needs to be optimized as much as possible to predict the space weather at Earth timely.

\begin{figure*}
    \centering
    \gridline{\fig{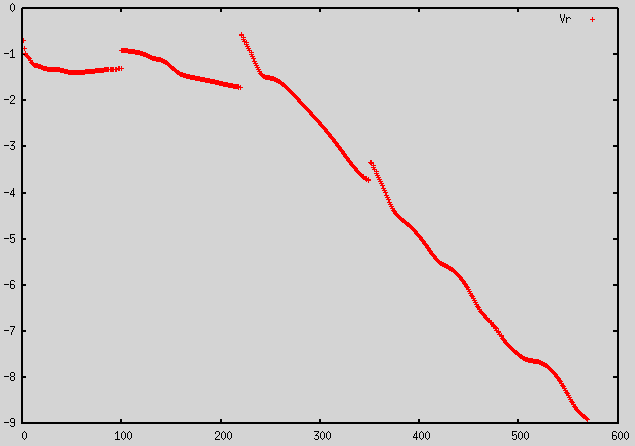}{0.45\textwidth}{(a)}
              \fig{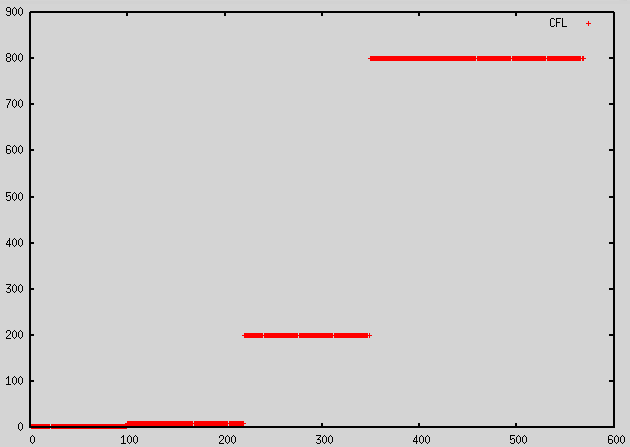}{0.45\textwidth}{(b)}}
    \caption{The convergence history for the quadrupole case. The number of iterations is given on the $x$-axis and the variable values on the $y$-axis. (a) shows the residual values for the $V_r$ component, where (b) shows the CFL function for the given simulation.}
    \label{fig:convergence}
\end{figure*}

The primary test cases were optimized and the times are compared to the corresponding simulations performed in Wind-Predict with the most similar setting between the two codes. As we solve the MHD equations with an implicit scheme in COCONUT, we have a freedom to increase the CFL values to much higher values than 1, resulting in much faster simulation run times. Different CFL conditions were examined to further optimize the run times, showing that step- rather than continuous CFL functions yield faster results. When increasing the CFL number, the simulation residuals also increase initially, while giving rise to a faster convergence for subsequent iterations as compared to the previous, lower CFL number. Fig.~\ref{fig:convergence} shows the relation between the variable residuals and the changes in the CFL function. At the steps where the CFL function increases, there is a discontinuous increase in the variable residuals as well. Therefore, an optimal solution is found to compensate the increase of the residuals at each step effectively. For the three cases under consideration (i.e. dipole, quadrupole and the minimum map) we have optimized the CFL conditions for the corresponding COCONUT simulations. The latter have all been run on the BrENIAC (Tier1) cluster of the Vlaams Supercomputing Center\footnote{\url{https://www.vscentrum.be/}}. To be sure that the comparison was architecture-independent, we also performed the Wind-Predict simulations using the same cluster.

To compare the solutions at the same state of convergence, we evaluated the residuals of different physical quantities and evaluated the time necessary to reach them. Residuals are computed using the following formula:
\begin{equation}
    \mathrm{res}(a) = \log\sqrt{\sum_i\left(a_i^t - a_i^{t+1}\right)^2},
\end{equation}
where $a$ is the physical quantity under consideration, $i$ is the spacial and $t$ the temporal index.

For the dipole magnetic field configuration a convergence of $10^{-3}$ in the radial velocity component was reached in 5.6 minutes on 84 processors for a mesh containing 332,800 cells. To reach the same convergence for the Wind-Predict model it takes 15 minutes on the same cluster, using the same configuration, i.e. improving run times by a factor of 2.7 already for such a simple idealized case.

The quadrupole magnetic field configuration simulations were also performed on the aforementioned mesh on 84 processors. A convergence of $10^{-3}$ in velocity is achieved in 11.9 minutes, while in the case of Wind-Predict this takes 17 minutes. This improvement factor of 1.43 shows that the implicit solver has a slightly lower level of efficiency for this slightly more complex topology.

We also investigated the simulation times for the GONG zero-corrected synoptic map of the 1st of August 2008. 
Due to the more complex structure of the map, we re-considered the residual values for a comparison. Due to the differences of the two codes (limiters, boundary conditions, meshes), it is expected that for the realistic and complex cases the convergence comparison would not be as trivial. In order to assess the code capabilities, we obtain the convergence for each code individually and then compare the performance.
Even though, the Sun is not very active during this period, the structures could not be resolved with the mesh used for the dipole and quadrupole cases, why we used a more refined mesh, with 1.9M cells. The simulation was performed on 196 cores in parallel. With COCONUT the desired convergence level, $10^{-3}$ for the pressure variable, was achieved in 87.5 min. The analogous Wind-Predict simulation reached the target convergence in 16~hours, thus giving a notable speed-up factor of 11 showing even more the advantages of our implicit solver, especially for more complex and realistic configurations. 

We also added the time required for the same synoptic map, but with a different pre-processing: here the maximum degree chosen for the spherical harmonics decomposition of the surface field $\ell_{max}$ is set to 30 instead of 15. This means more detailed structures at the surface of the star, which proves more challenging for the computation. We can see indeed that COCONUT is barely affected by the change in resolution of the input synoptic map, converging in 86.8 min, which is very similar to the time for $\ell_{max}=15$. Wind-Predict on the other hand has a lot of trouble converging: even with reverting to its usual boundary conditions, it takes about 50 hours to reach complete convergence. This gives a remarkable speed-up factor of 35, stressing once again the advantages of an implicit solver for complex configurations. The exact study of the influence of $\ell_\mathrm{max}$ on the final coronal solution will be performed in a follow-up paper ; for the moment, the reader can check appendix \ref{app:lmax} for more information about the $\ell_{max}=30$ run and its differences with $\ell_\mathrm{max}=15$.

Please see table \ref{tab:times} to see an overview of this convergence time benchmark.

\begin{table*}[]
    \centering
    \begin{tabular}{c|c||c|c|c|c|c}
        Code & Case & Number of elements & Number of processors & Highest CFL & Iterations & Time (minutes) \\ \hline \hline
        COCONUT & Dipole & 332 800 & 84 & 5000 & 137 & 5.6 \\ \hline
        Wind-Predict & Dipole & 320 000 & 84 & 0.3 & 80445 & 15.0 \\ \hline \hline
        COCONUT & Quadrupole & 332 800 & 84 & 300 & 290 & 11.9 \\ \hline
        Wind-Predict & Quadrupole & 320 000 & 84 & 0.3 & 94310 & 17.0 \\ \hline \hline
        COCONUT & GONG ($\ell_\mathrm{max}=15$) & 1.9 $10^6$ & 196 & 2000 & 1397 & 87.5 \\ \hline
        Wind-Predict & GONG ($\ell_\mathrm{max}=15$) & 2.0 $10^6$ & 196 &0.3 & 163768 & 960 \\ \hline \hline
        COCONUT & GONG ($\ell_\mathrm{max}=30$) & 1.9 $10^6$ & 196 & 2000 & 1528 & 86.8 \\ \hline
        Wind-Predict & GONG ($\ell_\mathrm{max}=30$) & 2.0 $10^6$ & 196 & 0.3 & 607988 & 3040 \\ \hline \hline
    \end{tabular}
    \caption{Convergence times for the comparison between COCONUT and Wind-Predict. We indicate for each run the code used, the simulation case, the number of cells of the employed mesh, the number of processors used on the BrENIAC cluster, the final CFL number when reaching the required convergence, the number of iterations and thele time (in minutes) needed to reach the same level of convergence. We also include the timing for more resolved input synoptic map ($\ell_{max}=30$ instead of $\ell_{max}=15$).}
    \label{tab:times}
\end{table*}

\section{Discussion \& Conclusions}
The solar corona is extensively studied for the sake of increasing our understanding of the fundamental coronal physics and addressing the still unresolved problem of its heating as well as out of necessity to predict eruptive events and track the propagation of transients through interplanetary space to shield our sensitive communication systems and prevent harm to aerospace and aviation. The coronal magnetic field can be modeled to varying degree of approximation of which we present in detail the potential field source-surface method, which we employ ourselves as model initialization. State of the art MHD models of the corona have been developed since the 1970s, but not before the mid-90s, models have been incorporating observational field data of the photosphere as boundary conditions which allowed validation and adaptation of the models with eclipse white-light images. We present a novel coronal MHD model based on a FV, time-implicit backward Euler scheme of the COOLFLuiD platform, developed as part of a broader heliospheric model of EUHFORIA, replacing the formerly used empirical WSA model. Our coronal model was recently validated against the explicit-time scheme polytropic MHD code Wind-Predict of similar physical scope and coupled to EUHFORIA's heliospheric model at 0.1~AU. Providing one of the first benchmarking procedures to validate a polytropic wind model, we compare the two codes for simple configurations (hydrodynamic, dipolar and quadrupolar wind) before demonstrating its ability to use magnetograms as boundary conditions. We compare quantitatively the two models to assess the impact of the various numerical options chosen on the final solution, and show the impact of the polar boundary condition, mesh and flux limiter. This shows that the exercise of comparing numerical models is a very complicated one, even with adjusting the boundary conditions, solvers and resolution to match as closely as possible. Ultimately, the best way to discriminate models remain the comparison with observations, as we show by comparing with eclipse white-light images, EUV maps for coronal hole boundaries and typical WSA models for the position of the current sheet. Our new model using the COOLFluiD framework shows good agreement with all three metrics, although a better quantification is required to understand how to improve the model. While our polytropic coronal model is still comparatively at an early stage of development and lags behind in physical detail, its numerical scheme, allowing for much faster MHD relaxations than explicit schemes, will be a key advantage in its role as an integrated part of a space weather prediction tool, as demonstrated by the time convergence benchmark, showing that the implicit solver is about 3 times faster for simple configurations, and up to 35 times faster for realistic configurations compared to an explicit solver. 
We plan to release soon a follow-up publication assessing the performances of the code for maximum of activity, as well as providing further comparisons with observations to further validate the  results. The impact of the pre-processing of the input magnetogram, as well as its impact for the coupling with EUHFORIA will also be discussed in follow-up publications. We will in particular focus on the choice of $\ell_{max}$ to quantify in details how this impacts the distribution of fast/slow wind at 0.1 AU. Last but not least, a resistive MHD version with more physical detail such as heat conduction and empirical heating terms is currently under development and will be validated and benchmarked in the near future. This will allow us to include more physics and discuss in more details the large-scale structures of the solar wind, as the combination of rotation and heating will allow for the description of SIRs (Stream-Interacting Regions). We are also experimenting with more accurate numerical schemes (e.g.\ HLLD \citep{MIYOSHI2005315}) or numerical methods (r-adaptation \citep{BENAMEUR2021107700}) to improve the convergence time of the code, thus making it a necessary tool for space-weather in the future years. Comparison with in-situ data will have to wait until a better heating is implemented in the code.

\begin{acknowledgments}
Please note that this article is a shared co-first authorship between the two first authors, as they have contributed equally to this work. 
This work has been granted by the AFOSR basic research initiative project FA9550-18-1-0093. The authors are grateful to Jon A.~Linker, Pete Riley, Zoran Miki\'{c}, Roberto Lionello, Tibor Török, Cooper Downs and Ronald M.~Caplan from Predictive Science Inc. for sharing their expertise in the field of coronal modeling and provision of simulation data for our model validation.
This project has also received funding from the European Union’s Horizon 2020 research and innovation program under grant agreement No.~870405 (EUHFORIA 2.0).
F.Z. is supported by a postdoctoral mandate from KU Leuven Internal Funds  (PDMT1/21/028).
These results were also obtained in the framework of the projects
C14/19/089  (C1 project Internal Funds KU Leuven), G.0D07.19N  (FWO-Vlaanderen), SIDC Data Exploitation (ESA Prodex-12), and Belspo projects BR/165/A2/CCSOM and B2/191/P1/SWiM.
The resources and services used in this work were provided by the VSC (Flemish Supercomputer Centre),
funded by the Research Foundation - Flanders (FWO) and the Flemish Government.
This work utilizes GONG data from NSO, which is operated by AURA under a cooperative agreement with NSF and with additional financial support from NOAA, NASA, and USAF.
\end{acknowledgments}

\facility{Vlaams Supercomputing Centrum}

\software{COOLFluiD \citep{Lani2005,Lani2006,Kimpe2005}, Pluto \citep{Mignone2007,Mignone2011}, ParaView \citep{Ahrens2005}}

\appendix

\section{Potential field extrapolation}
\label{sec:potential-field-extrapolation}
The spherical harmonics represent a complete set of orthonormal eigenfunctions of the angular part of the Laplacian operator giving rise to a harmonic expansion when solving Laplace's equation $\nabla^2 \phi(r,\theta,\varphi) = 0$ in spherical coordinates. The general solution is well known e.g. from the treatment of the time independent Schr\"odinger equation in elementary quantum mechanics:
\begin{equation}
  \phi(r,\theta,\varphi) = \sum_{\ell=0}^\infty \sum_{m=-\ell}^\ell \left( A_\ell^m r^\ell + \frac{B_\ell^m}{r^{\ell+1}} \right) Y_\ell^m(\theta,\varphi).
\end{equation}
As we base our data-driven model on a magnetogram map of radial magnetic field values $B_r(R_\odot,\theta,\varphi)$, we can constrain the general solution by the Neumann condition $\nabla \phi(R_\odot,\theta,\varphi) = -B_r(R_\odot,\theta,\varphi)$ at the inner boundary $\partial \varOmega_\mathrm{i} = \{(r,\theta,\varphi)|r=R_\odot\}$. Expanding the map in spherical harmonics,
\begin{equation}
  B_r(R_\odot,\theta,\varphi) = \sum_{\ell=0}^\infty \sum_{m=-\ell}^\ell C_\ell^m Y_\ell^m(\theta,\varphi)
\end{equation}
with yet to be determined expansion coefficients $C_\ell^m$ and substituting $\nabla \phi = \sum_{\ell,m} [\ell A_\ell^m R_\odot^{\ell-1} - (\ell+1)B_\ell^m/R_\odot^{\ell+2}] Y_\ell^m$ gives for the expansion coefficients
\begin{equation}\label{eqn:Clm}
  C_\ell^m = - A_\ell^m \ell R_\odot^{\ell-1} + \frac{\ell+1}{R_\odot^{\ell+2}} B_\ell^m.
\end{equation}
At the outer boundary a spherical, so-called source-surface \citep{Altschuler1969} is defined in a large enough distance to safely assume that the magnetic field piercing the surface is purely radial, $\vek B(R_\mathrm{S},\theta,\varphi) = B_r(R_\mathrm{S},\theta,\varphi)\,\hat{\vek e}_r \Leftrightarrow \phi(R_\mathrm{S},\theta,\varphi) = A_\ell^m R_\mathrm{S}^\ell + B_\ell^m R_\mathrm{S}^{-(\ell+1)} = 0$. This allows us to express expansion coefficients $A_\ell^m$ in terms of coefficients $B_\ell^m$, i.e. $A_\ell^m = - R_\mathrm{S}^{-(2\ell+1)} B_\ell^m$ and, after substitution into Eq.~(\ref{eqn:Clm}), gives a relation between coefficients $B_\ell^m$ and $C_\ell^m$,
\begin{equation}
    B_\ell^m = \frac{C_\ell^m R_\odot^{\ell+2}}{1+\ell+\ell R_\mathrm{S}^{-(2\ell+1)} R_\odot^{2\ell+1}},
\end{equation}
effective reducing the number of expansion coefficients to be computed to one. Using the orthonormality relation
\begin{equation}
    \int \mathrm{d}\varOmega\, Y_{\ell_1}^{m_1\ast}(\theta,\varphi) Y_{\ell_2}^{m_2}(\theta,\varphi) = \delta_{\ell_1 \ell_2} \delta_{m_1 m_2}
\end{equation}
they are evaluated by numerical integration of
\begin{equation}
  \left. C_\ell^m = \int_{\mathcal S} \mathrm{d}\varOmega \, Y_\ell^{m*}(\theta,\varphi) B_r(\theta,\varphi)\right|_{r = R_\odot} = \left.\int_{0}^{2\pi} \mathrm{d} \varphi \int_0^\pi \mathrm{d} \theta \sin \theta \, Y_\ell^{m*}(\theta,\varphi) B_r(\theta,\varphi)\right|_{r = R_\odot}.
\end{equation}
For the scalar potential, one finds
\begin{equation}
  \phi(r,\theta,\varphi) = -\sum_{\ell=0}^\infty \sum_{m=-\ell}^\ell \left( r^\ell - \frac{R_\mathrm{S}^{2\ell+1}}{r^{\ell+1}} \right) \frac{C_\ell^m}{\ell R_\odot^{\ell-1} + (\ell+1) R_\mathrm{S}^{2\ell+1}/R_\odot^{\ell+2}} Y_\ell^m(\theta,\varphi)
\end{equation}
and for the magnetic field $\vek B = -\partial_r \phi(\vek r)\,\hat{\vek e}_r - (\nicefrac 1 r) \partial_\theta \phi\,\hat{\vek e}_\theta - (\nicefrac 1 {r \sin \theta}) \partial_\varphi \phi\,\hat{\vek e}_\varphi$ accordingly
\begin{equation}\label{eqn:spherical-harmonics-Br}
  B_r(r,\theta,\varphi) = \sum_{\ell=0}^\infty \sum_{m=-\ell}^\ell \left( \frac{R_\odot}{r} \right)^{\ell+2} \frac{\ell+1+\ell \left( \nicefrac r {R_\mathrm{S}} \right)^{2\ell+1}}{\ell+1+\ell \left( \nicefrac{R_\odot}{R_\mathrm{S}} \right)^{2\ell+1}} C_l^m Y_\ell^m(\theta,\varphi),
\end{equation}
\begin{equation}\label{eqn:spherical-harmonics-Btheta}
  B_\theta(r,\theta,\varphi) = \sum_{\ell=0}^\infty \sum_{m=-\ell}^\ell \left( \frac{R_\odot}{r} \right)^{\ell+2} \frac{\left( \nicefrac r {R_\odot} \right)^{2\ell+1}-1}{\ell \left( \nicefrac{R_\odot}{R_\mathrm{S}} \right)^{2\ell+1} + \ell + 1} C_\ell^m \pdq{Y_\ell^m(\theta,\varphi)}{\theta},
\end{equation}
\begin{equation}\label{eqn:spherical-harmonics-Bphi}
  B_\varphi(r,\theta,\varphi) = - \frac{1}{r \sin \theta} \pdq{\phi}{\theta} = \sum_{\ell=0}^\infty \sum_{m=-\ell}^\ell \frac{\mathrm{i} m}{\sin \theta} \left( \frac{R_\odot}{r} \right)^{\ell+2} \frac{\left( \nicefrac r {R_\odot} \right)^{2\ell+1}-1}{\ell \left( \nicefrac{R_\odot}{R_\mathrm{S}} \right)^{2\ell+1} + \ell + 1} C_\ell^m Y_\ell^m(\theta,\varphi).
\end{equation}
Another commonly used PFSS model implementation distributed by SolarSoft, offering further features such as field-line tracing, visualization and coronal hole detection has been made available for the Interactive Data Language (IDL) by \citep{Schrijver2003}.

\section{Sensitivity to numerical parameters}
\label{app:num}

In this appendix, we want to discuss a bit further the sensitivity of our two coronal models to numerical parameters. Indeed, we have demonstrated that numerical implementation can have a strong impact of the final wind solution, but we wish to stress out that it does not affect the overall stability of the models. We demonstrate this for the most simple case which is the magnetic dipole.

\subsection{Sensibility of COCONUT to divergence cleaning method}
\label{app:num_cf_ch}

\begin{figure}
    \centering
    \gridline{\fig{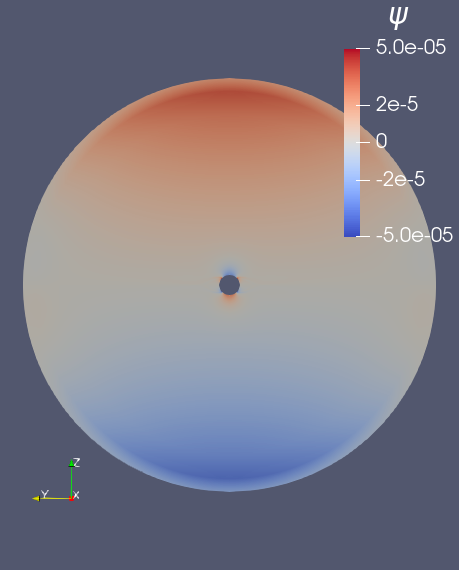}{0.24\textwidth}{(a)}
              \fig{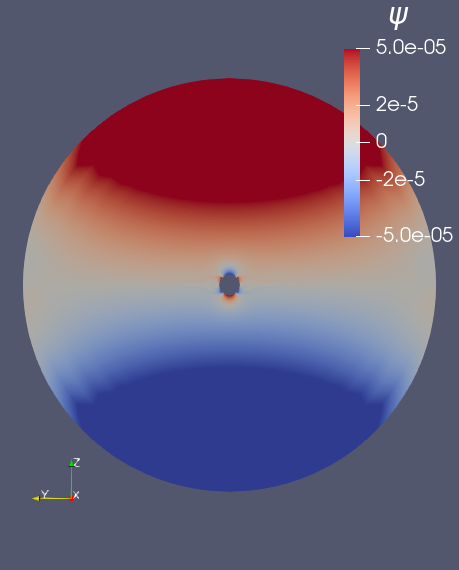}{0.24\textwidth}{(b)}
              \fig{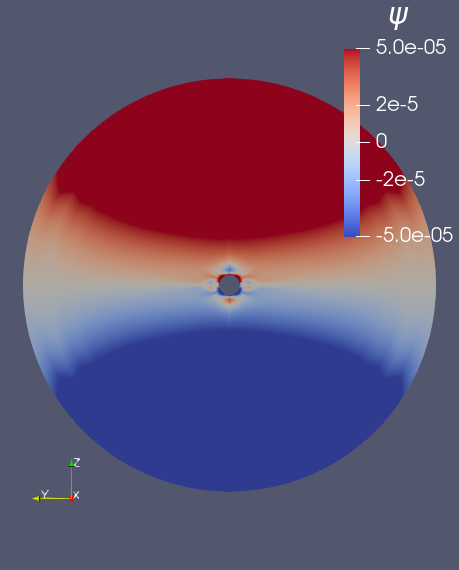}{0.24\textwidth}{(c)}
              \fig{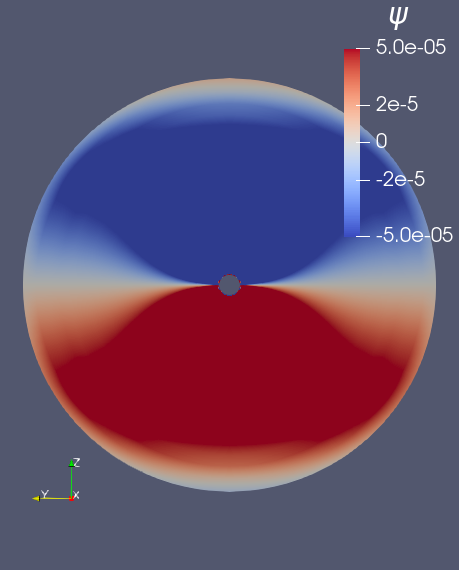}{0.24\textwidth}{(d}}
    \gridline{\fig{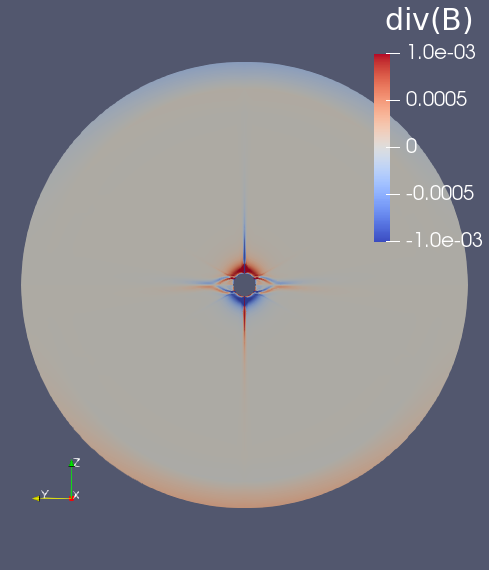}{0.24\textwidth}{(e)}
              \fig{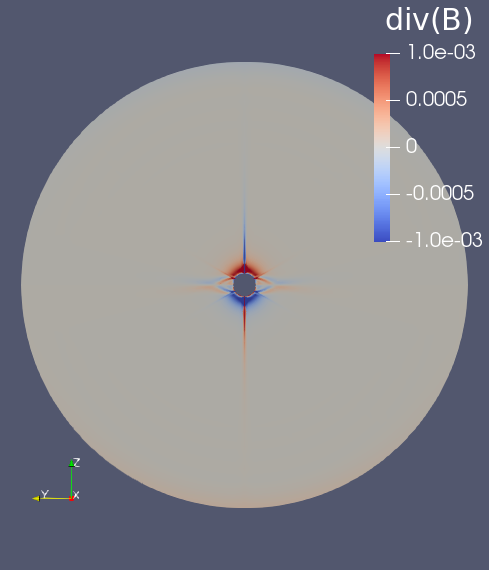}{0.24\textwidth}{(f)}
              \fig{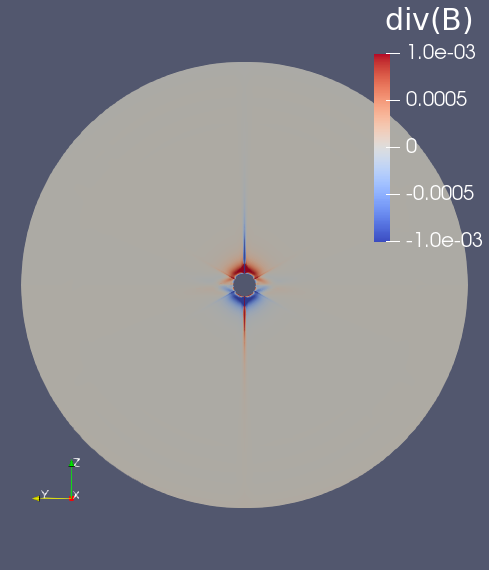}{0.24\textwidth}{(g)}
              \fig{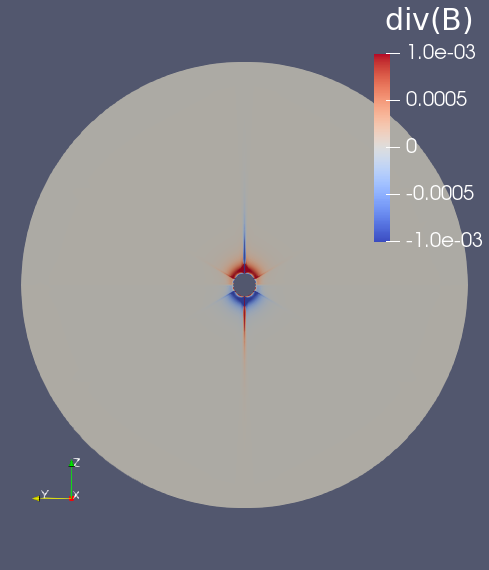}{0.24\textwidth}{(h}}
    \gridline{\fig{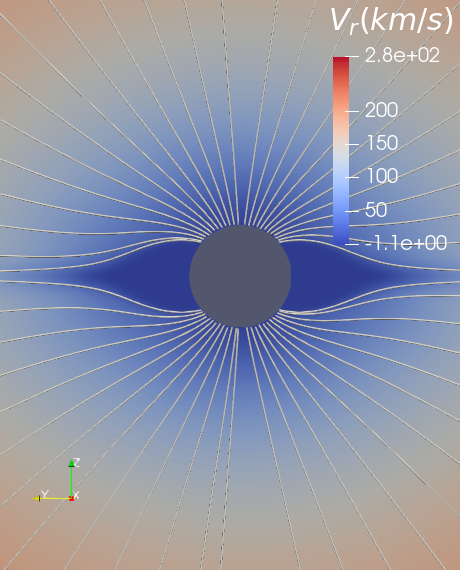}{0.24\textwidth}{(i)}
              \fig{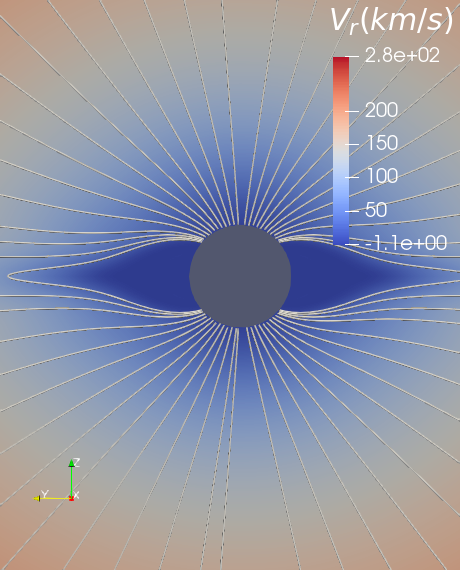}{0.24\textwidth}{(j)}
              \fig{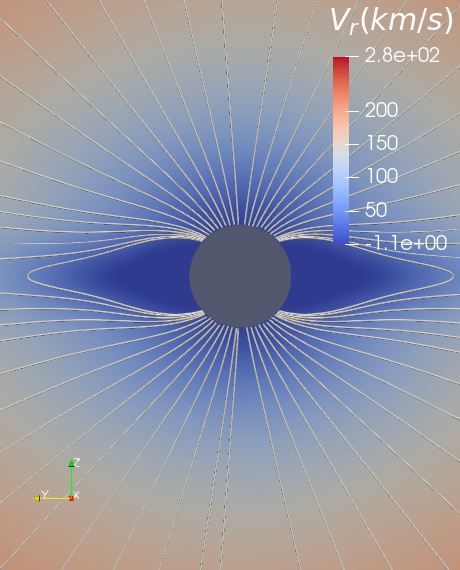}{0.24\textwidth}{(k)}
              \fig{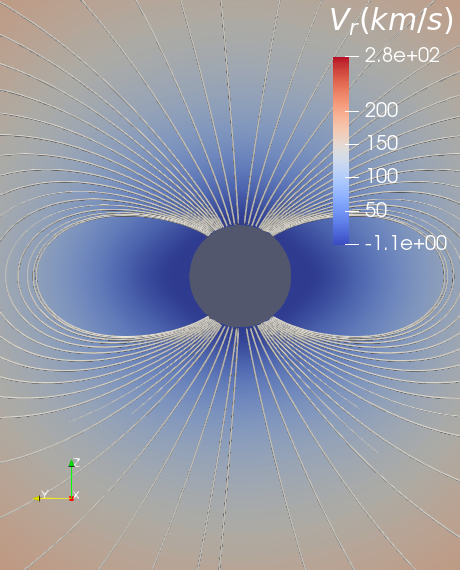}{0.24\textwidth}{(l)}}
    \caption{Impact of the value of the parameter $c_h$ on the efficiency of the divergence cleaning method for the dipole case. The first row shows the Lagrangian multiplier $\psi$ for $c_h=0.1$ (a), $c_h=1$ (b), $c_h=10$ (c), $c_h=100$ (d), with a colorscale adjusted between $-5.0 \ 10^{-5}$ and $5 \ 10^{-5}$. The second row shows the divergence of the magnetic field for $c_h=0.1$ (e), $c_h=1$ (f), $c_h=10$ (g), $c_h=100$ (h). The third row shows the corresponding radial velocity in km/s (in colors) and magnetic field lines (in white) for $c_h=0.1$ (i), $c_h=1$ (j), $c_h=10$ (k), $c_h=100$ (l).}
    \label{fig:ch_cf}
\end{figure}

As explained in section \ref{sec:FVM}, we have a free parameter for COCONUT for the divergence cleaning method, which is the speed of propagation of errors $c_h$. This parameter is fixed for Wind-Predict, with a value which depends on the CFL criteria and the current time step. We have estimated it at around 0.3 for simple simulations (dipole, quadrupole) and 1 for more complex data-driven cases. We want to demonstrate here that the value of the parameter $c_h$ can affect slightly the shape of the streamers, but not the overall divergence cleaning, if contained within the correct range of values. For this, we have tried several values, from 0.1, to 1 (the one used in the paper), 10 and 100. The results are displayed in figure \ref{fig:ch_cf}. The first row shows the Lagrangian multiplier $\psi$ output by the simulation. If the divergence was perfectly zero, $\psi$ would be as well, so regions of high $\psi$ show where the deviation from zero-divergence is the strongest. To better compare the plots, the colorscale is set between $-5 \times 10^{-5}$ and $5 \times 10^{-5}$ for all figures of the first row. As $c_h$ is increasing, $\psi$ is increasing as well. The regions with the highest values are located near the solar surface (as expected due to the intense magnetic field) but also at the outer boundary conditions near the poles. To explore this further, the second row shows the divergence of the magnetic field (with the color table fixed within -0.001 and 0.001), while the third row shows the corresponding radial velocity in km/s (in colors) and magnetic field lines (in white). For all models, the divergence is mostly zero, although there is a 0.1\% error close to the star surface. We can see that when the parameter $c_h$ is too small (at 0.1, see (a)), the divergence cleaning errors are not propagated fast enough out of the domain, thus resulting in divergence errors at the outer boundary. As we increase $c_h$ (see (b), (c), (d)), other errors start to disappear, such as those close to the edges of the current sheet. However, we can see on the second row that when we increase $c_h$, we also prevent the proper opening of the streamers. This results in a dipolar shape for case $c_h=100$ (see (h)) (which was chosen intentionally to be extreme), but also edges that are less sharp in case $c_h=10$ (see (g). Thus, the chosen value of $c_h=1$ (see (b) and (f)) is the best compromise between accuracy for the divergence and the shape of the streamers.

\subsection{Sensibility of Wind-Predict to limiters}
\label{app:num_wp_lim}

\begin{figure}
    \centering
    \gridline{\fig{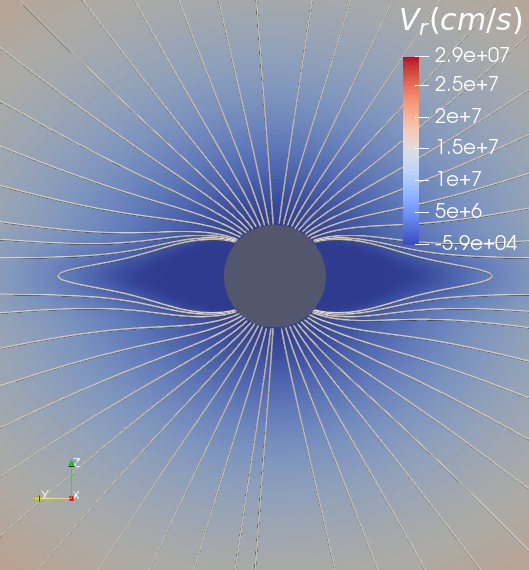}{0.24\textwidth}{(a)}
              \fig{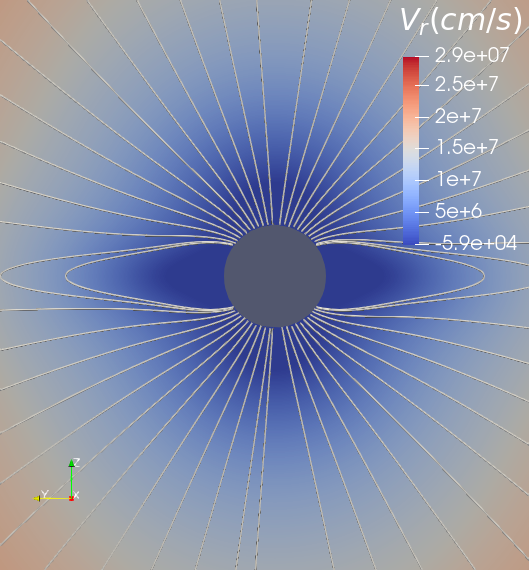}{0.24\textwidth}{(b)}
              \fig{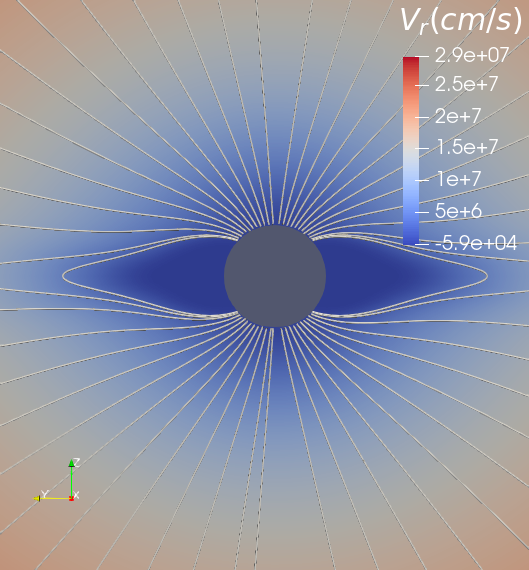}{0.24\textwidth}{(c)}
              \fig{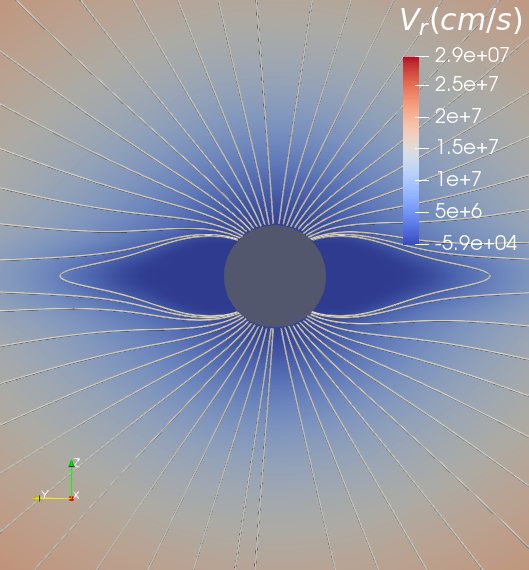}{0.24\textwidth}{(d}}
    \gridline{\fig{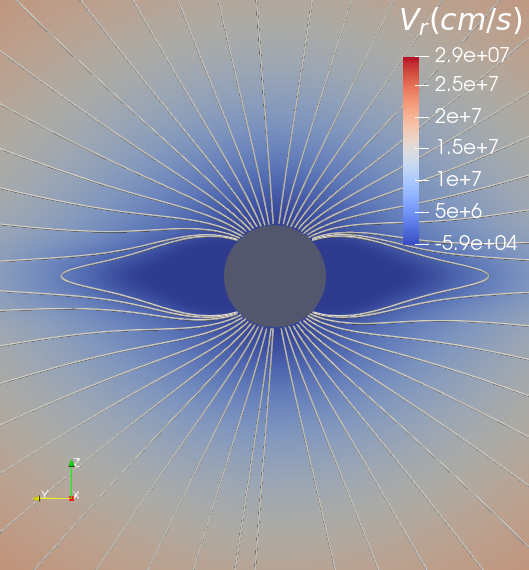}{0.24\textwidth}{(e)}
              \fig{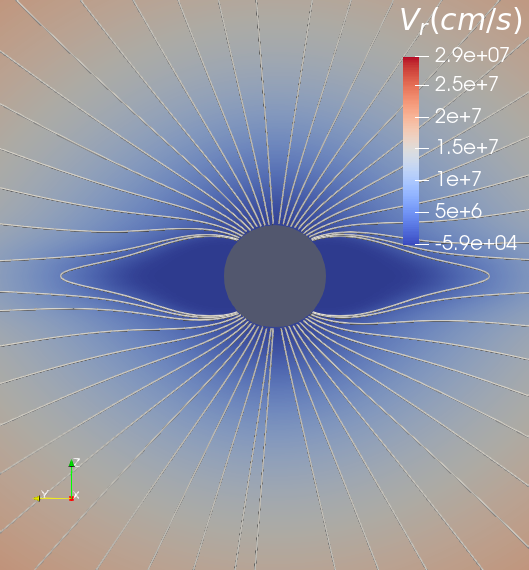}{0.24\textwidth}{(f)}
              \fig{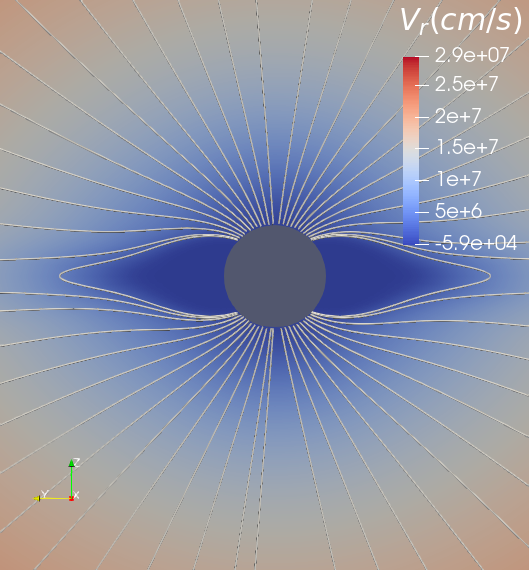}{0.24\textwidth}{(g)}}
    \caption{Impact of the choice of the limiter for the dipole case in Wind-Predict. We have explored all the available limiters in version 4.2 of the PLUTO code \citep{Mignone2007}, which are: monotonized central difference limiter (a), 1st order reconstruction (b), minmod limiter (c), OSPRE limiter (d), umist limiter (e), van Albada limiter function (f) and harmonic mean limiter of van Leer (g). For each plot, we show the radial velocity $V_r$ in cm/s (in colors) and the magnetic field lines (in white).}
    \label{fig:lim_wp}
\end{figure}

Here we want to discuss the impact of the choice of the limiter on the final solution from Wind-Predict. Wind-Predict is based on the PLUTO code, which offers 7 different options for the spatial reconstruction (6 different limiters): monotonized central difference limiter (a) (which is the one we have used for the results in this paper), 1st order reconstruction (b), minmod limiter (c), OSPRE limiter (d), umist limiter (e), van Albada limiter function (f) and harmonic mean limiter of van Leer (g). For each plot, we show the radial velocity $V_r$ in cm/s (in colors) and the magnetic field lines (in white). We can see that the 1st order reconstruction (see (b)) is the one with the most differences, because the other limiters are 2nd order reconstructions. Other limiters have a good agreement, although we can see differences in the position of the edges of the streamers (because of their numerical diffusivity). This also results in differences in the end velocity (which is the maximum velocity) with a difference up to 7\%. This is why the limiter can affect the comparison of our simulations. Unfortunately it was impossible to remove completely the limiter for Wind-Predict because of the initial transient, so in these conditions we went for the least diffusive limiter which is shown in figure (a) (monotonized central difference limiter).

\subsection{Sensitivity of both COCONUT and Wind-Predict to $\ell_\mathrm{max}$}
\label{app:lmax}

\begin{figure}[!t]
    \centering
    \gridline{\fig{eclipse_obs}{0.32\textwidth}{(a)}
              \fig{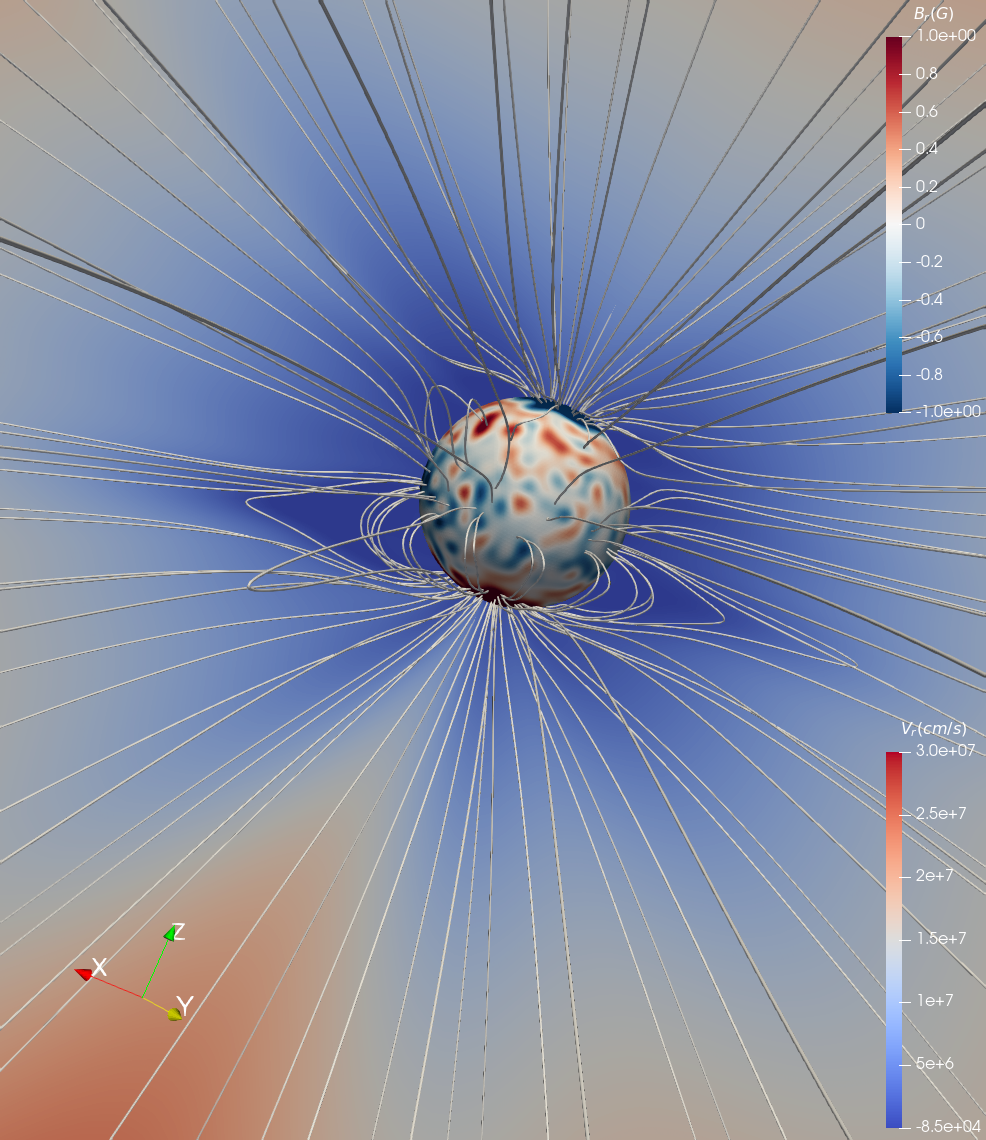}{0.33\textwidth}{(b)}
              \fig{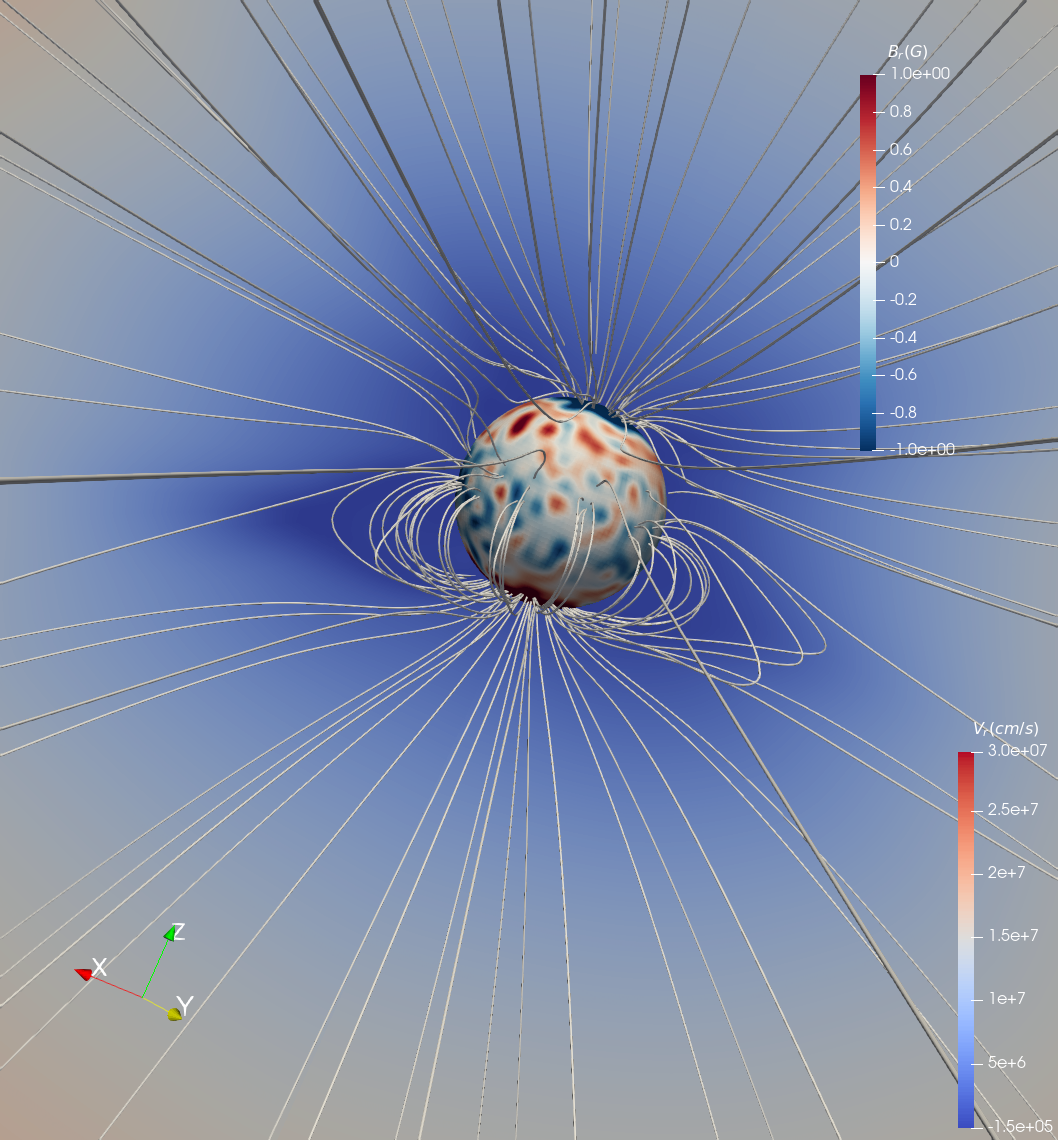}{0.355\textwidth}{(c)}}
    \gridline{\fig{ch_detection}{0.28\textwidth}{(d)}
              \fig{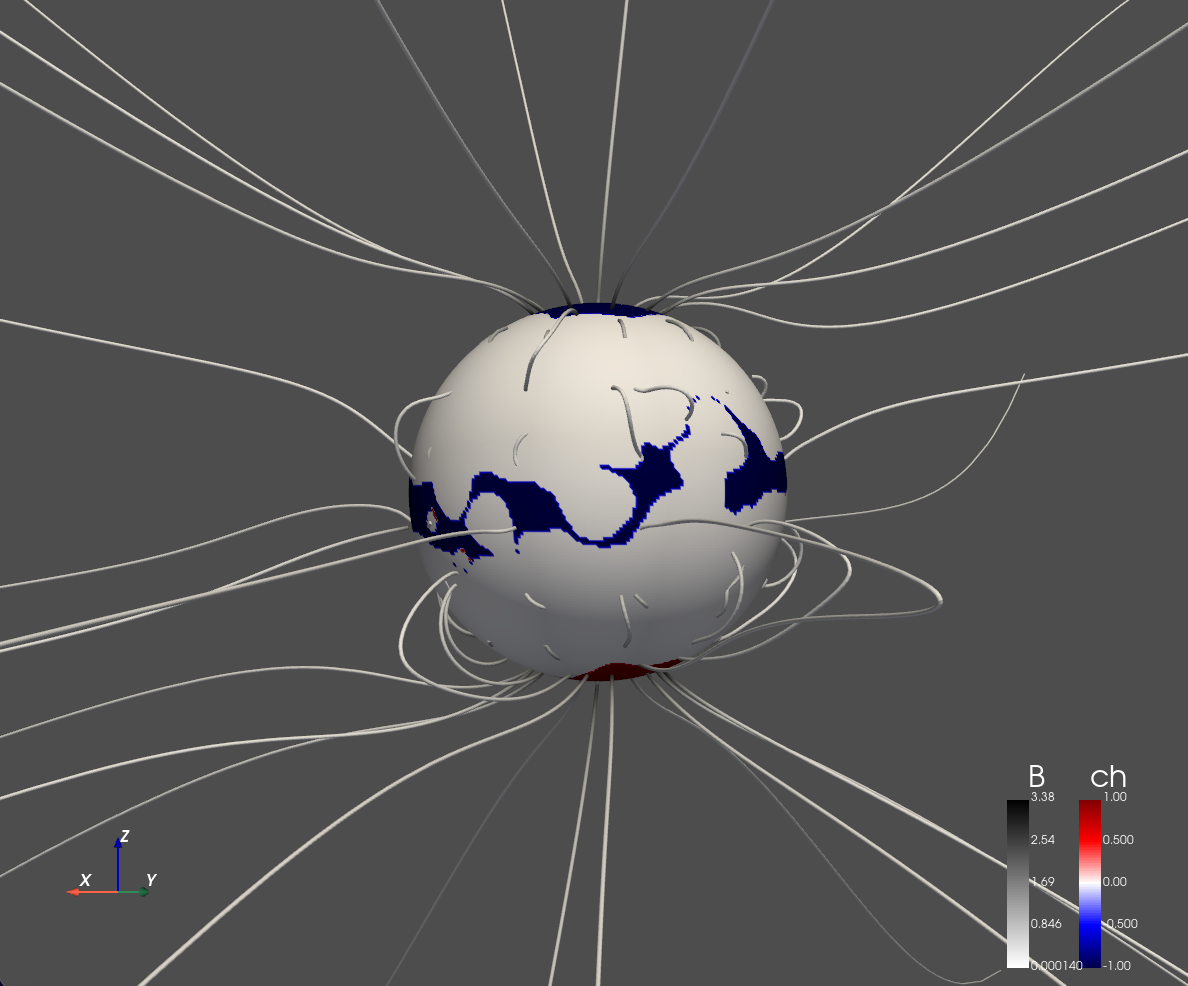}{0.33\textwidth}{(e)}
              \fig{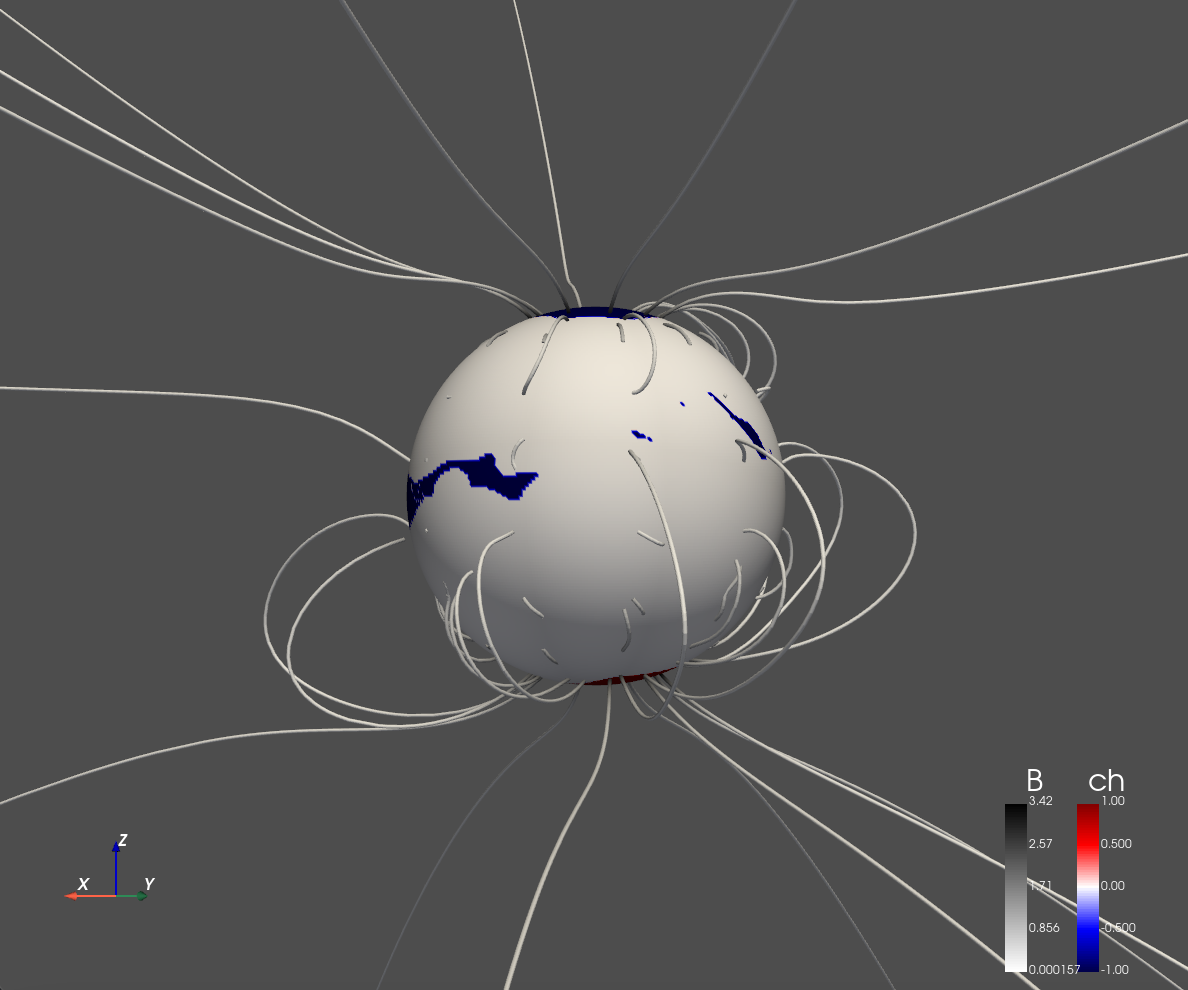}{0.33\textwidth}{(f)}}
    \gridline{\fig{euhforia_mag_wsa}{0.33\textwidth}{(g)}
              \fig{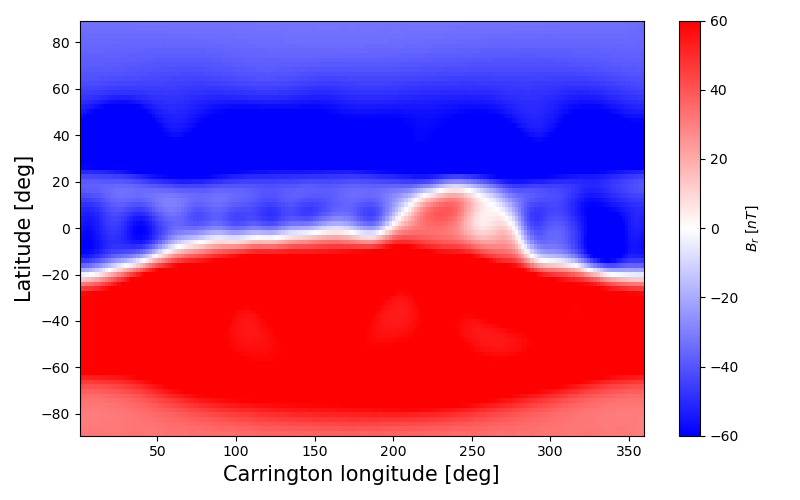}{0.33\textwidth}{(h)}
              \fig{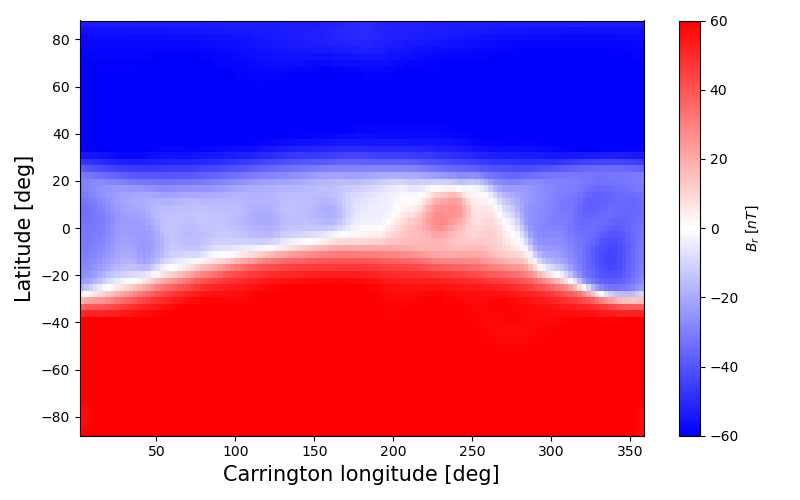}{0.33\textwidth}{(i)}}
    \caption{Comparison with observations for the case $\ell_\mathrm{max}=30$ for the GONG synoptic map of the 1st of August 2008. The first column presents the observations of that day in white-light (a), EUV (d) and the WSA model at 0.1 AU (as seen in previous figures). The second column (figures (b), (e) and (h)) shows the corresponding results for COCONUT. The third column (figures (c), (f) and (i)) shows the corresponding results for Wind-Predict. More information about the figures details can be found in the description of figures \ref{fig:wl}, \ref{fig:ch} and \ref{fig:euhforia_bc}.}
    \label{fig:lmax30_obs}
\end{figure}

As discussed in the run-time benchmark, we have tested the capability of our new coronal model to handle more spatially resolved structures. As we are studying only a minimum of activity in this study, the impact of $\ell_{max}$ is more limited than it would be for a maximum of activity, but we still want to discuss it briefly in this appendix. Figure \ref{fig:lmax30_obs} shows the comparison of COCONUT and Wind-Predict with the same observations as before for the $\ell_{max}=30$ case. We can see in figures (b) and (c) that the streamers structure does not appear to change much, as it is mostly determined by lower modes. We can see however that the velocity changes, with a faster stream appearing at the southern pole for COCONUT. The impact is more visible for the coronal hole boundaries shown in figures (e) and (f). The two codes have an opposite behavior, which is probably linked to their different boundary conditions: the equatorial coronal holes are more connected in COCONUT, while they mostly disappear in Wind-Predict. This is consistent with the streamers seen above, where the distinction between the two streamers on the right is clearly visible in COCONUT, while the southern streamer takes over in Wind-Predict. 
Finally, the magnetic field to be coupled with EUHFORIA at 0.1 AU is not affected much by the change in $\ell_{max}$. The HCS is more detailed because of the increase of resolution in the magnetic structures described, as to be expected, but its global position remains the same. All these results show that the final wind solution's global features are not affected by the pre-processing of the input magnetic map. The increase in $\ell_{max}$ can however improve the description of more complex structures, such as the shape of detailed coronal holes. These results tend to be true at minimum of activity, but it is very likely that this conclusion will not hold up at maximum of activity. This will be explored more thoroughly in a follow-up paper.

\bibliography{corona}{}
\bibliographystyle{aasjournal}

\end{document}